\documentclass[aps,prb,reprint,showpacs,floatfix,superscriptaddress]{revtex4-1}

\usepackage{graphicx}
\usepackage{amsmath,amssymb,amsfonts}
\usepackage{multirow}
\usepackage[bookmarks=false]{hyperref}
\usepackage[flushleft]{threeparttable}
\hypersetup{
	breaklinks=true,
    unicode=false,          
    pdftoolbar=true,        
    pdfmenubar=true,        
    pdffitwindow=false,     
    pdfstartview={FitH},    
    pdftitle={Ce-Co-Cu-Fe-Ta},    
    pdfauthor={Taufour},     
    pdfsubject={CeCoMg},   
    pdfcreator={Taufour},   
    pdfproducer={Onyszczak}, 
    pdfkeywords={keyword1} {key2} {key3}, 
    pdfnewwindow=true,      
    colorlinks=true,       	
    linkcolor=black,          
    citecolor=black,    	    
    filecolor=black,   		    
    urlcolor=blue            
}

\begin{document}

\title{Single Crystal Permanent Magnet: Extraordinary Magnetic Behavior in the Ta,  Cu and Fe Substituted CeCo$_{5}$ Systems.}

\author{Tej N. Lamichhane}
\affiliation{The Ames Laboratory, U.S. Department of Energy, Iowa State University, Ames, Iowa 50011, USA}
\affiliation{Department of Physics and Astronomy, Iowa State University, Ames, Iowa 50011, USA}

\author{Michael T. Onyszczak}
\affiliation{Department of Physics and Astronomy, Iowa State University, Ames, Iowa 50011, USA}

\author{Olena Palasyuk}
\affiliation{The Ames Laboratory, U.S. Department of Energy, Iowa State University, Ames, Iowa 50011, USA}
\affiliation{Department of Material Science and Engineering, Iowa State University, Ames, Iowa 50011, USA}

\author{Saba Sharikadze}
\affiliation{Department of Physics and Astronomy, Iowa State University, Ames, Iowa 50011, USA}

\author{Tae-Hoon Kim}
\affiliation{The Ames Laboratory, U.S. Department of Energy, Iowa State University, Ames, Iowa 50011, USA}

\author{Matthew J. Kramer}
\affiliation{The Ames Laboratory, U.S. Department of Energy, Iowa State University, Ames, Iowa 50011, USA}
\affiliation{Department of Material Science and Engineering, Iowa State University, Ames, Iowa 50011, USA}

\author{R.W. McCallum}
\affiliation{The Ames Laboratory, U.S. Department of Energy, Iowa State University, Ames, Iowa 50011, USA}

\author{Aleksander L. Wysocki}
\affiliation{The Ames Laboratory, U.S. Department of Energy, Iowa State University, Ames, Iowa 50011, USA}

\author{Manh Cuong Nguyen}
\affiliation{The Ames Laboratory, U.S. Department of Energy, Iowa State University, Ames, Iowa 50011, USA}

\author{Vladimir P. Antropov}
\affiliation{The Ames Laboratory, U.S. Department of Energy, Iowa State University, Ames, Iowa 50011, USA}

\author{Tribhuvan Pandey}
\author{David Parker}
\affiliation{Oak Ridge National Laboratory, Oak Ridge, TN 37831}

\author{Sergey L. Bud'ko}
\author{Paul C. Canfield}
\affiliation{The Ames Laboratory, U.S. Department of Energy, Iowa State University, Ames, Iowa 50011, USA}
\affiliation{Department of Physics and Astronomy, Iowa State University, Ames, Iowa 50011, USA}

\author{Andriy Palasyuk}
\affiliation{The Ames Laboratory, U.S. Department of Energy, Iowa State University, Ames, Iowa 50011, USA}

\date{\today}

\begin{abstract}

To reduce material and processing costs of commercial permanent magnets and to attempt to fill the empty niche of energy products, 10 -- 20 MGOe, between low-flux (ferrites, alnico) and high-flux (Nd$_2$Fe$_{14}$B- and SmCo$_5$-type) magnets, we report synthesis, structure, magnetic properties and modeling of Ta, Cu and Fe substituted CeCo$_{5}$. Using a self-flux technique, we grew single crystals of $\textbf{I}$ -- Ce$_{15.1}$Ta$_{1.0}$Co$_{74.4}$Cu$_{9.5}$, $\textbf{II}$ -- Ce$_{16.3}$Ta$_{0.6}$Co$_{68.9}$Cu$_{14.2}$, $\textbf{III}$ -- Ce$_{15.7}$Ta$_{0.6}$Co$_{67.8}$Cu$_{15.9}$, $\textbf{IV}$ -- Ce$_{16.3}$Ta$_{0.3}$Co$_{61.7}$Cu$_{21.7}$ and $\textbf {V}$ -- Ce$_{14.3}$Ta$_{1.0}$Co$_{62.0}$Fe$_{12.3}$Cu$_{10.4}$. X-ray diffraction analysis (XRD) showed that these materials retain a CaCu$_5$ substructure and incorporate small amounts of Ta in the form of ``dumb-bells", filling the 2$\textit{e}$ crystallographic sites within the 1D hexagonal channel with the 1$\textit{a}$ Ce site, whereas Co, Cu and Fe are statistically distributed among the 2$\textit{c}$ and 3$\textit{g}$ crystallographic sites.  Scanning electron microscopy, energy dispersive X-ray spectroscopy (SEM-EDS) and scanning transmission electron microscopy (STEM) examinations provided strong evidence of the single-phase nature of the as-grown crystals, even though they readily exhibited significant magnetic coercivities of $\sim$1.6 -- $\sim$1.8 kOe caused by Co-enriched, nano-sized, structural defects and faults that can serve as pinning sites. Heat treatments at 1040 $^\circ$C for 10 h and a hardening at 400 $^\circ$C for 4 h lead to the formation of a so-called ``composite crystal" with a bimodal microstructure that consists of a Ta-poor matrix and Ta-rich laminal precipitates. Formation of the ``composite crystal" during the heat treatment creates a 3D array of extended defects within a primarily single grain single crystal, which greatly improves its magnetic characteristics. Possible causes of the formation of the ``composite crystal" may be associated with Ta atoms leaving matrix interstices at lower temperatures and/or matrix degradation induced by decreased miscibility at lower temperatures. Fe strongly improves both the Curie temperature and magnetization of the system resulting in ($\textit{BH}$)$_{max.}$$\approx$13 MGOe at room temperature.	
\end{abstract}

\maketitle
\section{Introduction}
\begin{figure*}[t]
	\includegraphics[width=15.5cm,height=8.0cm]{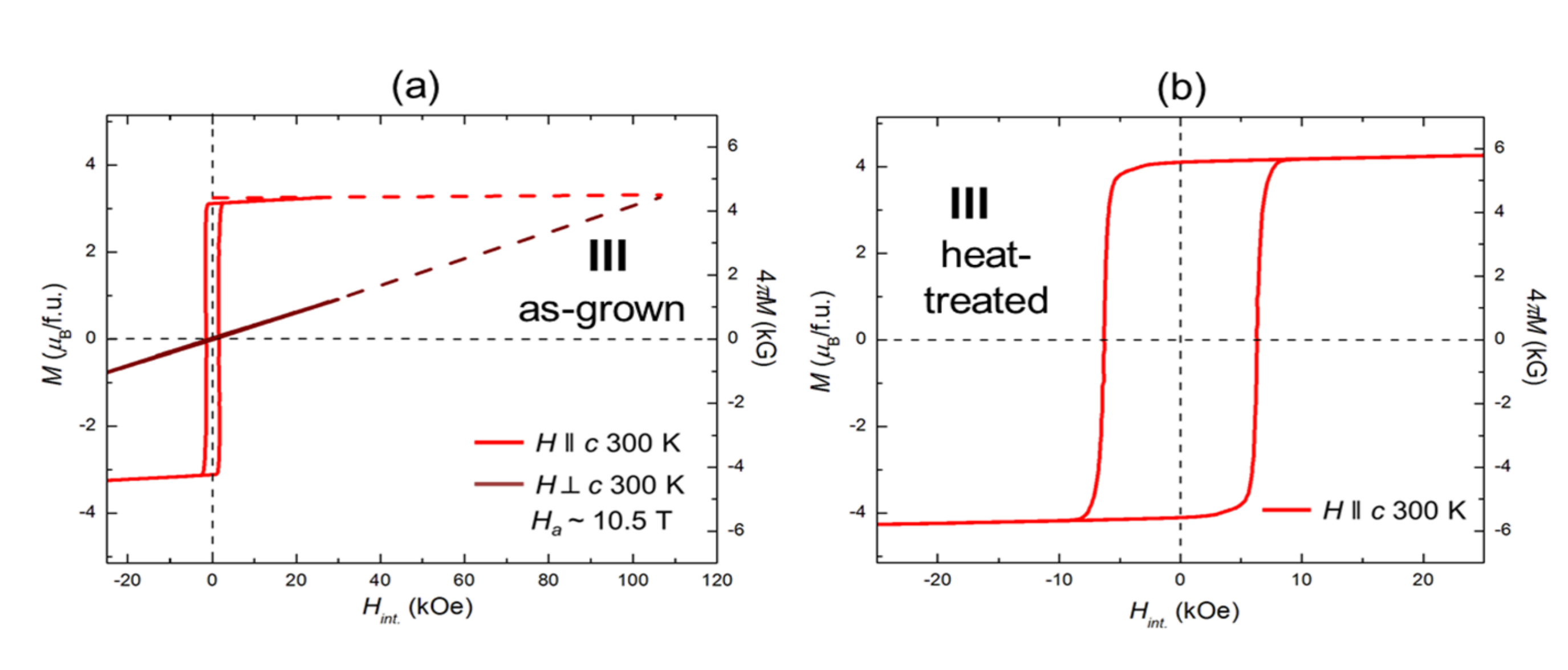}
	\caption{(a) -- anisotropic field dependent magnetization of the as-grown Ce$_{15.7}$Ta$_{0.6}$Co$_{67.8}$Cu$_{15.9}$ (sample $\textbf{III}$) at 300 K for applied field along and perpendicular to the crystallographic axis. The inferred anisotropy field, $\textit{H$_{a}$}$, is also shown. (b) -- after heat treatment, the magnetic hysteresis loop of the same sample along the easy magnetization axis [001].}
	\label{Motivation}
\end{figure*}

To find new economical alternatives to commercial, high flux, permanent magnets, we focused on the Ta, Cu, and Fe substituted CeCo$_{5}$ system which, unlike typical commercial magnetic grades with critical rare earths (Nd, Sm, Dy, etc.) utilizes widely available and more affordable Ce. \cite{1Haxel,2Patnaik2002,3Xie2014} The incorporation of Ce into magnets, instead of critical elements, may significantly reduce the price and supply-chain dependence of commercial magnets. Despite the mixed Ce$^{3+}$/Ce$^{4+}$ valency problem, typically adverse for the magneto-crystalline anisotropy, there are recent experimental efforts on the Nd$_{2}$Fe$_{14}$B (2:14:1) system showing that Ce-substitutes can compete with commercial high-flux grades at lower material costs. \cite{Herbst1985,4Pathak,5Pathak,Susner2017} Similarly, efforts on the Ce-containing SmCo$_{5}$ (1:5) and Sm$_{2}$Co$_{17}$ (2:17) systems showed that satisfactory cost-to-performance balances suitable for modern rare earth criticalities and market demands are expected.\cite{Strnat1991,Senno1974,6Kramer,7McCallum} Therefore, we believe that Cu and Fe substituted CeCo$_{5}$ systems require a new and deeper examination.\cite{Nesbitt1969,8Nesbitt,9Tawara} After being understood and optimized these Ce-based systems may compete on both material-processing-cost and properties levels as so-called ``gap magnets", performing in the gap of magnetic energies, between  10 -- 20 MGOe, which currently exists between the rare-earth-free alnico and ferrite grades and the sintered 1:5 and 2:14:1 magnetic grades which contain critical rare earths.\cite{Coey2012}

Despite previous extensive explorations, the intrinsic properties of the CeCo$_{5}$ system has not been fully or systematically established,\cite{,8Nesbitt,9Tawara,10Senno,11Nesbitt,12Cullen,13Sherwood,14Tawara,15Chin,16Nesbitt,17Leamy,18Nesbitt,19Khan,20Arbuzov1974,21Arbuzov1975,22Arbuzov1975,23Arbuzov1977,24Arbuzov1977,25Arbuzov1977,26Inomata,27Tawara,28Labulle,29Labulle,30Girodin}   and the metallurgy related to the magnetic pinning/coercivity mechanism is not fully understood. Although anisotropy characterization is best obtained from single crystals, single crystal growth reports in Cu or Fe substituted CeCo$_{5}$ systems are scarce and limited to several Bridgman type attempts  \cite{15Chin,16Nesbitt}  in the vicinity of the composition $\sim$CeCo$_{3.5}$Fe$_{0.5}$Cu.\cite{31ASMFCeCo} 

In this paper we  report the successful self-flux growth \cite{32CanfieldFisk,33Canfieldbook} of five representatives of Ta, Cu and Fe substituted CeCo$_{5}$ followed by characterization of their structural and magnetic properties. We study the phenomenon of pronounced magnetic coercivity in the ``as-grown'' crystals and its further development during the heat treatment as illustrated in [Fig.~\ref{Motivation}]. Sub-grain phase segregation creates the necessary conditions for magnetic domain pinning. We also discuss the possible ways to improve, manipulate, and control the system in an attempt to increase its magnetic characteristics in conjunction with first principles DFT calculations and multiscale modeling.  

\section{Experimental}
\begin{figure}[!h]
	\includegraphics[scale =0.32]{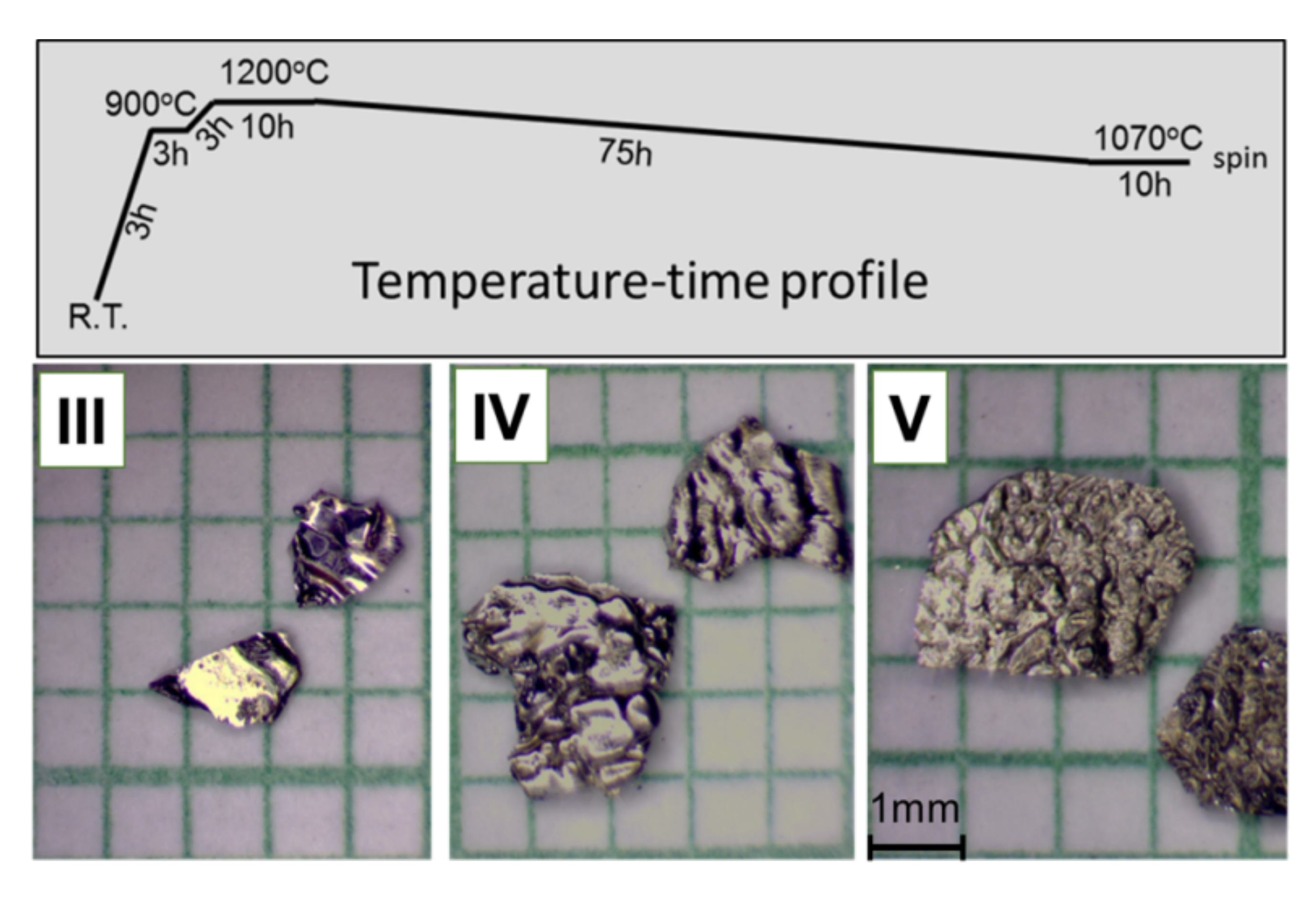}
	\caption{Temperature program used for single crystal growths and general look of self-flux grown plate-like crystals of $\textbf{III}$ -- Ce$_{15.7}$Ta$_{0.6}$Co$_{67.8}$Cu$_{15.9}$, $\textbf{IV}$ -- Ce$_{16.3}$Ta$_{0.3}$Co$_{61.7}$Cu$_{21.7}$ and $\textbf{V}$ -- Ce$_{14.3}$Ta$_{1.0}$Co$_{62.0}$Fe$_{12.3}$Cu$_{10.4}$ (for details see Table I).}
	\label{FIG1}
\end{figure}

 \textbf{A}. \textbf{Synthesis}
 Single crystals were grown via the solution growth method  \cite{32CanfieldFisk,33Canfieldbook}. The reaction metals (Ce (99.99\%), Cu (99.95\%) from Ames Laboratory MPC and Co (99.95\%) from Alfa Aesar) were placed into 3-capped Ta containers\cite{Canfield2001JCG} welded under an Ar atmosphere, which then were sealed into fused silica tubes and placed into a high-temperature box furnace. The furnace was heated from near room temperature to 900 $^{\circ}$C over 3 hours, held at 900 $^{\circ}$C for 3 hours, heated to 1200 $^{\circ}$C over 3 more hours, and held at 1200 $^{\circ}$C for 10 hours. The furance was then cooled to 1070 $^{\circ}$C over 75 hours. At 1070 $^{\circ}$C the excess flux was decanted by centrifuging.\cite{32CanfieldFisk,33Canfieldbook} The exact temperature profile of the growths and pictures of the typical crystals are presented in Fig.~\ref{FIG1}.

\begin{table*}[]
	\centering
	\begin{threeparttable}
	\caption{Composition of single crystals and their lattice parameters as-grown and after the heat treatment}
	\label{Table 1}
	\begin{tabular}{|c|c|c|c|c|c|c|c|c|c|c|c|c|}
		\hline
		\multirow{3}{*}{\#} & \multicolumn{10}{c|}{EDS-composition, at. \%}                                                                                       & \multicolumn{2}{c|}{Lattice parameters}                                                                                                         \\ \cline{2-13} 
		& \multicolumn{2}{c|}{Ce} & \multicolumn{2}{c|}{Ta} & \multicolumn{2}{c|}{Co} & \multicolumn{2}{c|}{Fe} & \multicolumn{2}{c|}{Cu} & \multicolumn{2}{c|}{$\textit{a}$, $\textit{c}$, \AA; $\textit{V}$, \AA$^{3}$ }                                                                                                              \\ \cline{2-13} 
		& ag*        & ht**        & ag         & ht         & ag         & ht         & ag         & ht         & ag         & ht         & ag                                                                     & ht                                                                     \\ \hline
		\textbf{I}          & 15.1       & 16.1       & 1.0        & 0.6        & 74.4       & 73.6       & -          & -          & 9.5        & 9.8        & \begin{tabular}[c]{@{}c@{}}4.912(1)\\ 4.045(1)\\ 84.52(1)\end{tabular} & \begin{tabular}[c]{@{}c@{}}4.921(1)\\ 4.031(1)\\ 84.58(2)\end{tabular} \\ \hline
		\textbf{II}         & 16.3       & 16.2       & 0.6        & 0.4        & 68.9       & 69.4       & -          & -          & 14.2       & 14.0       & \begin{tabular}[c]{@{}c@{}}4.933(1)\\ 4.031(1)\\ 84.95(2)\end{tabular} & \begin{tabular}[c]{@{}c@{}}4.933(1)\\ 4.028(1)\\ 84.90(2)\end{tabular} \\ \hline
		\textbf{III}        & 15.7       & 15.8       & 0.6        & 0.1        & 67.8       & 67.1       & -          & -          & 15.9       & 17.1       & \begin{tabular}[c]{@{}c@{}}4.943(1)\\ 4.032(1)\\ 85.31(1)\end{tabular} & \begin{tabular}[c]{@{}c@{}}4.944(1)\\ 4.028(1)\\ 85.26(1)\end{tabular} \\ \hline
		\textbf{IV}         & 16.3       & 16.5       & 0.3        & 0.05       & 61.7       & 61.9       & -          & -          & 21.7       & 21.6       & \begin{tabular}[c]{@{}c@{}}4.950(1)\\ 4.033(1)\\ 85.57(2)\end{tabular} & \begin{tabular}[c]{@{}c@{}}4.954(1)\\ 4.028(1)\\ 85.61(2)\end{tabular} \\ \hline
		\textbf{V}          & 14.3       & 13.9       & 1.0        & 0.2        & 62.0       & 62.7       & 12.3       & 13.0       & 10.4       & 10.2       & \begin{tabular}[c]{@{}c@{}}4.922(1)\\ 4.075(1)\\ 85.50(2)\end{tabular} & \begin{tabular}[c]{@{}c@{}}4.924(1)\\ 4.071(1)\\ 85.48(2)\end{tabular} \\ \hline
	\end{tabular}
\begin{tablenotes}[1cm]
	\item  * - as grown, ** - heat-treated: 1040 $^\circ$C (10h) $\to$ [10 $^\circ$C/min] $\to$ 400 $^\circ$C  (8h)
\end{tablenotes}
\end{threeparttable}
\end{table*}

\textbf{B}. \textbf{Heat Treatment} 
After growth, some single crystals underwent identical, two-stage, heat treatments performed in a Dentsply Ceramico (Vulcan 3-Series) multi-stage programmable furnace, which included dwelling at 1040 $^\circ$C for 10 h, then cooling at a rate of 10~$^\circ$C/min to 400 $^\circ$C followed by dwelling at this temperature for next 8 h with a subsequent furnace cool to room temperature. We based this schedule on literature reports.  \cite{13Sherwood,15Chin,20Arbuzov1974,21Arbuzov1975,22Arbuzov1975,23Arbuzov1977,24Arbuzov1977} Different Cu contents may require slightly different temperature/time parameters for the best final magnetic characteristics, but the optimization of the heat treatment procedure is a subject of ongoing work. 

\textbf{C}. \textbf{Metallography and SEM/EDS Analysis} 
Samples for metallographic examination were placed in $\sim$1 inch diameter epoxy resin pucks, and polished with various grits of silicon carbide followed by a glycol-based, fine, polycrystalline, diamond suspension. Plate-like single crystals [Fig.  ~\ref{FIG1}] were mounted with their plates parallel to the polishing surface to allow for characterization along planes perpendicular to the crystals' [001] direction. Imaging studies  of single crystal samples were performed on an FEI Teneo field emission scanning electron microscope. Their compositions were determined via energy dispersive X-ray spectra obtained using an Oxford EDS/EBSD module averaging over 3-5 regions on their metallographicaly prepared surfaces [see Table ~\ref{Table 1}]. 

\textbf{D}. \textbf{TEM Characterization}
Cross sections from single crystal \textbf{III} were prepared using a dual-beam focused ion beam system (FEI Helios NanoLab G3 UC) with a lift-out approach. To reduce surface damage sustained during Ga ion milling, the final thinning and cleaning step were conducted using 5 kV and 2 kV for 5 min. The TEM analysis was performed on a Titan Themis (FEI) probe Cs-corrected TEM equipped with a Super-X EDS detector to characterize microstructure and elemental distribution.

\textbf{E}. \textbf{Powder and single crystal X-ray diffraction}
Polycrystalline powders were obtained by crushing the sample with an agate mortar and pestle. X-ray powder diffraction data were collected from the as-grown and heat-treated crystals. The measurements were performed using PANalytical X-Pert Pro (Co $\textit{K}$$_{\alpha}$ - radiation, $\lambda$ = 1.78897 \AA) and Bruker D8 Advance (Cu $\textit{K}$$_{\alpha}$ - radiation, $\lambda$ = 1.54056 \AA) diffraction systems.  Powdered samples were evenly dispersed on a zero-background Si-holder with the aid of a small quantity of vacuum grease. Diffraction scans were taken in the $\theta$/2$\theta$ mode with the following parameters: 2$\theta$ region: 20 -- 110$^\circ$, step scan: 0.02$^\circ$, counting time per step: 60 s. The FullProf Suite program package  \cite{35Rodriguez-Carvajal} was used for Rietveld refinement of the crystal structures. 

Single-crystal diffraction data were collected at room temperature using a Brucker SMART APEX II diffractometer (Mo $\textit{K}$$_{\alpha}$ - radiation ) equipped with a CCD area detector. Four sets of 360 frames with 0.5$^\circ$ scans in $\omega$ and exposure times of 10 -- 15 s per frame were collected. The reflection intensities were integrated using the SAINT subprogram in the SMART software package. \cite{36SMART} The space group was determined using the XPREP program and the SHELXTL 6.1 software package \cite{37SHELXTL}. Empirical absorption corrections were made using the SADABS program .\cite{38Blessing} Finally, each structure was solved by direct methods using SHELXTL 6.1 and refined by full-matrix least-squares on $\textit{F}$$_0$$^2$, with anisotropic thermal parameters and a secondary extinction parameter.

\textbf{F}. \textbf{Magnetic Properties Measurements}
Magnetic properties were obtained using a vibrating sample magnetometer in a cryogen-free VersaLab physical property measurement system (Quantum Design) with magnetic fields up to 3 T and temperatures in the 50 -- 350 K range using the standard option and 300 -- 1000 K range using the oven option. An alumina cement (Zircar) was used to hold the sample on the heater stick for the high-temperature measurements. The demagnetization factors are determined experimentally using the relation $\textit{H}$$_{int.}$ = $\textit{H}$ - $\textit{NM}$.\cite{Lamichhane2015,Lamichhane2018}

\section{Structure and Composition Analysis}
The Ce(Co$_{1-x-y}$Fe$_{x}$Cu$_{y}$)$_{5}$ system favors slightly Ce deficiency \cite{15Chin,16Nesbitt,17Leamy,18Nesbitt} and the appearance of transition metal, T, ``dumb-bells" may lead to structural transformations towards the 1:7 and 2:17 phases. Therefore, we use $\sim$1:5 designations for the general description of our reported systems.
\begin{center}
\textbf{A}. \textbf{SEM/EDS Examinations and Composition Analysis.}
\end{center}

\begin{figure*}[t]
	\includegraphics[width=15cm,height=9cm]{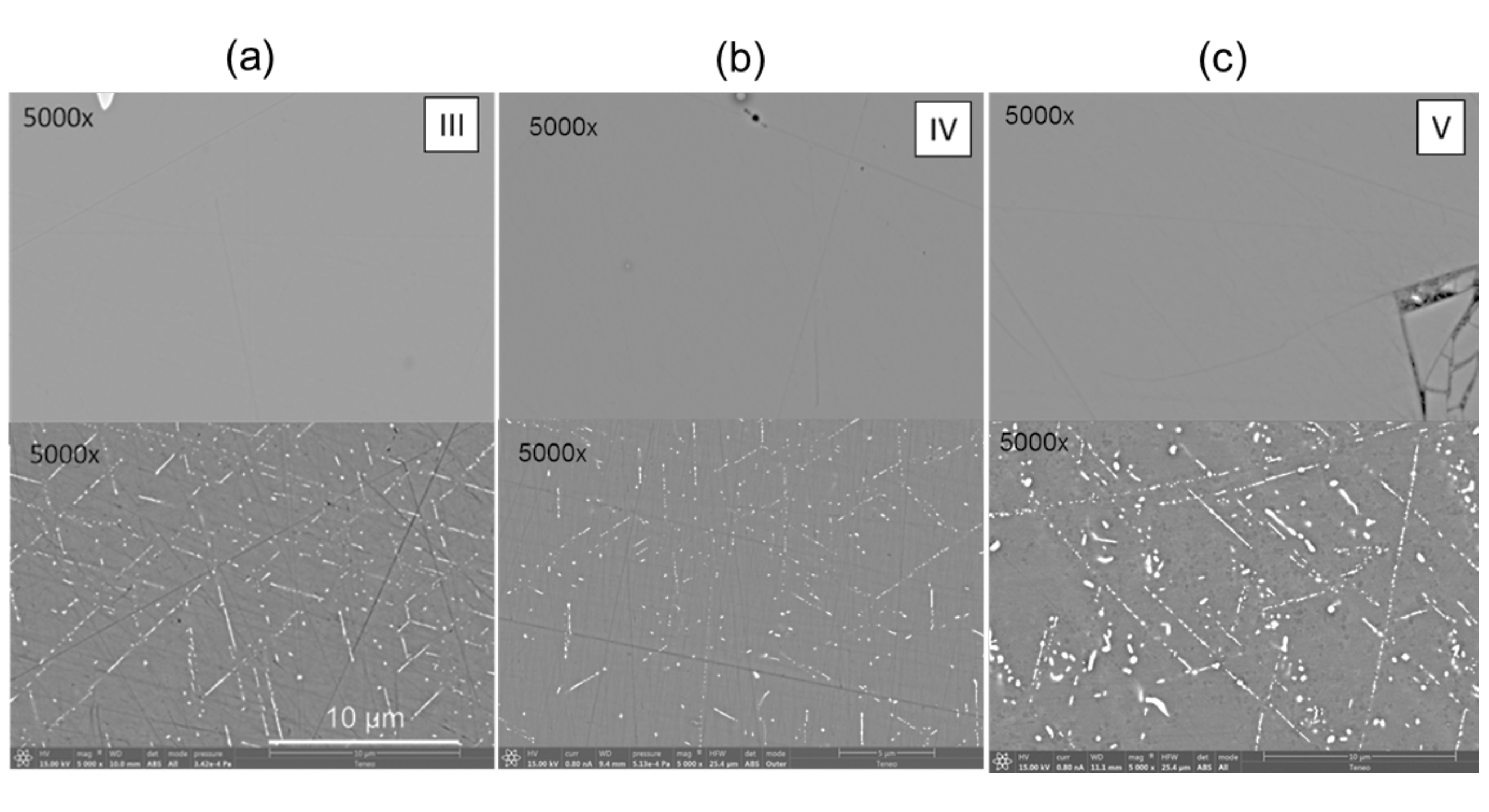}
	\caption{SEM backscattered electron images of samples (a) - $\textbf{III}$ -- Ce$_{15.7}$Ta$_{0.6}$Co$_{67.8}$Cu$_{15.9}$, (b) - $\textbf{IV}$ -- Ce$_{16.3}$Ta$_{0.3}$Co$_{61.7}$Cu$_{21.7}$ and (c) - $\textbf{V}$ -- Ce$_{14.3}$Ta$_{1.0}$Co$_{62.0}$Fe$_{12.3}$Cu$_{10.4}$ before (upper panels) and after (lower panels) heat treatment. All images were taken at a magnification 5000x and 15 kV.}
	\label{FIG2}
\end{figure*}

\begin{figure*}[!h]
	\includegraphics[width=15cm,height=10cm]{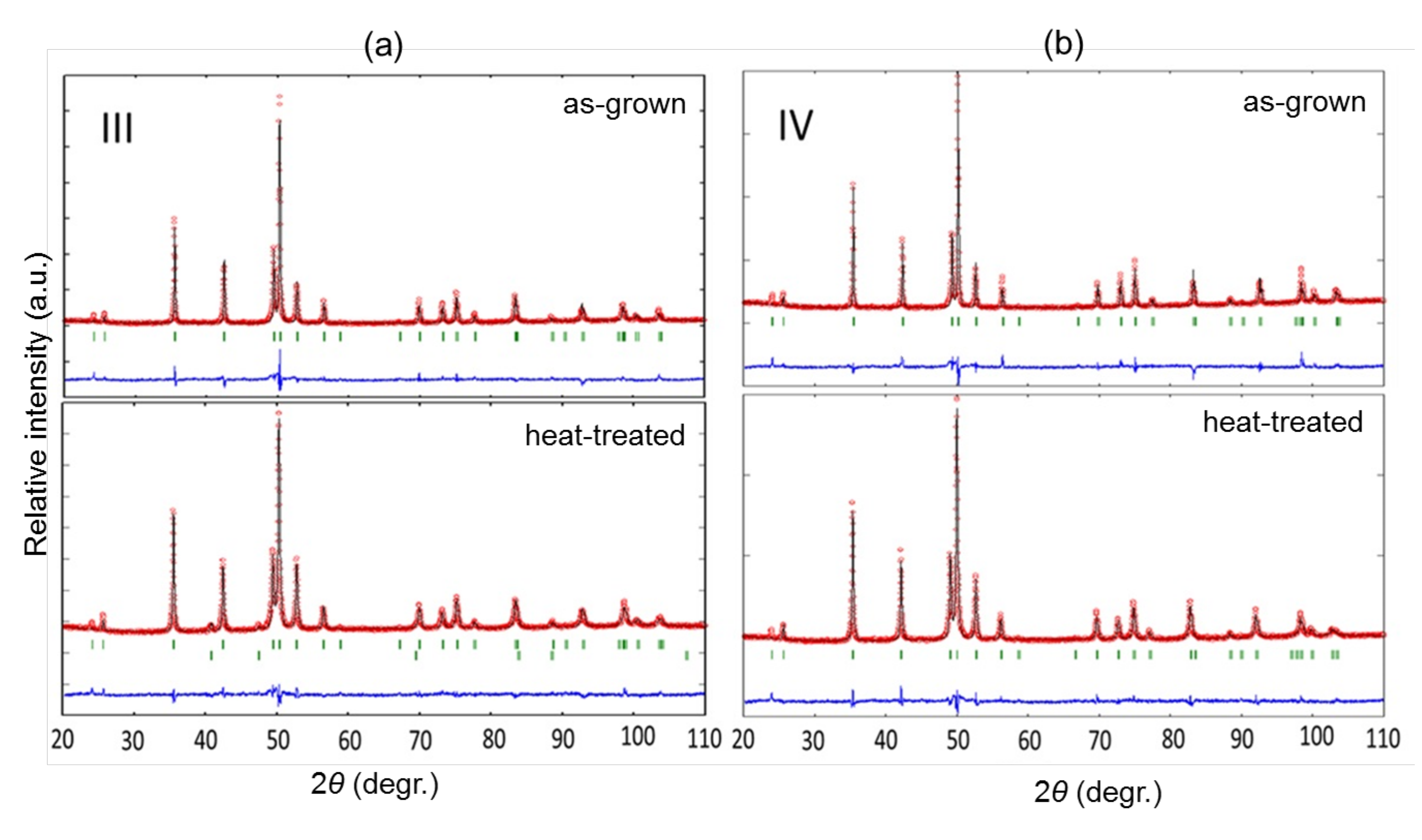}
	\caption{Powder X-ray patterns and Rietveld refinement results for (a) - $\textbf{III}$ -- Ce$_{15.7}$Ta$_{0.6}$Co$_{67.8}$Cu$_{15.9}$, (b) - $\textbf{IV}$ -- Ce$_{16.3}$Ta$_{0.3}$Co$_{61.7}$Cu$_{21.7}$ before (upper) and after (lower)  heat treatment. The observed profile is indicated by circles and the calculated profile by the solid line. Bragg peak positions are indicated by vertical tics, and the difference is shown at the bottom.}
	\label{FIG3}
\end{figure*}

\begin{figure*}[t]
	\includegraphics[width=16cm,height=9cm]{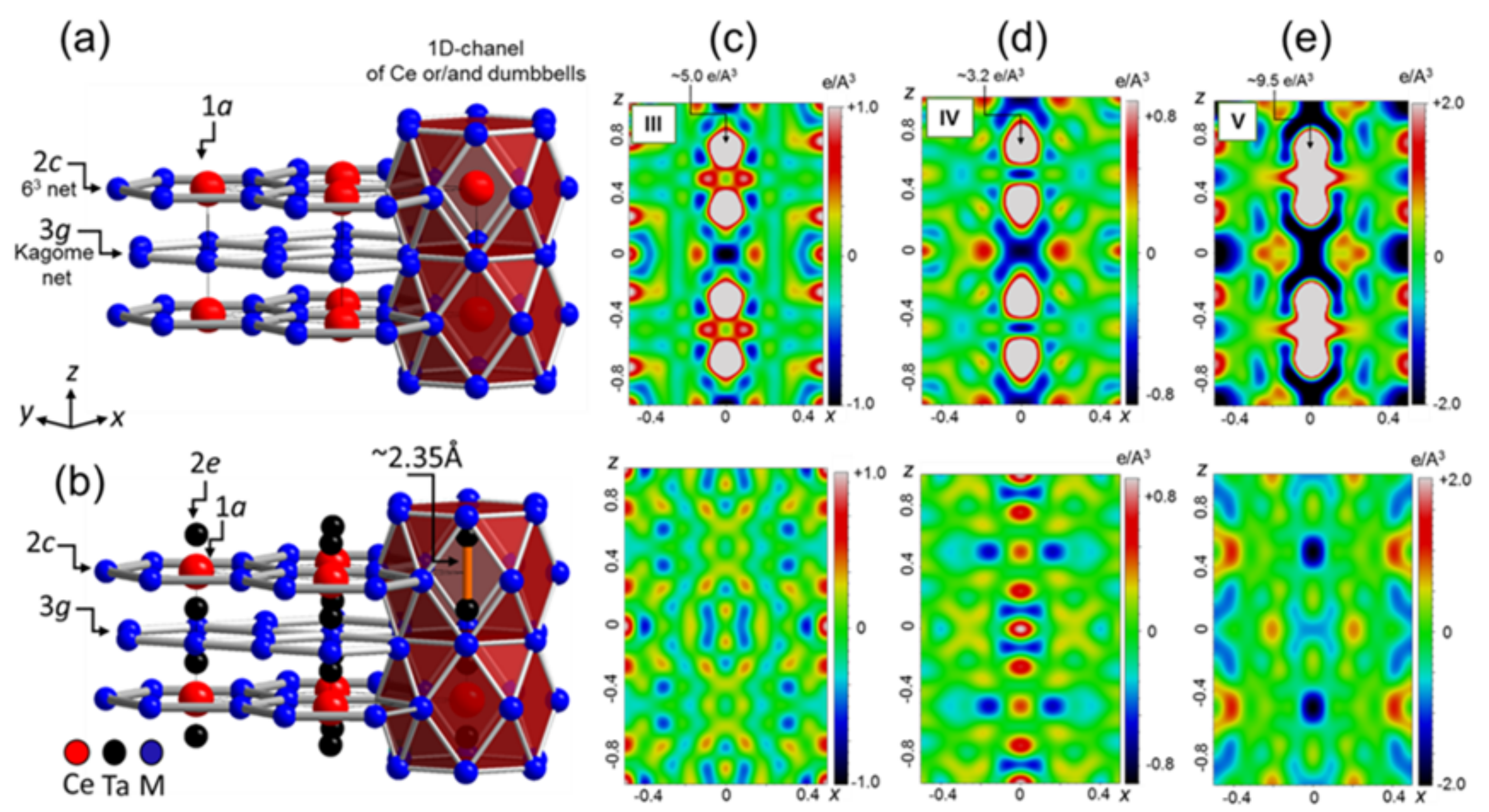}
	\caption{Single crystal refinement for $\textbf{III}$ -- Ce$_{15.7}$Ta$_{0.6}$Co$_{67.8}$Cu$_{15.9}$, $\textbf{IV}$ -- Ce$_{16.3}$Ta$_{0.3}$Co$_{61.7}$Cu$_{21.7}$ and $\textbf{V}$ -- Ce$_{14.3}$Ta$_{1.0}$Co$_{62.0}$Fe$_{12.3}$Cu$_{10.4}$:  [110] views of (a) -- CaCu$_{5}$-type  and (b) -- TbCu$_{7}$-type  structures with and without Ta ``dumb-bells'', respectively and (c) - (e) -- difference electron density maps of structure solutions without ``dumb-bells'' (upper row) showing significant residual electron density peaks of $\sim$3 -- $\sim$10 $\textit{e}$/A$^{3}$ in 1D structural channels at  (0 0 $\textit{z}$) with $z\approx$0.3, and with ``dumb-bells'' (lower row) with significantly smaller residuals.}
	\label{FIG4}
\end{figure*}

The SEM backscattered electron images of the as grown crystals [Fig.~\ref{FIG2} (a-c), upper panels] display the uniformity of their polished $\sim$[001] surface (even at 30,000$\times$ magnification) which suggests a single-phase. Elemental EDS analysis [Table ~\ref{Table 1}] showed the Ce:Co/Cu ratios are close to the 1:5 stoichiometry with Cu contents increasing from $\sim$10 to $\sim$ 22 at. \%, corresponding to 12 -- 26 \% of Co/Cu substitution. 
With respect to Ce content, crystal $\textbf{I}$ and $\textbf{III}$ contain 15 --~15.7~at. \%, which is lower than the Ce content in \textbf{II} and \textbf{IV} and significantly lower than $\sim$16.7 at.~\% Ce content required for the exact 1:5 type stoichiometry. Also a minor presence of Ta (0.3 -- 1 at.\%) was detected in all five samples. The Ta content appears to be correlated to the Cu content as seen in [Table ~\ref{Table 1}]. The presence of Ta is explained by the slight dissolution of the inner walls of the Ta reaction container and diffusion of Ta atoms into the liquid at high temperatures. Since no Ta precipitation or segregation was observed in the SEM/EDS analysis of the as-grown crystals, we believe Ta is either being incorporated into the crystal structure as interstices or as uniformly distributed nano-scale precipitates. However the slight Ce depletion and the presence of Ta suggest the possibility of minor deviations from the classic CaCu$_{5}$-type crystal structure towards various channel disorders or ``dumb-bell" problems characterized elsewhere. \cite{39Bodak,40Tokaychuk,41Cerny} These deviations were accounted for in our structural models and refinements [Figs.~\ref{FIG3},~\ref{FIG4}].
 
The SEM back scattered electron images taken from the [001] surface of the heat treated crystals, [Fig.~\ref{FIG2} (a-c) lower panels], show degradation of the single phase crystal into  a bimodal microstructure consisting of a darker matrix and lighter laminas. These laminas follow the hexagonal symmetry of the original crystal crossing each other at 60$^\circ$  or 120$^\circ$ angles. The thickness of the laminar features is $\sim$0.05 -- 0.1 $\mu$m, and their lengths vary in the range $\sim$ 1 -- 10 $\mu$m. Distances between two laminas are $\sim$2 -- 3 $\mu$m. The elemental EDS analysis of the heat treated material [Table ~\ref{Table 1}] indicates the segregation of Ta-rich phases into the laminar features, whereas the matrix material becomes practically Ta-free in the Cu-richest crystal $\textbf{IV}$.  

\begin{center}
	\textbf{B}. \textbf{X-Ray Analysis and Crystal Structure Determination.}
\end{center}
\begin{table*}[b]
	\centering
	\begin{threeparttable}
		\caption{Single crystal and refinement data for $\textbf{III}$ -- Ce$_{15.7}$Ta$_{0.6}$Co$_{67.8}$Cu$_{15.9}$, $\textbf{IV}$ -- Ce$_{16.3}$Ta$_{0.3}$Co$_{61.7}$Cu$_{21.7}$ and $\textbf{V}$ -- Ce$_{14.3}$Ta$_{1.0}$Co$_{62.0}$Fe$_{12.3}$Cu$_{10.4}$}
		\label{Table 2}
		\begin{tabular}{|l|l|l|l|}
			\hline
			crystal                                & \textbf{III}             & \textbf{IV}              & \textbf{V}                     \\ \hline
			EDS composition                        & Ce$_{0.94}$Ta$_{0.04}$Co$_{4.06}$Cu$_{0.94}$ & Ce$_{0.99}$Ta$_{0.00}$Co$_{3.70}$Cu$_{1.30}$ & Ce$_{0.86}$Ta$_{0.06}$Co$_{3.72}$Fe$_{0.73}$Cu$_{0.62}$ \\ \hline
			refined composition                    & Ce$_{0.98}$Ta$_{0.02}$Co$_{4.25}$Cu$_{0.75}$ & Ce$_{0.99}$Ta$_{0.01}$Co$_{3.68}$Cu$_{1.32}$ & Ce$_{0.94}$Ta$_{0.06}$Co$_{3.71}$Fe$_{0.75}$Cu$_{0.54}$ \\ \hline
			formula mass                           & 439.12                   & 441.26                   & 425.66                         \\ \hline
			Space group, $\textit{Z}$                         & $\textit{P}$6/$\textit{mmm}$; 1                & $\textit{P}$6/$\textit{mmm}$; 1                & $\textit{P}$6/$\textit{mmm}$; 1                      \\ \hline
			$\textit{a}$ (\AA)                                  & 4.946(1)                 & 5.010(1)                 & 5.005(1)                       \\ \hline
			$\textit{c}$ (\AA)                                  & 4.038(1)                 & 4.075(1)                 & 4.131(1)                       \\ \hline
			$\textit{V}$ (\AA$^{3}$)                                 & 85.57(2)                 & 88.60(2)                 & 89.63(2)                       \\ \hline
			$\textit{d$_{c}$}$ (Mg/ m$^{3}$)                            & 8.52                     & 8.27                     & 7.90                           \\ \hline
			$\mu$ (mm$^{-1}$; Mo $\textit{K}$$_{\alpha}$)abs. coef.          & 37.85                    & 37.08                    & 37.56                          \\ \hline
			reflns. collected/ $\textit{R$_{int}$}$                & 1631/0.025               & 761/0.028                & 770/0.025                      \\ \hline
			ind. data/ restrains/ params.          & 79/0/12                  & 66/0/13                  & 67/0/12                        \\ \hline
			GoF ($\textit{F$^{2}$}$)                               & 1.222                    & 1.262                    & 1.256                          \\ \hline
			$\textit{R1}$/ $\textit{wR}$2 {[}I \textgreater 2$\sigma$(I){]} & 0.018/ 0.041             & 0.016/ 0.034             & 0.030/ 0.074                   \\ \hline
			$\textit{R}$1/ $\textit{wR}$2 {[}all data{]}                 & 0.021/ 0.041             & 0.015/ 0.033              & 0.030/ 0.074                   \\ \hline
			Largest diff peak/ hole ($\textit{e}$/\AA$^{3}$)         & 0.080/ -0.74             & 0.052/ -0.59             & 1.45/ -1.13                    \\ \hline
		\end{tabular}
	\end{threeparttable}
\end{table*}

Powder and single crystal X-ray analyses were performed to determine the structure of crystals $\textbf{I}$ -- $\textbf{V}$. Rietveld fitting of the powder X-ray pattern taken from the as-grown, crushed and thoroughly powdered, single crystals of $\textbf{I}$ -- $\textbf{V}$ showed that all Bragg reflections were well indexed within the CaCu$_{5}$-type structure ($\textit{hP6}$,  $\textit{P}$6/$\textit{mmm}$), providing a strong  argument for the single-phase nature of the as-grown crystals in agreement with our SEM observations [Fig. ~\ref{FIG2}]. To address the EDS-observed Ta presence and Ce deficiency, especially in the as-grown crystals $\textbf{I}$, $\textbf{III}$, and $\textbf{V}$ [see Table ~\ref{Table 1}], we considered known structural derivatives of CaCu$_{5}$. \cite{39Bodak} These derivatives are typically observed in binary and ternary rare-earth -- transition metal systems near the $\sim$1:5 and $\sim$2:17 stoichiometries and result from the replacement of rare-earth atoms by pairs of transition metal atoms. The CaCu$_{5}$ substructure can be retained if the replacement is fully random, as in TbCu$_{7}$ \cite{42Buschow}, but may be transformed into various superstructures, such as Th$_{2}$Zn$_{17}$ \cite{43Makarov}, Th$_{2}$Ni$_{17}$ \cite{44Florio}, etc., if the substitution is ordered.  A third possibility comes as combination of ordered and disordered rare-earth  -- ``dumb-bell" substitutions which are contained in a superstructure, e.g., LuFe$_{9.5}$ \cite{45Givord} and PrFe$_{7}$. \cite{46BUSCHOW} We tried Rietveld refinements with structural models allowing the presence of Ta but the clear indexing of Bragg reflections within the parent, CaCu$_{5}$-type, 1:5 structure [Fig.~\ref{FIG3}] indicates a minor and random distribution of Ta. 

We allowed Ce/Ta or T/Ta (T = Co, Cu, Fe) statistical mixings on the 1$\textit{a}$, 2$\textit{c}$, and 3$\textit{g}$ sites with and without an under-occupancy of Ce on the 1$\textit{a}$ site. The substitution of Ce atoms by T -- T ``dumb-bells" was introduced into the model as an independent crystallographic 2$\textit{e}$ site (0 0 $\textit{z}$) with $\textit{z}$ = $\sim$0.3. The last model represents a small departure from the CaCu$_{5}$ structure towards the closely related TbCu$_{7}$ structure with slight Ce/``dumb-bell" substitution within the hexagonal 1D channel [see Fig.~\ref{FIG4}]. The Ce/Ta and T/Ta mixings did not produce satisfactory fits, significantly increasing the residuals and showing unreasonable isotropic temperature parameters, whereas the ``dumb-bell" refinements  were insensitive to small amounts of T -- T (T = Ta, Co, Cu and/or Fe) pairs and were comparable to those without any Ta, and were proportional to the EDS- determined Co/Cu mixings on the 2$\textit{c}$ and 3$\textit{g}$ sites, suggesting minimal disorder. Although present powder X-ray refinements did not clearly address Ta occupation, they clearly determined lattice parameters as well as demonstrated phase content and purity of the material before and after the heat treatment. Data are presented in [Fig.~\ref{FIG3} (a, b). upper panels] for the crushed, as-grown, crystals of $\textbf{III}$ and $\textbf{IV}$. Phase analysis of powder X-ray patterns taken from crushed heat treated crystals of $\textbf{III}$ [Fig.~\ref{FIG3} (a, b) lower panels] revealed clear presence of Ta-like impurities ($\textit{Fm}$-3$\textit{m}$, $\textit{a}$ = 4.446(1) \AA) confirming the EDS findings [Fig.~\ref{FIG2}], whereas in $\textbf{IV}$ Ta was not detected in the X-ray pattern. 

Single crystal X-ray diffraction of the as-grown crystals showed poor quality of the crystals suggesting crystal intergrowth, twining, residual stress or stacking faults effects.
\begin{table}[!h]
	\centering
	\begin{threeparttable}
	\caption{Atomic coordinates, Equivalent Isotropic Displacement Parameters (\AA$\times$ 10$^{3}$), and Site Occupancy Factors Refined for $\textbf{III}$ -- Ce$_{15.7}$Ta$_{0.6}$Co$_{67.8}$Cu$_{15.9}$, $\textbf{IV}$ -- Ce$_{16.3}$Ta$_{0.3}$Co$_{61.7}$Cu$_{21.7}$ and $\textbf{V}$ -- Ce$_{14.3}$Ta$_{1.0}$Co$_{62.0}$Fe$_{12.3}$Cu$_{10.4}$}
	\label{Table 3}
	\begin{tabular}{|l|l|l|l|l|l|l|l|}
		\hline
		atom & WP & \textit{x} & \textit{y} & \textit{z}                                                             & \textit{U$_{eq.}$}                                                 & SOF                                                                                      & \#                                                            \\ \hline
		Ce   & 1$\textit{a}$ & 0          & 0          & 0                                                                      & \begin{tabular}[c]{@{}l@{}}15(1)\\ 15(1)\\ 15(1)\end{tabular} & \begin{tabular}[c]{@{}l@{}}0.977(2)\\ 0.988(2)\\ 0.941(3)\end{tabular}                   & \textbf{\begin{tabular}[c]{@{}l@{}}III\\ IV\\ V\end{tabular}} \\ \hline
		Ta   & 2$\textit{e}$ & 0          & 0          & \begin{tabular}[c]{@{}l@{}}0.282(6)\\ 0.287(9)\\ 0.293(5)\end{tabular} & \begin{tabular}[c]{@{}l@{}}15(1)\\ 15(1)\\ 9(1)\end{tabular}  & \begin{tabular}[c]{@{}l@{}}0.012(2)\\ 0.006(2)\\ 0.030(3)\end{tabular}                   & \textbf{\begin{tabular}[c]{@{}l@{}}III\\ IV\\ V\end{tabular}} \\ \hline
		M1$^{a}$  & 2$\textit{c}$ & 2/3        & 1/3        & 0                                                                      & \begin{tabular}[c]{@{}l@{}}14(1)\\ 14(1)\\ 24(1)\end{tabular} & \begin{tabular}[c]{@{}l@{}}1.00 Co\\ 0.33(7) Cu\\ 1.00 Co\end{tabular}                   & \textbf{\begin{tabular}[c]{@{}l@{}}III\\ IV\\ V\end{tabular}} \\ \hline
		M2   & 3$\textit{g}$ & 1/2        & 0          & 1/2                                                                    & \begin{tabular}[c]{@{}l@{}}10(1)\\ 9(1)\\ 24(1)\end{tabular}  & \begin{tabular}[c]{@{}l@{}}0.25(2) Cu\\ 0.22(5) Cu\\ 0.25(5) Fe/ 0.18(4) Cu\end{tabular} & \textbf{\begin{tabular}[c]{@{}l@{}}III\\ IV\\ V\end{tabular}} \\ \hline
	\end{tabular}
\begin{tablenotes}[]
	\item  $^{a}$The atomic symbol ``M'' stands for Co/Cu or Co/Fe/Cu mixed occupancy  
\end{tablenotes}
\end{threeparttable}
\end{table}
These defects were very apparent on Laue frames from numerous ($>$10) specimens of $\textbf{I}$ in form of strong streaking, doubling of the reflections, sometimes presence of the Debye rings. However, these effects diminished in $\textbf{II}$ and were practically absent in $\textbf{III}$ and $\textbf{IV}$ allowing structural characterization of the as-grown crystals of $\textbf{III}$ -- $\textbf{V}$ [see Table ~\ref{Table 2}].
Single crystal structure solutions of $\textbf{III}$ -- $\textbf{V}$ confirmed their CaCu$_{5}$ substructure [see Table ~\ref{Table 2}, ~\ref{Table 3}]. However, disorder was detected within the 1D hexagonal channels, as seen in the residual electron density peaks of $\sim$5.2, $\sim$3.2 and $\sim$9.5 e/\AA$^{3}$  at (0 0 $\textit{z}$), $\textit{z}\approx$0.295  for $\textbf{III}$, $\textbf{IV}$ and $\textbf{V}$, respectively. Only by filling the 2$\textit{e}$ site with the heaviest and largest available pair, Ta -- Ta, we were able to reach satisfactory refinement. The $\textit{R}$1/$\textit{wR}$2 residuals dropped by 50 -- 70 \% in comparison to the solutions without Ta and showed minimal fluctuations of the rest electron density in the final fits. Fig. ~\ref{FIG4} showes the differential Fourier maps for $\textbf{III}$ -- $\textbf{V}$ with and without the ``dumb-bell" disorder. One significant deficiency of the solutions is the interatomic T -- T distances of $\sim$2.35 \AA, which is typical for Co -- Co, Co -- Cu and  Co -- Fe pairs but is extremely short for Ta -- Ta.
\begin{figure*}[b]
	\includegraphics[width=12.8cm,height=8.8cm]{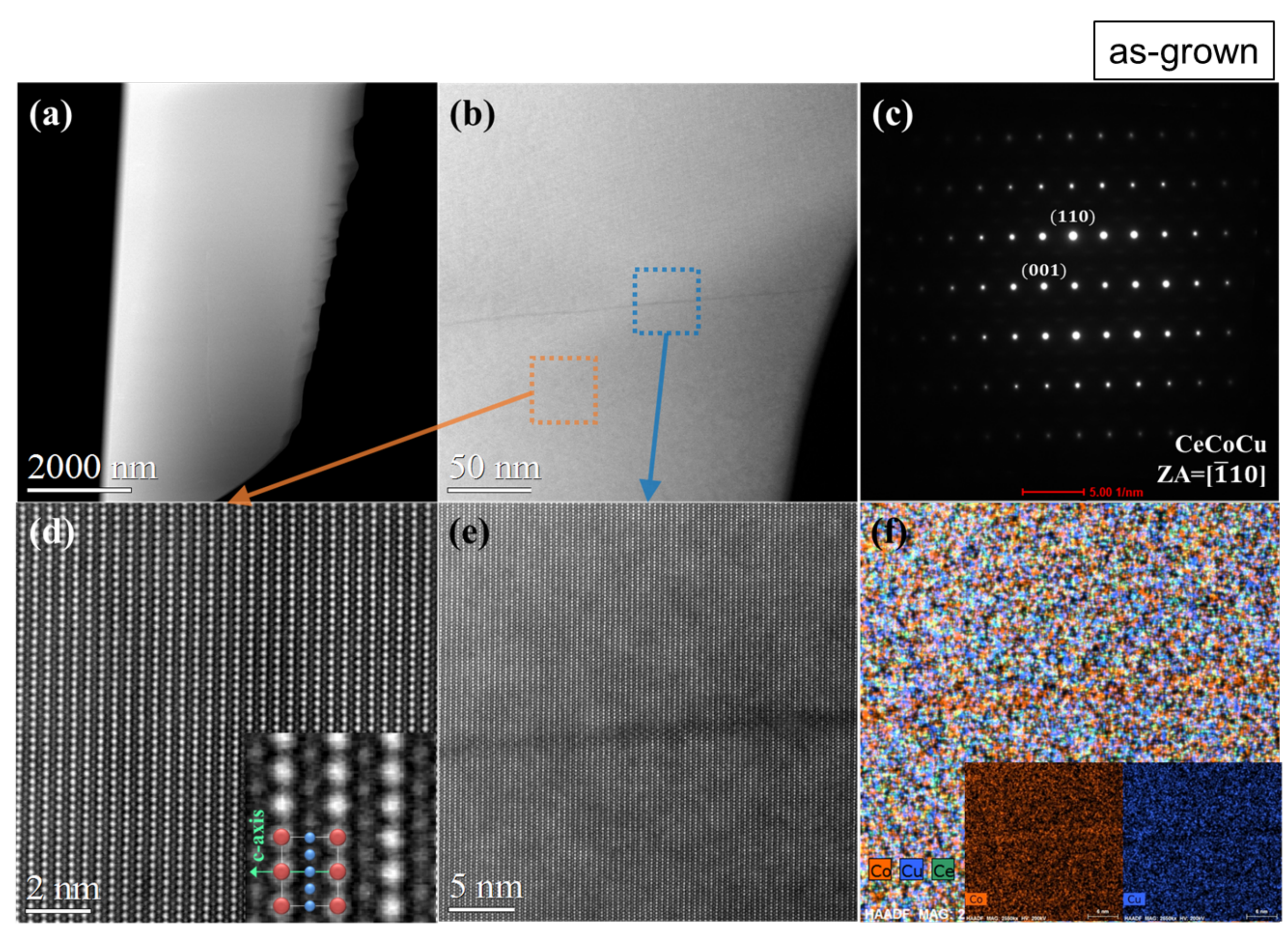}
	\caption{(a) - HAADF STEM image of as-grown $\textbf{III}$ -- Ce$_{15.7}$Ta$_{0.6}$Co$_{67.8}$Cu$_{15.9}$ showing the overall microstructure, (b) - enlarged HAADF image where shows a dark-contrast line, (c ) - diffraction pattern taken from the region shown (b) including the matrix and the dark line, (d)  - high resolution STEM image taken from orange-boxed area in (b) under [1-10] zone axis. The inset at bottom right is an enlarged atomic image with atomic model of hexagonal 1:5 Ce/Co/Cu phase. The bright dots and dark dots in the images correspond to atomic columns of Ce and (Co, Cu) elements, respectively, (e) - enlarged image of blue-boxed area in (b) and dark line in single crystalline phase is shown clearly, (f) - EDS elemental mapping of (e) clearly showing Co enrichment in the line, the small Co and Cu elemental maps-insets are presented for contrasting observation of Cu depletion in the same line.}
	\label{TEM_FIG1}
\end{figure*} 
However, the ``dumb-bell" configuration with large and heavy atoms similar to Ta is not unprecedented and was reported for similar structure of CeFe$_{10}$Zr$_{0.8} ($\textit{d}$_{(\text{Zr - Zr})}\approx$~2.65 \AA). \cite{47Zhou} However, the stoichiometry of $\textbf{V}$ shows significant deviation from the ideal 1:5 stoichiometry. The content of 1D channels (Ce plus the Ta -- Ta pairs) does not reach the expected 16.7 at.\%, meaning that some of T atoms must participate in the channel disorder. 
\begin{center}
\textbf{C}. \textbf{Transmission Electron Microscopy.}
\end{center}
\begin{figure*}[t]
	\includegraphics[width=13.0cm,height=9.0cm]{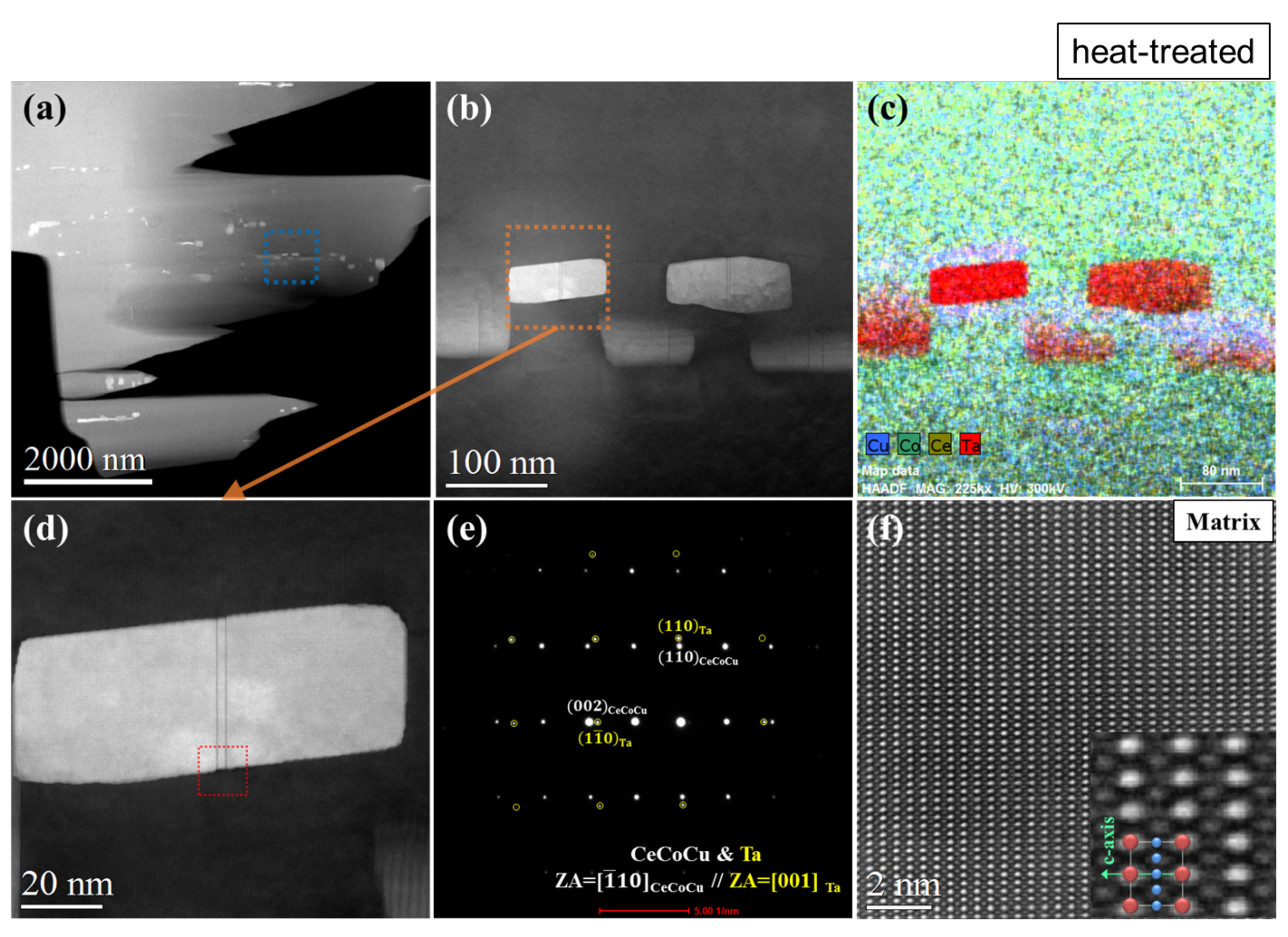}
	\caption{(a) - HAADF image of heat treated $\textbf{III}$ -- Ce$_{15.7}$Ta$_{0.6}$Co$_{67.8}$Cu$_{15.9}$ showing the overall microstructure, (b) - enlarged image of blue-boxed area in (a), (c) - EDS elemental mapping corresponding to (b). The bright regions are Ta-rich and considered as Ta precipitate, (d) - enlarged image of orange-boxed area in (b), (e)  - diffraction pattern taken from (d) including the matrix and the Ta precipitate, (f) - high resolution STEM image taken from the matrix in (d) under [1-10] zone axis.}
	\label{TEM_FIG2}
\end{figure*} 
\begin{figure}[t]
	\includegraphics[scale =0.3]{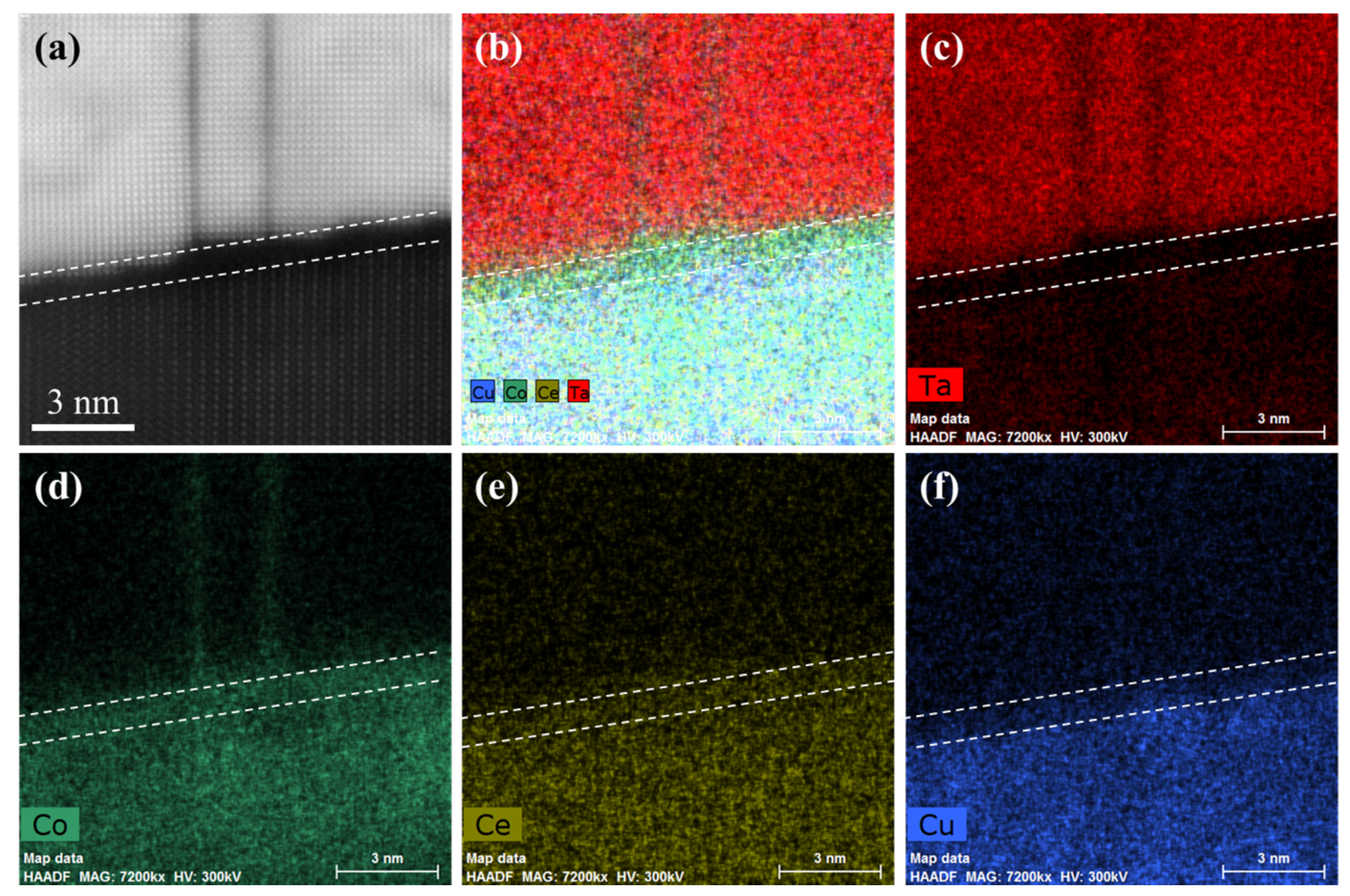}
	\caption{(a) - high resolution HAADF image of the interface between the matrix and the Ta precipitate taken from red-boxed area in Fig.~\ref{TEM_FIG2} (d) and (b - f) - corresponding EDS elemental mapping results. The white lines indicate the same position in each image.}
	\label{TEM_FIG3}
\end{figure}
Fig.~\ref{TEM_FIG1} (a) is a high-angle-annular-dark-field (HAADF) scanning transmission electron microscope (STEM) image of an as-grown sample $\textbf{III}$ showing the overall microstructure. The entire region consists a single crystalline phase. Fig.~\ref{TEM_FIG1} (b) is an enlarged HAADF image which shows a dark-contrast line. It was the only feature that can be found in the entire scan area. Fig.~\ref{TEM_FIG1} (c) is a diffraction pattern taken from the region shown in Fig.~\ref{TEM_FIG1} (b) including the matrix and the dark line. It clearly shows the single crystalline 1:5 phase. It seems that the dark line region has the same crystal structure and it is not a precipitate which would have made additional diffracted spots in Fig.~\ref{TEM_FIG1} (c). Fig.~\ref{TEM_FIG1} (d) is a high resolution STEM image taken from orange-boxed area in Fig.~\ref{TEM_FIG1} (b) under [1-10] zone axis. The inset at bottom right is an enlarged atomic image with an atomic model of hexagonal 1:5 phase. The bright dots and dark dots in the images correspond to atomic columns of Ce and Co/Cu elements, respectively. Fig.~\ref{TEM_FIG1} (e) is an enlarged image of blue-boxed area in Fig.~\ref{TEM_FIG1} (b) and dark line in single crystalline phase is shown clearly. Fig.~\ref{TEM_FIG1} (f) is an EDS elemental mapping of Fig.~\ref{TEM_FIG1} (e). The chemical contrast between the matrix and the dark line is observed. The EDS result shows the dark line is Co-rich and Cu-deficient. 

Fig.~\ref{TEM_FIG2} (a) is a HAADF image of an annealed sample showing the overall microstructure. Many bright areas were observed unlike the unanneald sample shown in before in Fig.~\ref{TEM_FIG1}. Fig.~\ref{TEM_FIG2} (b) is an enlarged image of the blue-boxed area in Fig.~\ref{TEM_FIG2} (a) and Fig.~\ref{TEM_FIG2} (c) is the EDS elemental mapping corresponding to Fig.~\ref{TEM_FIG2} (b). The bright regions in Fig.~\ref{TEM_FIG2} (b) are Ta-rich and considered as Ta precipitates. Additionally, a few dark lines are observed in the Ta precipitate. The difference in brightness of precipitates is attributed to the difference in the thickness of each precipitate. Fig.~\ref{TEM_FIG2} (d) is an enlarged image of the orange-boxed area in Fig.~\ref{TEM_FIG2} (b) and Fig.~\ref{TEM_FIG2} (e) is a diffraction pattern taken from Fig.~\ref{TEM_FIG2} (d) including the matrix and the Ta precipitates. Fig.~\ref{TEM_FIG2} (d) shows Ta precipitates coherently embedded by epitaxial precipitation and the corresponding diffraction pattern shows the epitaxial relationship between the matrix and Ta precipitate. The orientation relation was observed as follows:
(110) CeCoCu // (110) Ta;
(002) CeCoCu // (1-10) Ta;
and [1-10] CeCoCu // [001] Ta.
Fig.~\ref{TEM_FIG2} (f) is a high resolution STEM image taken from the matrix in Fig.~\ref{TEM_FIG2} (d) under [1-10] zone axis. It is the same as that seen in Fig.~\ref{TEM_FIG1} (d). The inset at bottom right is an enlarged atomic image with an atomic model of hexagonal 1:5 Ce/Co/Cu phase. The bright dots and dark dots in the images correspond to atomic columns of Ce and Co/Cu elements, respectively. Fig.~\ref{TEM_FIG3} shows high resolution HAADF images of the interface between the matrix and the Ta precipitate taken from red-boxed area in Fig.~\ref{TEM_FIG2} (d) and corresponding EDS elemental mapping results [Fig.~\ref{TEM_FIG2} (b-f)]. The white dashed lines indicate the same position in each image. Although Cu-rich and Co-deficient region was observed near the precipitate [Fig.~\ref{TEM_FIG2} (c)], there was also Co, Ce-rich and Cu-deficient interface between the matrix and Ta precipitate. The dark lines in the Ta precipitate turned out to be Co-rich. Considering EDS maps at the interface and near the precipitate, it is assumed that Co was infiltrated into the precipitate [Fig.~\ref{TEM_FIG2} (d)], and Co became deficient near the precipitate with relative Cu-rich as a result. As will be discussed below the high resolution TEM results will be returned to, for both as-grown and heat-treated samples of crystal $\textbf{III}$, to elucidate the pinning mechanism that leads to significant coercivities in the crystals.
\section{Magnetic Properties}
\begin{center}
	\textbf{A}. \textbf{As-grown crystals. Curie temperature, magnetocrystalline anisotropy field and energy density.}
\end{center}

\begin{figure}[t]
	\includegraphics[width=9cm,height=7.5cm]{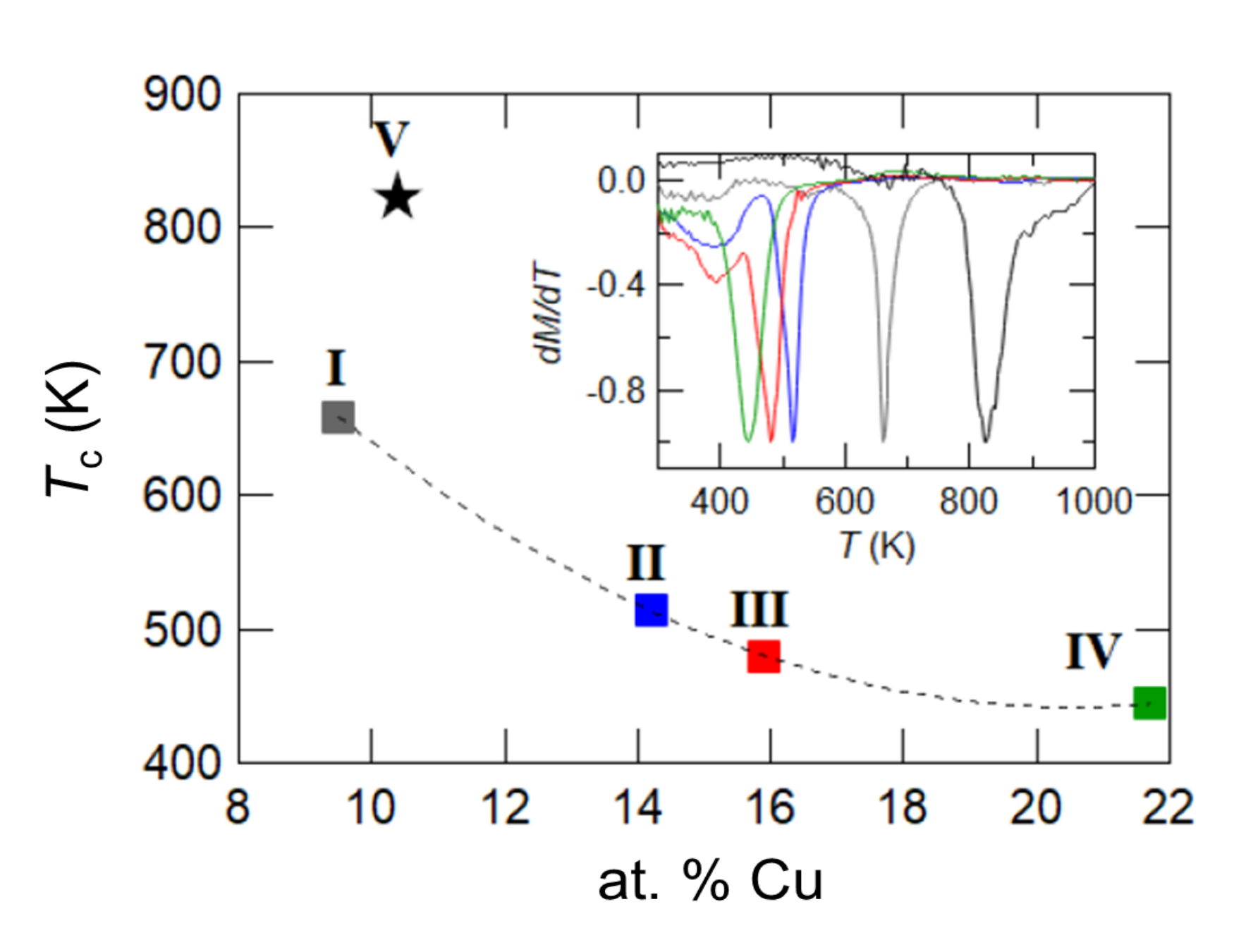}
	\caption{Curie temperatures for the as-grown $\textbf{I}$ -- Ce$_{15.1}$Ta$_{1.0}$Co$_{74.4}$Cu$_{9.5}$, $\textbf{II}$ -- Ce$_{16.3}$Ta$_{0.6}$Co$_{68.9}$Cu$_{14.2}$, $\textbf{III}$ -- Ce$_{15.7}$Ta$_{0.6}$Co$_{67.8}$Cu$_{15.9}$, $\textbf{IV}$ -- Ce$_{16.3}$Ta$_{0.3}$Co$_{61.7}$Cu$_{21.7}$ and $\textbf {V}$ -- Ce$_{14.3}$Ta$_{1.0}$Co$_{62.0}$Fe$_{12.3}$Cu$_{10.4}$ inferred from the peaks in derivative of magnetization with respect to temperature, i.e. $\textit{dM/dT}$ obtained for each crystal (see inset). Magnetization data were obtained under magnetic field of 0.01 T.}
	\label{Tc}
\end{figure}

Fig.~\ref{Tc} presents the Curie temperatures for samples $\textbf{I}$ -- $\textbf{V}$ as inferred from the peak in $\textit{dM/dT}$  shown in the inset. The Curie temperatures $\textit{T$_{c}$}$ estimated by the maximum in the derivative correspond closely to the \textit{T$_{c}$} derived via the more accurate Arrot plot method (see below).  The $\textit{T$_{c}$}$-value decreases rapidly with increasing Cu content for Fe-free samples $\textbf{I}$ -- $\textbf{IV}$. This is consistent with the early report \cite{30Girodin} and indicates weakening in the ferromagnetic exchange interactions within Co sublattice due to the introduction of nonmagnetic Cu. In contrast, the Fe-doped crystal $\textbf{V}$ shows remarkable improvement of $\textit{T$_{c}$}$, increasing by over 150 K to $\sim$820 K, a value that is significantly higher than the $\textit{T$_{c}$}$ = 653 K of parent CeCo$_5$.\cite{51BARTASHEVICH} Using band structure analysis, we find that Fe-doping of  CeCo$_5$ and Ce(Co,Cu)$_5$, increases the ordering energy $\Delta E = E_{NM}-E_{FM}$ (NM -- non-magnetic and FM -- ferromagnetic states), as well as the total magnetic moment of the systems (see Section $\textbf{V}$ below). This leads to the remarkable increase of the Curie temperature and saturation magnetization. 

\begin{figure}[!h]
	\includegraphics[width=8.5cm,height=7.1cm]{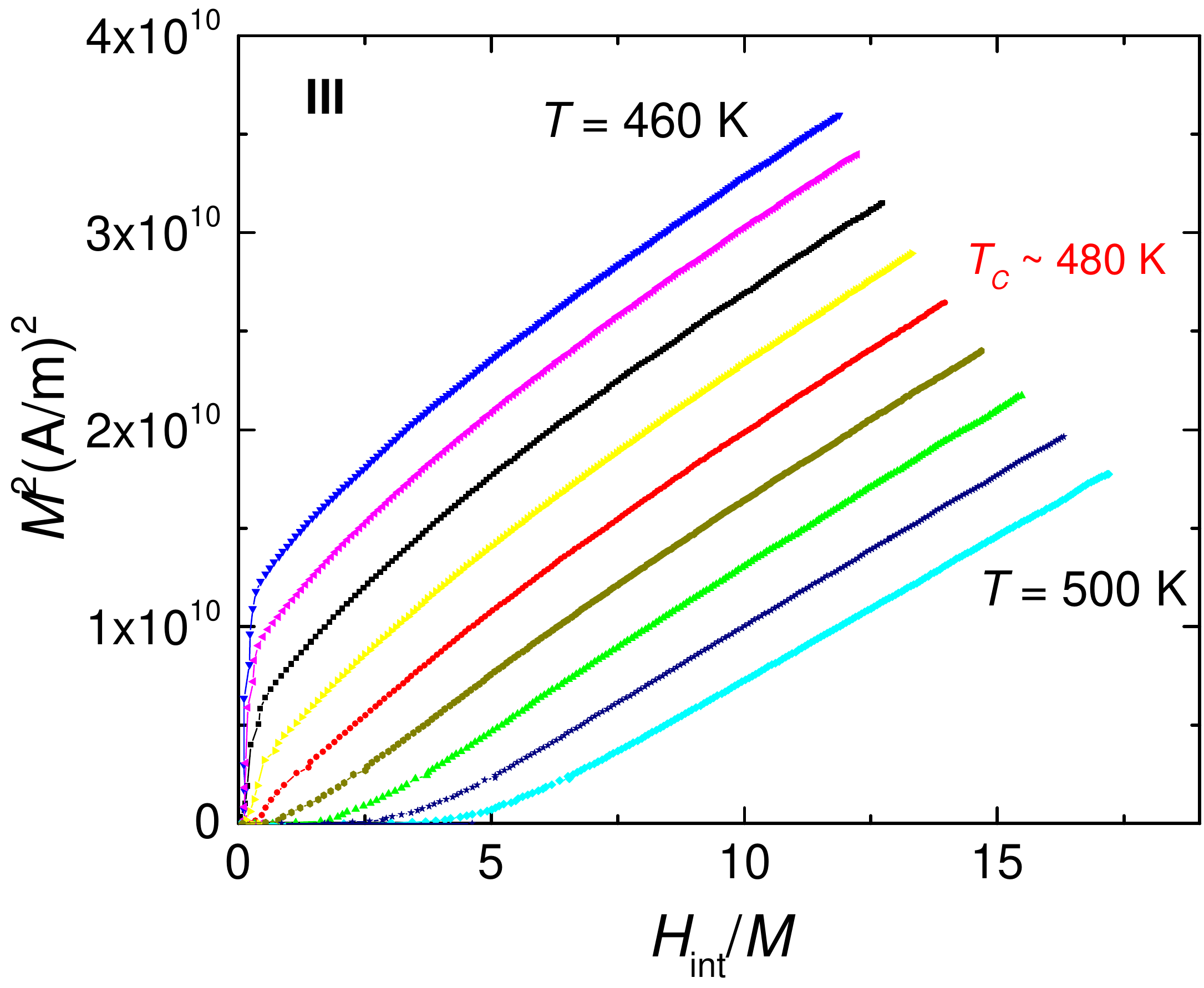}
	\caption{Arrott plot for $\textbf{III}$ - Ce$_{15.7}$Ta$_{0.6}$Co$_{67.8}$Cu$_{15.9}$ with isotherms taken in 5.0 K intervals as indicated in the graph. The Curie temperature is $\sim480$~K as inferred from the plot since the isotherm is closest to linear and passes through origin.}
	\label{ArrottPlot}
\end{figure}

To improve the precision of the $\textit{T$_{c}$}$ determination, we prepared an Arrot plot analysis for $\textbf{III}$  using isotherms between 460~K to 500~K ~[Fig.~\ref{ArrottPlot}] . The Curie temperature for $\textbf{III}$ is estimated to be $\sim480$~K, since the isotherm at that temperature is closest to a straight line and passes through the origin. 

\begin{figure}[!h]
	\includegraphics[width=8.7cm,height=7.0cm]{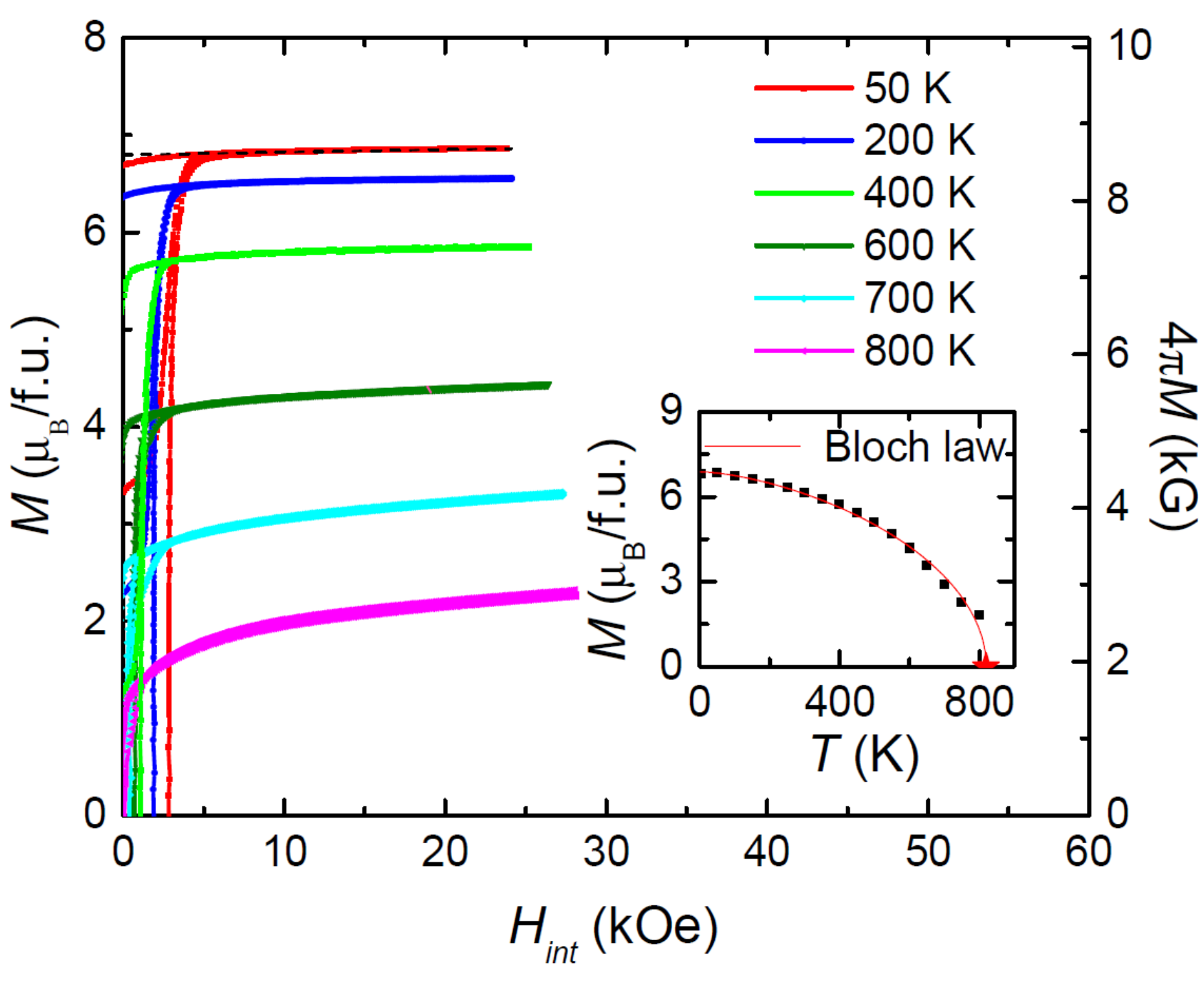}
	\caption{Representative $\textit{M(H)}$ isoterms for the as-grown crystal $\textbf {V}$ -- Ce$_{14.3}$Ta$_{1.0}$Co$_{62.0}$Fe$_{12.3}$Cu$_{10.4}$. In the inset - spontaneous magnetizations for each tempreature inferred from the extrapolation of the linear regions of the $\textit{M(H)}$ back to $\textit{H}$ = 0. Red star shows extrapolated $\textit{T}$$_c$ value following Bloch law $\textit{M(T)}$ = $\textit{M}$(0)(1-($\textit{T/T$_{c}$}$)$^{3/2}$).}
	\label{FIG11}
\end{figure}

In Fig.~\ref{FIG11} we show representative $\textit{M}$($\textit{H}$) isoterms for sample $\textbf{V}$. In the inset we plot the spontaneous magnetization value for each temperature inferred from the extrapolation of the linear region of the $\textit{M}$($\textit{H}$) back to $\textit{H}$ = 0. As can be seen, these data suggest a $\textit{T$_{c}$}$ $\sim$ 820 K (estimated by generalized Bloch law fitting of spontaneous magnetization), in good agreement with Fig.~\ref{Tc}

\begin{figure}[!h]
	\includegraphics[width=9cm,height=13cm]{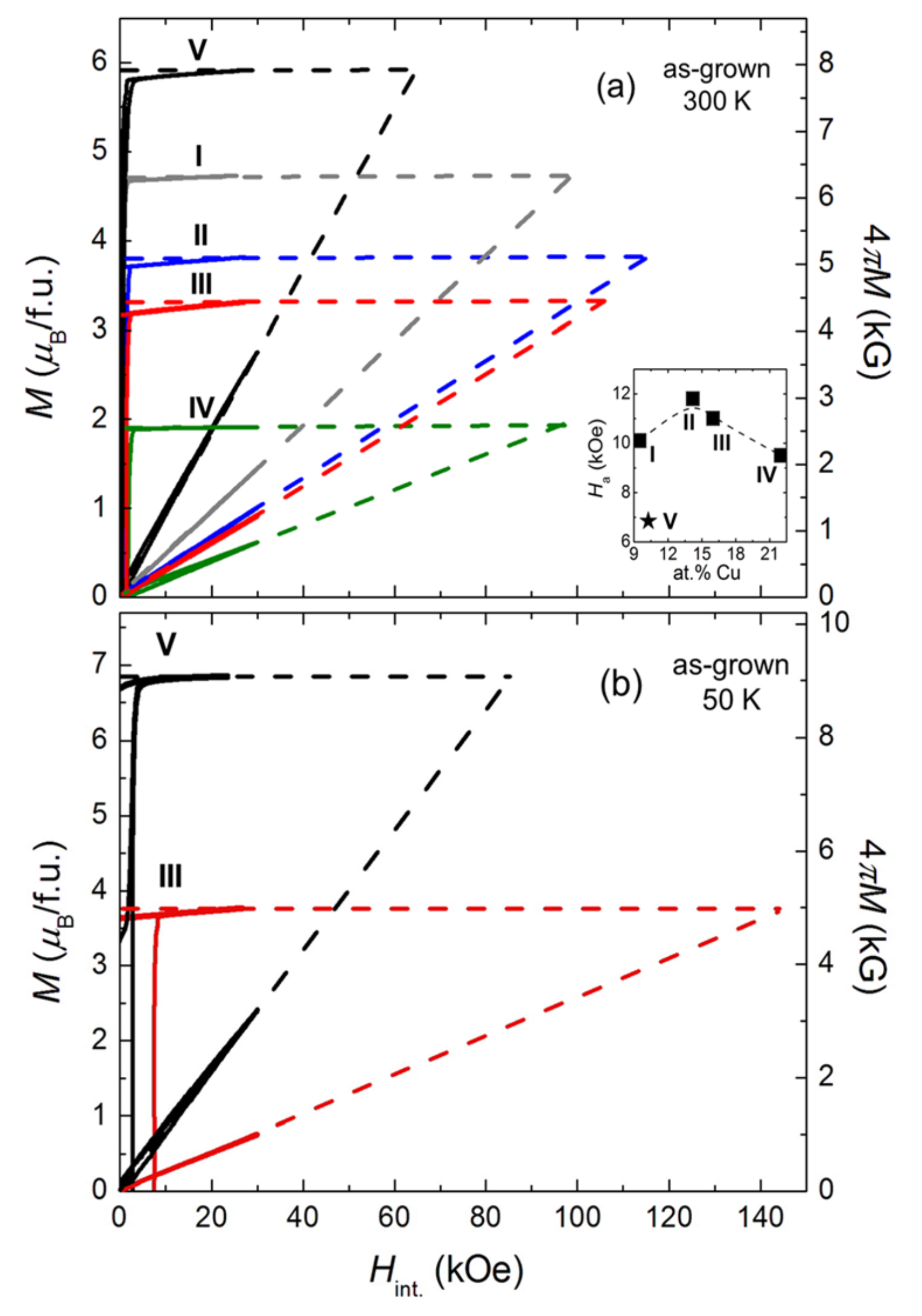}
	\caption{(a) - anisotropic field-dependent magnetization along easy and hard axes of the as-grown crystals $\textbf{I}$ -- Ce$_{15.1}$Ta$_{1.0}$Co$_{74.4}$Cu$_{9.5}$, $\textbf{II}$ -- Ce$_{16.3}$Ta$_{0.6}$Co$_{68.9}$Cu$_{14.2}$, $\textbf{III}$ -- Ce$_{15.7}$Ta$_{0.6}$Co$_{67.8}$Cu$_{15.9}$, $\textbf{IV}$ -- Ce$_{16.3}$Ta$_{0.3}$Co$_{61.7}$Cu$_{21.7}$ and $\textbf {V}$ -- Ce$_{14.3}$Ta$_{1.0}$Co$_{62.0}$Fe$_{12.3}$Cu$_{10.4}$ at 300 K. Inset in the lower-right corner - dependence of the anisotropy field $\textit{H$_{a}$}$ $\textit{vs.}$ Cu concentration, (b) - anisotropic field-dependent magnetization along easy and hard axes of the as-grown crystals  $\textbf{III}$ -- Ce$_{15.7}$Ta$_{0.6}$Co$_{67.8}$Cu$_{15.9}$ and $\textbf {V}$ -- Ce$_{14.3}$Ta$_{1.0}$Co$_{62.0}$Fe$_{12.3}$Cu$_{10.4}$ at 50 K.}
	\label{Anisotropy_I_II_IV}
\end{figure}
\begin{figure}[!h]
	\includegraphics[width=8.7cm,height=7.0cm]{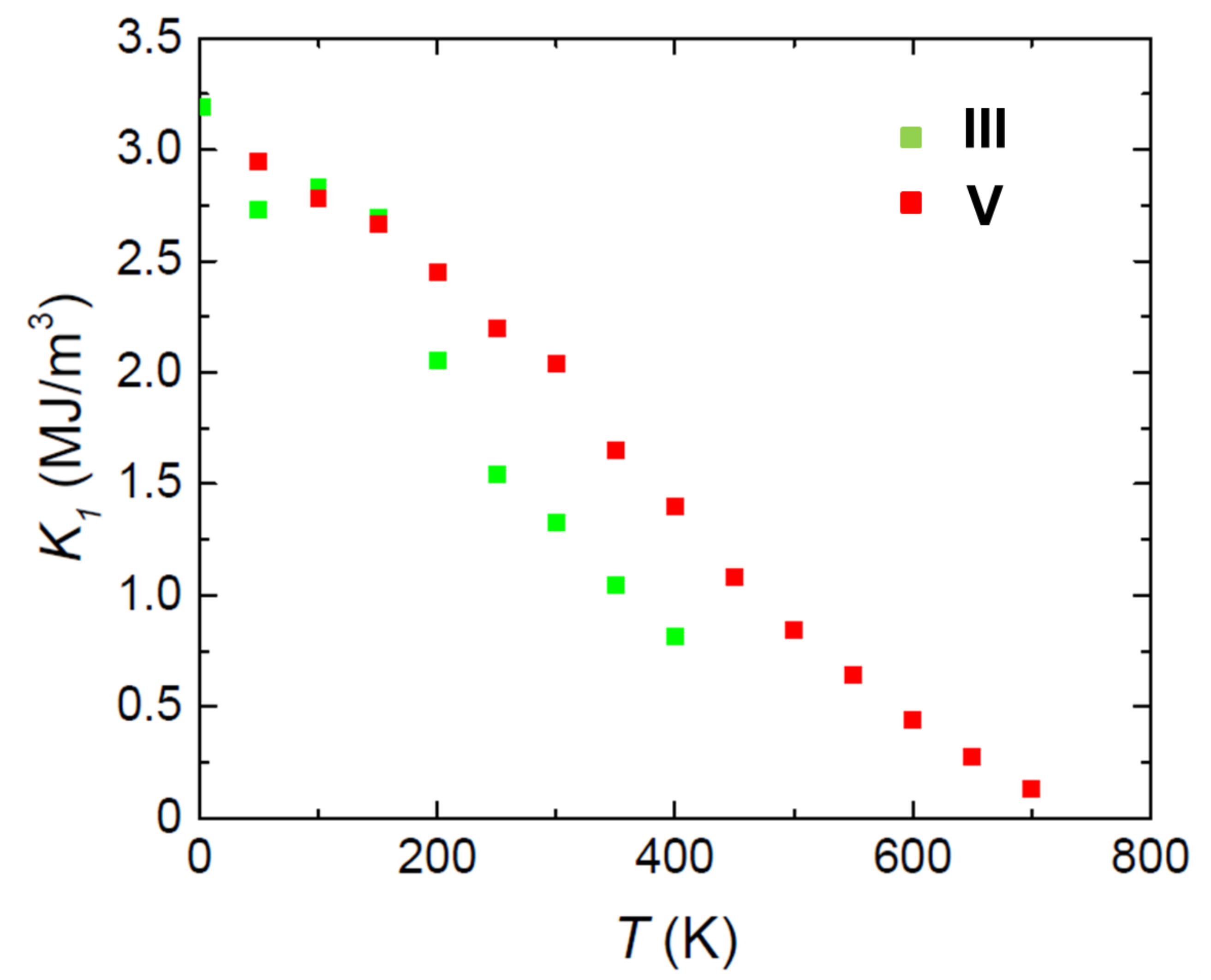}
	\caption{Comparison of temperature dependent magnetocrystalline anisotropy energy density of  $\textbf{III}$ - Ce$_{15.7}$Ta$_{0.6}$Co$_{67.8}$Cu$_{15.9}$  and $\textbf{V}$ - Ce$_{14.3}$Ta$_{1.0}$Co$_{62.0}$Fe$_{12.3}$Cu$_{10.4}$.}
	\label{K1}
\end{figure}
Fig.~\ref{Anisotropy_I_II_IV} (a) shows the magnetocrystalline anisotropy field, $\textit{H}$$_a$, at room temperature for all as-grown crystals $\textbf{I}$ -- $\textbf{V}$; the low temperature estimations of $\textit{H}$$_a$ were done for crystals $\textbf{III}$ and $\textbf{V}$ and are presented in Fig. ~\ref{Anisotropy_I_II_IV}b. The anisotropy field was determined by the high-field, linear extrapolation of the field-dependent moment along the easy [001]  and hard ($\textit{H}$ $\perp$ [001]) axes. The room temperature $\textit{H}$$_a$ for the Fe-free, as-grown  crystals $\textbf{I}$ -- $\textbf{IV}$ exhibit a maximum anisotropy field of  $\sim$118 kOe in crystal $\textbf{II}$. The addition of Fe shows a detremental influence on the magnetocrystalline anisotropy, (in Fe-doped $\textbf{V}$ the anisotropy field drops to $\sim$65 kOe [see inset in Fig.~\ref{Anisotropy_I_II_IV}a], but the spontaneous magnetization increases by $\sim$30 \% compared to crystals with similar Cu contents). Low temperature measurements estimate the spontaneous magnetization for crystals $\textbf{III}$ and $\textbf{V}$ to be $\sim$3.7 and $\sim$6.8 $\mu$$_{B}$/f.u., respectively. 

The temperature dependent magnetocrystalline anisotropy energy density was measured using the Sucksmith-Thompson method\cite{Sucksmith362, VTaufourMnBi2015,Lamichhane2018} by using the hard axis magnetization isotherms for crystals $\textbf{III}$ and $\textbf{V}$ [Fig. ~\ref{K1}]. 

\begin{figure}[t]
	\includegraphics[width=8.8cm,height=5.4cm]{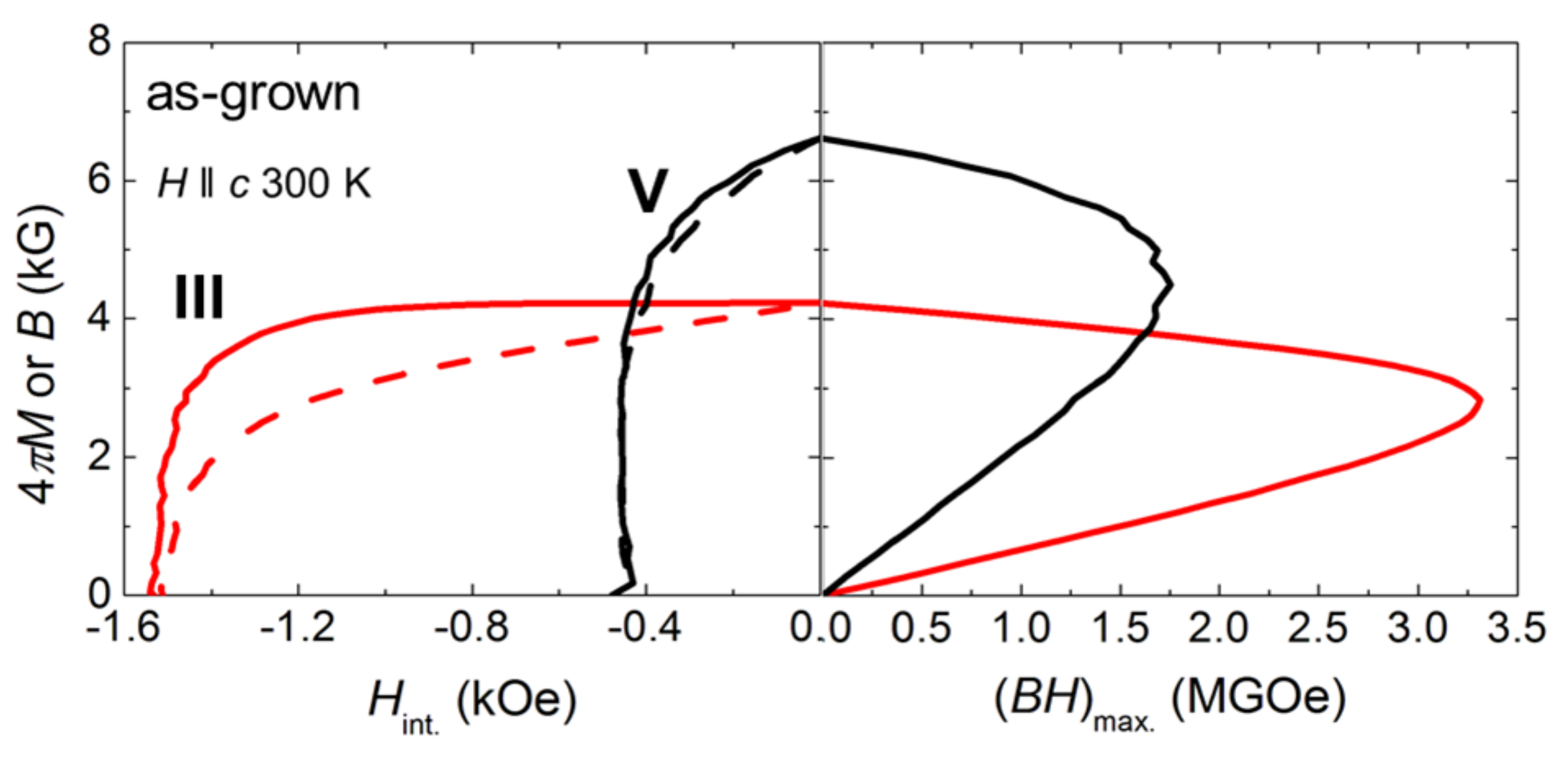}
	\caption{Room temperature second quadrant magnetic hysteresis loops for the as-grown crystals $\textbf{III}$ -- Ce$_{15.7}$Ta$_{0.6}$Co$_{67.8}$Cu$_{15.9}$ and $\textbf{V}$ - Ce$_{14.3}$Ta$_{1.0}$Co$_{62.0}$Fe$_{12.3}$Cu$_{10.4}$, 4$\pi$$\textit{M}$ indicated as solid line and $\textit{B}$ as a dashed line (left pannel). Estimation of the energy products ($\textit{BH}$)$_{max.}$ (right pannel).}
	\label{Energy_ag}
\end{figure}

Interestingly, the as-grown single crystals showed magnetic hysteresis when measured along the easy axis of magnetization [001]. For example, crystal $\textbf{III}$ exhibited a hysteresis (see Fig.~\ref{Motivation}) which reached $\textit{H}$$_c$ $\approx$~1.6 kOe and $\textit{B}$$_r$ $\approx$ 4.2 kG, $\textit{M}$$_s$  $\approx$ 4.2 kG and ($\textit{BH}$)$_{max.}$~$\approx$~3.5~MGOe [Fig.~\ref{Energy_ag}], which is comparable to most of anisotropic sintered alnico grades.\cite{alnico} This is remarkable considering the common belief that the appearance of the coercivity is a result of the extrinsic properties, e.g., development of proper microstructure for strong magnetic domain pinning, and this is generally not associated with a single phase single crystal as determined by the SEM and XRD examinations [Figs.~\ref{FIG2},\ref{FIG3}], which did not reveal any elemental precipitations, segregations, or any microstructure on their corresponding length-scales. 
\begin{figure}[t]
	\includegraphics[width=8.7cm,height=7cm]{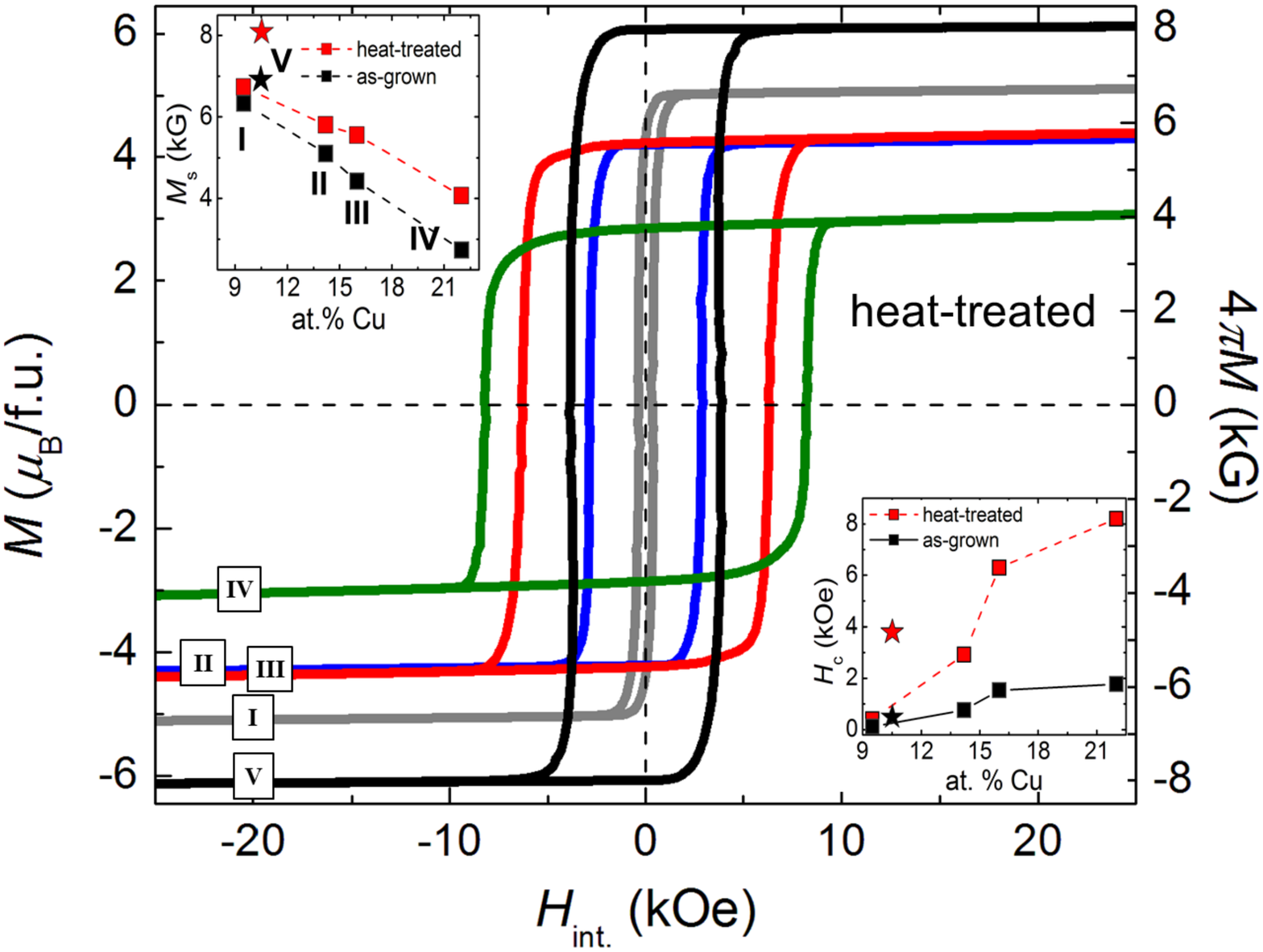}
	\caption{Magnetic hysteresis loops of the heat-treated crystals $\textbf{I}$ -- Ce$_{15.1}$Ta$_{1.0}$Co$_{74.4}$Cu$_{9.5}$, $\textbf{II}$ -- Ce$_{16.3}$Ta$_{0.6}$Co$_{68.9}$Cu$_{14.2}$, $\textbf{III}$ -- Ce$_{15.7}$Ta$_{0.6}$Co$_{67.8}$Cu$_{15.9}$, $\textbf{IV}$ -- Ce$_{16.3}$Ta$_{0.3}$Co$_{61.7}$Cu$_{21.7}$ and $\textbf {V}$ -- Ce$_{14.3}$Ta$_{1.0}$Co$_{62.0}$Fe$_{12.3}$Cu$_{10.4}$ at 300 K. Upper-right inset -- dependence of the spontaneous magnetization $\textit{M$_{s}$}$ $\textit{vs.}$ Cu concentration for the as-grown and heat treated crystals. Lower-right inset -- dependence of the coercivity $\textit{H$_{c}$}$ $\textit{vs.}$ Cu concentration for the as-grown and heat-treated crystals.}
	\label{LOOPS}
\end{figure}
\begin{figure}[t]
	\includegraphics[width=8.8cm,height=5.4cm]{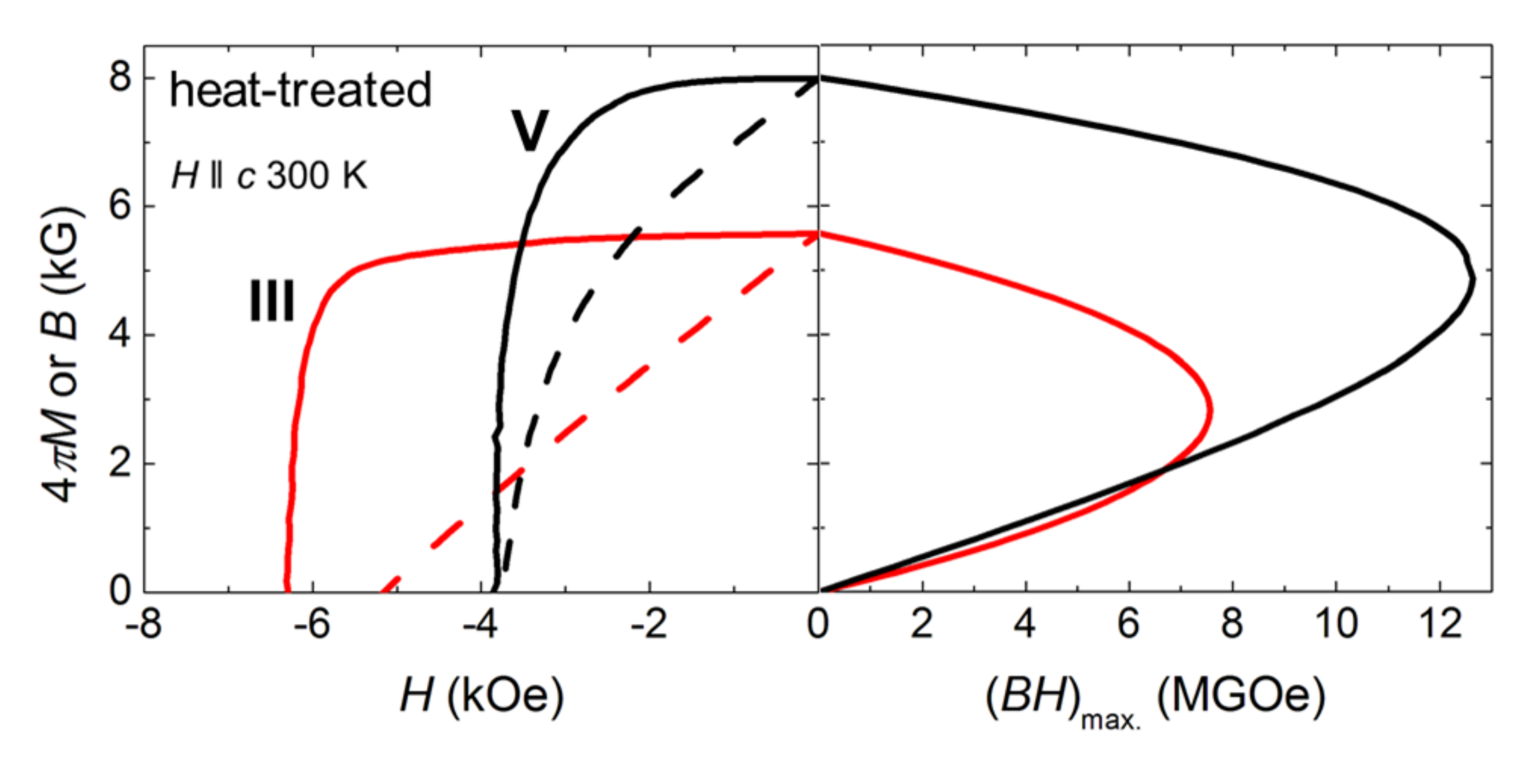}
	\caption{Room temperature second quadrant magnetic hysteresis loops for the heat-treated crystals $\textbf{III}$ -- Ce$_{15.7}$Ta$_{0.6}$Co$_{67.8}$Cu$_{15.9}$ and $\textbf{V}$ - Ce$_{14.3}$Ta$_{1.0}$Co$_{62.0}$Fe$_{12.3}$Cu$_{10.4}$, 4$\pi$$\textit{M}$ indicated as solid line and $\textit{B}$ as a dashed line (left pannel). Estimation of the energy products ($\textit{BH}$)$_{max.}$ (right pannel).}
	\label{Energy_ht}
\end{figure} 

The detailed high resolution STEM examination of the as-grown sample $\textbf{III}$ [Fig.~\ref{TEM_FIG1}] showed the basic uniformity and integrity with small Co-enriched and Cu depleted regions/stripes coherently dispersed throughout the matrix. Unfortunately, the size of these stripes did not allow for EDS composition determination or structural analysis. However, based on previous reports \cite{15Chin,16Nesbitt,17Leamy}, we assume that these are embryonic structural defects caused by stacking faults compensating for various channel disorders within the material. These may also be the nucleation sites for the decomposition and/or miscibility gap as suggested in previous literature. \cite{30Girodin}

\begin{center}
	\textbf{B}. \textbf{Heat-treated crystals. Coercivity, pinning and magnetic energy.}
\end{center}
After heat treatment, the crystals $\textbf{I}$ -- $\textbf{IV}$ show significantly increased magnetic hystereses with a monotonic increase of coercivity, $\textit{H}_c$, and a linear decrease of spontaneous magnetization $\textit{M}_s$ with increasing Cu content [Fig.~\ref{LOOPS}]. For example, the magnetic characteristics of $\textbf{III}$ change as follows: significant increase of $\textit{H}_c$ from $\sim$1.6 to $\sim$6.3 kOe with an increase of $\textit{B}_r$ ($\textit{M}_s$) from $\sim$4.2 (4.2) to $\sim$5.3 (5.7) kG, resulting in ($\textit{BH}$)$_{max.}$ of $\sim$7.8 MGOe [Fig.~\ref{Energy_ht}]. In addition to the conspicuous increase in magnetic hysteresis, there is a note-worthy increase in saturated magnetization of the heat-treated samples [Fig.~\ref{Motivation} and Fig.~\ref{LOOPS}, upper inset]. This increase in magnetic properties correlates with the appearance of the Ta-rich precipitates [see SEM images above, Fig.~\ref{FIG2}]. The STEM analysis confirmed that these are 90 - 95 \% pure rectangular blocks of Ta (according to diffraction patterns and elemental analysis), and their interfaces were coherent with the matrix material. However, high magnification TEM EDS maps [Fig. ~\ref{TEM_FIG2} (c), Fig. ~\ref{TEM_FIG3}] observed a Cu-deficient and Co-enriched layer at the interface of the precipitates and the matrix, and Co was detected in precipitates as lines, which somewhat resemble observations of rare Co-enriched and Cu-depleted lines in the as-grown STEM examination [Fig.~\ref{TEM_FIG1}]. 

These results suggest that the high coercivity may be explained by the Ta-rich precipitates serving as pinning sites and can be described using a simple domain pinning model. Typically, the coercive force is inversely proportional to the saturation magnetization for a particular magnetocrystalline energy ($\textit{H}_c$ = $\sqrt{\textit{AK}}$/$\textit{M$_s$l}$, where $\textit{A}$~-- exchange constant, $\textit{K}$ -- magnetocrystalline anisotropy, $\textit{M}_s$ -- saturation magnetization and $\textit{l}$ -- the distance between the precipitates). \cite{15Chin,16Nesbitt,17Leamy}  According to the equation, by increasing the amount of pinning precipitates we decrese the volume fraction of the matrix material and magnetization $\textit{M}_s$ of the system. Also the distances $\textit{l}$ between the precipitates become shorter. As a result, the coercivity $\textit{H}_c$  increases. Thus, the $\textit{H}_c$ of our crystals should be directly proportional to the Ta content. However we observe the inverse proportionality: total Ta content monotonically decreases in crystals $\textbf{I}$ through $\textbf{IV}$ [Table ~\ref{Table 1}], whereas the coercivity monotonically increases [Fig.~\ref{LOOPS}]. 

In contrast, the $\textit{H}_c$ increase corelates directly with increasing Cu content [Fig.~\ref{LOOPS}, see both insets], also following the proposed precipitation coercivity mechansim (see equation above). Pinning of magnetic domains should occur on the precipitates, amount of which is regulated by Cu rather than Ta content. However we do not observe precipitates that are clearly associated with Cu, except Cu-depleted regions observed in STEM experiments [Fig. ~\ref{TEM_FIG2} (c), Fig. ~\ref{TEM_FIG3}]. 

Therefore, we consider the Ta-rich precipitates as a secondary effect, which decorates the extended 3D defects and structural imperfections that originate from Cu depleted and Co enriched lines observed in the as-grown crystals [Fig.~\ref{TEM_FIG1}] and consequently develop into the regions between Ta-rich precipitates and matrix in the thermally aged crystals [Fig. ~\ref{TEM_FIG2} (c), Fig. ~\ref{TEM_FIG3}]. The amount of these imperefections must increase with increasing Cu content and lead to incresed coercivity.

According to the literature, coercivity in the Cu and Fe substituted CeCo$_5$ is casued by fine preciptates which originate from partial matrix decomposition similar to eutectoidal, observed in pure CeCo$_{5}$. \cite{15Chin,16Nesbitt,17Leamy} Whereas in the Cu substituted CeCo$_{5}$ the pronounced coercivity is related to  a miscibility gap with a critical temperature close to 800 $^\circ$C.\cite{30Girodin} Both observations support the idea of intragranual domain pinning on extended 3D deffects created as results of matrix phase transformations during the heat-treatment (hardening) of the samples. In the first case, the precipitated 2:17 phase being less anisotropic than matrix serves as a pinning site and contributes sligtly to increase of magnetization. In the second, because of decreased miscibility at lower temperatures, two phases with similar Cu/Co ratios and different Curie temperatures exist. The Cu-poor phase supposedly serves as a pinning precipitate with increased magnetization, and the Cu-richer phase contributes towards the higher anisotropy matrix. One indirect confirmation of such mechanism is observed in present Fe-free crystals \textbf{I} -- \textbf{IV}, which show atypical and increasing magnetization after the heat treatment [see the left inset in Fig.~\ref{LOOPS}]. This suggests a change of the magnetic nature of the matrix. However, this must occur with a minimal composition change as no significant difference in compositions were detected before and after the heat treatment [Table ~\ref{Table 1}]. With the addition of Fe, the decomposition process complicates, and besides the miscibility gap, the precipitation of the very stable 2:17 Ce/Co/Fe phase is possible.\cite{15Chin,16Nesbitt,17Leamy} This however was not clearly confirmed in present Fe-doped crystal $\textbf{V}$. Current SEM/EDS examinations of $\textbf{V}$ show a microstructure similar to the Fe-free crystals $\textbf{III}$ and $\textbf{IV}$ [Fig.~\ref{FIG2}]. After the heat treatment the 2:17 phase was not observed. 

Another explanation of pronounced increse in magnetization after the heat tretment may be associated with removal of Ta from the matrix material. Please note that the increase in magnetization is most pronounced in $\textbf{IV}$ with most complete removal of Ta after the heat treatment [see Table ~\ref{Table 1}]. One possible explanation for the surprisingly large impact of the removal of Ta on magnetic properties of our CeCo$_5$-based material is as follows. Previous theoretical work \cite{Nordstroem1990} finds that CeCo$_5$ is surprisingly near to a non-magnetic state, based on Stoner physics, despite its substantial Curie point. We suggest that Ta may locally drive the system toward a non-magnetic or less-magnetic state, so that its removal may restore or enhance magnetic character locally.  Further theoretical work would clearly be needed to address this notion, and it may well be difficult to account quantitatively for the observed magnitude of the behavior - $\sim$25 percent increase in magnetization for a removal of 0.3 atomic percent of Ta. Nevertheless, systems near a magnetic instability can exhibit a disproportionate response to small impurity concentrations, as in paramagnetic Fe impurities having huge effective moments in a Pd-Rh matrix \cite{Manuel1963}, and we suggest the possibility of similar behavior here.

\section{Theoretical calculations}

There are two main outstanding questions associated with extraordinary magnetic nature of the Cu and Fe substituted CeCo$_5$: $\textit{i}$ - strong improvement of both Curie temperature and magnetization with addition of Fe and  $\textit{ii}$ - high coercivity that is driven primarely by Cu regulated intragranual pinning mechanism rather than strong magnetocrystalline anisotropy.  We address these in the next two chapters through theoretical calcualtions and multiscale modeling.

\begin{center}
	\textbf{A}. \textbf{Increase of Curie temperature in Fe-doped samples.}
\end{center}

To understand the observed magnetic behavior and increase in Curie point with Fe substitution, first principles calculations for CeCo$_{5}$ and CeCo$_{4}$Cu were performed using the density functional theory as implemented in the WIEN2K code\cite{48Blaha}. Calculations were performed using the experimental lattice parameters. In this structure there all internal coordinates are symmetry-dictated, so no internal coordinate optimization was performed. The LAPW sphere radii were set to 2.4 Bohr for Ce and 2.0 Bohr for Co and Cu. In addition to ensure the well convergence of the basis set $\textit{R}$$\textit{k}$$_{max}$ = 9.0, was used $\textit{R}$ and $\textit{k}$$_{max}$ are the smallest LAPW sphere radius and interstitial plane-wave cutoff, respectively.) All the calculations are performed by assuming collinear spin arrangements. The magnetic anisotropy energy (MAE) is obtained by calculating the total energies of the system with spin obit coupling (SOC) as $\textit{K}$ = $\textit{E}$$_{[110]}$ - $\textit{E}$$_{[001]}$, where $\textit{E}$$_{[110]}$  and $\textit{E}$$_{[001]}$,  are the total energies for the magnetization oriented along the $\textit{a}$ and $\textit{c}$ directions, respectively.  For MAE calculations the convergence with respect to $\textit{K}$-points was carefully checked.   All the MAE results reported in this paper correspond to 2000 reducible $\textit{K}$-points in the full Brillouin zone. To correctly treat the strong interactions between the Ce-$\textit{f}$ electrons, the Hubbard $\textit{U}$ correction was applied, with $\textit{U}$$_{Ce}$ = 3.0 eV. For the DFT+$\textit{U}$ calculations, the standard self-interaction correction (SIC)\cite{49Anisimov,50Liechtenstein} method was used. 

\begin{figure}[!h]
\includegraphics[width=8.8cm,height=5.7cm]{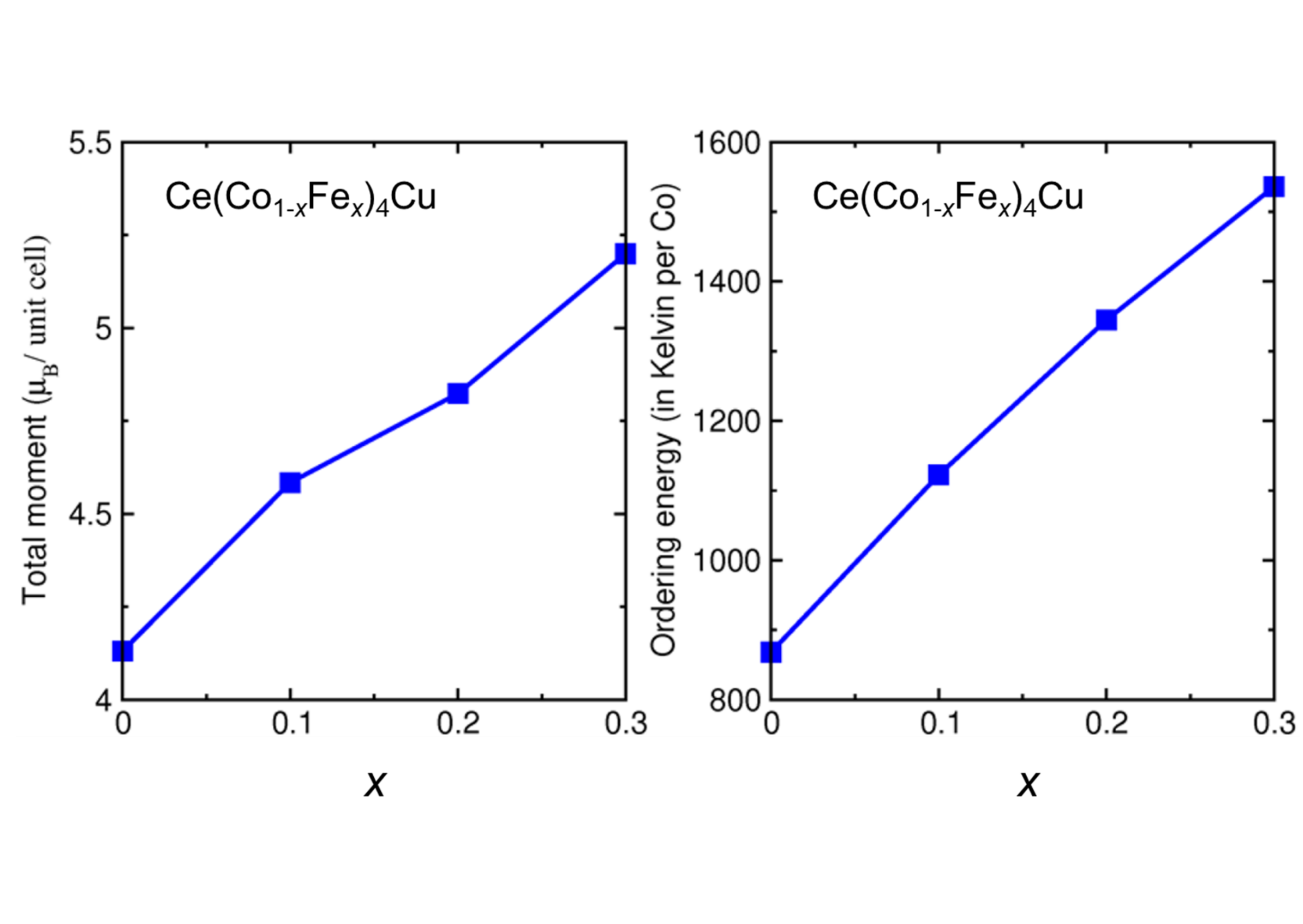}
\caption{The energy difference between nonmagnetic state and ground state (Ferromagnetic state) in Kelvin on per Co/Fe basis as a function of Fe doping for Ce(Co$_{1-x}$Fe$_{x})$$_{4}$Cu.}
\label{FIG10}
\end{figure}

For modeling of CeCo$_{4}$Cu, Cu was substituted in the Co hexagonal ring (2$\textit{c}$ Co-site), as our calculations find this location for Cu to be energetically favorable (relative to the 3$\textit{g}$ site) by some 30 meV/Cu. Fe alloying in CeCo$_{4}$Cu was realized within the virtual crystal approximation (VCA). The calculated magnetic behavior for CeCo$_{5}$ is in good agreement with the experimental measurements with a total magnetization of 6.8 $\mu$$_{B}$ per unit cell. The calculated spin moment on each 2$\textit{c}$-Co atoms was 1.42 $\mu$$_{B}$ whereas the moment on 3$\textit{g}$-Co atoms is 1.5 $\mu$$_{B}$. This is accompanied by a small orbital moment of $\sim$ 0.13 $\mu$$_{B}$.  The calculated Ce spin moment is -0.71 $\mu$$_{B}$. Upon Cu substitution (for CeCo$_{4}$Cu) the moment on Co atoms is reduced to 1.18 and 1.40 $\mu$$_{B}$ on Co-2$\textit{c}$ and Co-3$\textit{g}$ site, respectively. However on Fe allloying in CeCo$_{4}$Cu (CeCo$_{4-x}$Fe$_{x}$Cu) the total magnetization in the unit cell increases linearly with Fe doping as shown in Figure ~\ref{FIG10}a. The calculated MAE of $\sim$ 3.17 MJ/m$^{3}$ without including Hubbard $\textit{U}$ parameter ($\textit{U}$ = 0) is small compared to the experimental MAE of ~10.5 MJ/m$^{3}$.\cite{51BARTASHEVICH} However a GGA+$\textit{U}$ calculation with $\textit{U}$$_{Ce}$ as 3.0 eV gives a MAE of ~9.0 MJ/m$^{3}$ in excellent agreement with experimental value. 

The most remarkable observation of experimental measurements is the increase in Curie temperature of CeCo$_{5}$ by alloying with Cu and Fe. We explain this observation
using two methods, one more roughly qualitative, the other more quantitative.  For the first method, we note that for a local moment magnetic material, the Curie point is ultimately controlled by the interatomic exchange interactions, which are often determined by an effective mapping of the first-principles-calculated energies of various magnetic configurations to a Heisenberg-type model.  However, the magnetic configurations considered here (using the parameters above) all converged instead to the spin-polarized case with all Co spins aligned in the ferromagnetic fashion. These calculations suggest that the magnetism in this material is of itinerant type.

For such a system the calculation of the Curie point is more involved.  For this first qualitative approach we therefore limit ourselves to a simple consideration of the magnetic ordering or formation energy - the energy difference $\Delta E$ =  $\textit{E}$$_{NM}$-$\textit{E}$$_{FM}$  where $\textit{E}$$_{NM}$ is the energy of a non-magnetic configuration and $\textit{E}$$_{FM}$ is the energy of the ferromagnetic configuration. While this energy contains contributions from both the {\it intra}-atomic Hund's rule coupling (which does not determine Curie points) and the {\it inter}-atomic exchange coupling (which does determine Curie points), it is plausible that within a given alloy system, such as CeCo$_{4-x}$Fe$_{x}$Cu, the {\it trend} of the Curie point with alloying is generally captured by the {\it trend} of this energy.  For example, the general quenching of 3$d$ orbital moments in magnetic systems away from the atomic limit indicates, as expected, that the atomic Hund's rule coupling is not the only relevant interaction here. 

In order to get some insight into the Curie temperature, we have therefore calculated this energy for Ce(Co$_{1-x}$Fe$_{x}$)$_{4}$Cu on a per Co/Fe basis. This, along with the associated magnetic moment, is plotted in Figure ~\ref{FIG10} as a function of Fe doping.  One observes a substantial ordering energy increase (from $\sim$~900 K to $\sim$ 1500 K) with increasing Fe doping from $\textit{x}$ = 0 to $\textit{x}$ = 0.3, along with a substantial magnetic moment increase.  Both these results are consistent with the experimental observation of increased Curie point with Fe alloying. From a theoretical standpoint, it is noteworthy that the ordering energy is as high as 1500 K for $\textit{x}$ = 0.3.  This large energy does suggest the possibility of some local character with increased Fe content in Ce(Co$_{1-x}$Fe$_{x}$)$_{4}$Cu.  Note that a previous theoretical treatment of CeCo$_{5}$ \cite{53Nordstroem1990} found evidence, as we do here, for itinerant character in CeCo$_{5}$, so that increased Fe contents in these materials are of both theoretical and experimental interest.

Our second method of calculating the Curie point of this system is more quantitative.  While the increase of magnetization and a corresponding magnetic formation energy (Fig. \ref{FIG10})  with the addition of Fe atoms to CeCo$_5$ is generally expected, the observed increase of Curie temperature $T_{\mathrm{C}}$  needs more detailed explanation. To describe theoretically the dependence of the $T_{\mathrm{C}}$  on concentration of Fe we employed the traditional mean field approximation (MFA)
for systems with non equivalent magnetic atoms in a primitive cell. In MFA, $T_{\mathrm{C}}$ of the system with $N$ nonequivalent
magnetic atoms is calculated as the largest solution of the equation
\begin{equation}
\mathop{\rm det}[T_{nm}-T\delta_{nm}]=0,  \label{tcmfa}
\end{equation}
where $n$ and $m$ are the indices of the non-equivalent magnetic
sublattices, $T_{nm}=\frac{2}{3} J_{mn}^{0}$, and $J_{mn}^{0}$ is an
effective interaction of an atom from sublattice $n$ with all other atoms
from the sublattice $m$. The exchange coupling parameters $J_{mn}^{0}$ have been obtained
using the multiple scattering formalism expression obtained in Ref.\cite{JIJ} and extensively tested in Ref.\cite{rev}. The corresponding dependence on concentration has been described using the coherent potential approximation (see details in Ref.\cite{CPA}).

The calculations for pure CeCo$_5$  revealed that the value of $T_{\mathrm{C}}$  is determined by the strong ferromagnetic nature of the Co-Co interactions and the absolute value of the exchange parameters decays quickly with increasing interatomic distance so that the main contribution to $T_{\mathrm{C}}$ coming from the interaction between atoms lying in a distance of two first neighboring shells. The average value of the nearest neighbor coupling $J_{Co-Co}$ is around 15 meV.  The contribution from Ce atoms is weak and negative. 

The obtained MFA value of $T_{\mathrm{C}}$  = 790K in CeCo$_5$  shows the typical overestimation of experimental $T_{\mathrm{C}}$  in this classical spin approach. The addition of Fe atoms shows an interesting development of exchange coupling. First, we notice the appearance of strong and ferromagnetic Fe-Co coupling $J_{Fe-Co}$=21meV. Second, the addition of Fe atoms increases the magnetic moments on Co atoms by approximately 0.1 $\mu_B$ with a corresponding increase of Co-Co exchange coupling as well. Overall this effect leads to nearly 20\%  increase of $T_{\mathrm{C}}$  with a maximum of 950K at around x=0.2-0.23 qualitatively confirming the experimentally observed trend.  We find that the further addition of Fe atoms is detrimental for the ferromagnetism in these alloys and $T_{\mathrm{C}}$  starts to decrease. Finally, we also find that a theoretical large increase of $T_{\mathrm{C}}$  by nearly 35\% can be obtained when replacing Ce by Y atoms in CeCo$_5$ alloys.

\begin{center}
	\textbf{B}. \textbf{Multiscale modeling.}
\end{center}

In this section we present theoretical studies of hysteretic behavior of Cu- and Ta-doped CeCo$_5$ crystals in order to better understand the mechanism of coercivity in these systems. The physics of magnetic hysteresis involves multiple length scales and is controlled both by intrinsic and extrinsic properties of magnets. Therefore, our method is based on a multiscale approach that combines first principles electronic structure calculations, micromagnetic models, and statistical macromagnetic simulations.

Electronic structure calculations describe material behavior on subatomic length scales and allow us to evaluate intrinsic properties like spontaneous magnetization and magnetocrystaline anisotropy energy (MAE). We used density functional theory (DFT) within the rotationally invariant DFT+$\textit{U}$ method\cite{50Liechtenstein} and the PBE approximation to the exchange-correlation functional.\cite{54perdew1996generalized} We used $\textit{U}$=2 eV and $\textit{J}$=0.8 eV for Co 3\emph{d} elecrons which was demonstrated to provide good description of magnetic properties for LaCo$_5$ and YCo$_5$ materials.\cite{Antropov2018} The Kohn-Sham equations were solved using the projector-augmented wave method \cite{55Bloechl1994} as implemented in VASP code.\cite{56Kresse1996,57Kresse1999} We used a 1$\times$1$\times$2 supercell with respect to the primitive unit cell for CeCo$_5$. For Brillouin zone sampling the Monkhost-Pack scheme\cite{58Monkhorst1976} was used with a dense 16$\times$16$\times$10 k-mesh. The energy cutoff for the plane wave expansion was set to 320 eV. The crystal structures were fully relaxed until forces acting on each atom were smaller than 0.01 eV/\AA\ and all stress tensor elements were smaller than 1 kbar. The MAE was evaluated using the force theorem as a total energy difference between states with magnetization aligned along [100] and [001] directions. 

\begin{figure}[t]
\includegraphics[scale =0.3]{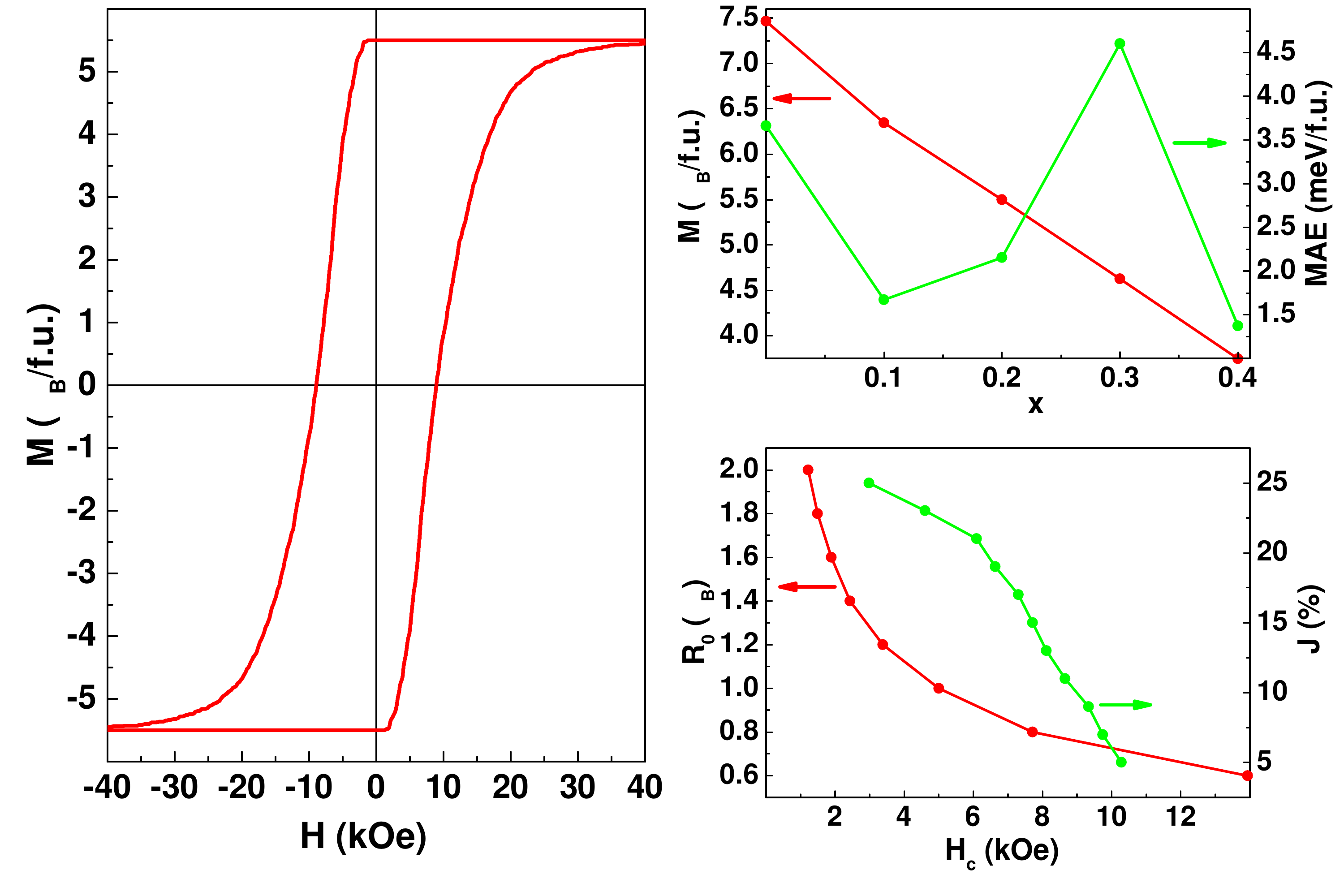}
\caption{(Left) hysteresis loop of Ce(Co$_{0.8}$Cu$_{0.2}$)$_5$ system calculated using $\textit{J}$=15\% and $\textit{R}_{0}=0.8\delta_{B}$. (Right top) calculated spontaneous magnetization and MAE of Ce(Co$_{1-x}$Cu$_x$)$_5$ as a function of Cu concentration. (Right bottom) coercivity of Ce(Co$_{0.8}$Cu$_{0.2}$)$_5$ system as a function of $\textit{R}_{0}$ and $\textit{J}$ parameters. $\textit{J}$ ($\textit{R}_{0}$) dependence was evaluated for fixed $\textit{R}_{0}$=0.8 ($\textit{J}$=15\%).}
\label{Theory}
\end{figure}

Figure \ref{Theory} (top right) shows the calculated spontaneous magnetization and MAE for Ce(Co$_{1-x}$Cu$_x$)$_5$ as a function of Cu concentration. We assumed that Cu atoms occupy 3$\textit{g}$ atomic sites and we chose lowest energy atomic configurations that correspond to Cu atoms occupying the same atomic layers. As seen, the spontaneous magnetization decreases with Cu concentration since Cu atoms have negligible moments as expected from their 3$\textit{d}^{10}$ nominal configuration. This is in agreement with experimental results shown in Fig. \ref{LOOPS}. Regarding MAE, our calculations show that it has a complex nonmonotonous dependence on Cu concentration. In particular, whereas small Cu additions decrease MAE, for larger concentrations MAE shows a maximum as a function of x. This behavior is consistent with experimental results for anisotropy field as a function of Cu concentration shown in Fig. \ref{Anisotropy_I_II_IV} except that, experimentally, the MAE maximum is observed at lower Cu concentrations and the corresponding MAE value is lower than the one for pure CeCo$_5$. This is probably caused by the fact that the configuration of Cu atoms in real samples differ somehow from the lowest energy configurations used in calculations.

In order to study hysteresis process, in addition to the knowledge of the calculated-above, intrinsic parameters, we need to also specify the microstructural features of the system at the nanometer and micron length scales. As discussed in the previous sections, SEM and STEM measurements indicate that a network of planar defects is present in the single crystal samples. For the as-grown crystals these defects form Co-enriched and Cu-depleted laminar regions, which after heat treatment, become nucleation points for Ta-rich planar precipitates. Clearly, these extended defects play a crucial role in establishing coercivity in the system since they can efficiently pin the reversed magnetic domains preventing them from expanding  over the entire crystal. In fact, as seen in Fig. \ref{FIG2}, the crystal can be viewed as a collection of blocks (of the size of several microns) that are, to a large degree, magnetically becoupled by the planar defects. In our model we consider an idealized version of such microstructure in the form of a simple cubic lattice of identical cubic micron-size Ce(Co$_{1-x}$Cu$_x$)$_5$  blocks. We assume periodic boundary conditions in a closed magnetic circuit setup so that there is no demagnetization field. Since the decoupling of blocks by the planar defects is not perfect, we introduce parameter $\textit{J}$ which represents the probability on neighboring blocks being exchange coupled. In addition, blocks interact by magnetostatic interaction that was described using the Ewald technique as described in Ref. \onlinecite{60Wysocki2017}. Similarly to the approach in Ref. \onlinecite{61Blank1991}, we assume that each block has a number of magnetically soft defects (e.g. Co precipitates) with the exponentially distributed sizes $f\left(R\right)=\frac{1}{{R}_{0}}{e}^{-R/{R}_{0}}$ where the parameter $\textit{R}_{0}$ is the characteristic defect size. Assuming well-separated spherical defects, the switching field of each block can be determined by micromagnetic calculations using intrinsic parameters calculated using first priniciples calculations above. The hysteresis loop is then calculated by starting from the saturated state and gradually decreasing the external magnetic field. For each value of the external field the system magnetization is determined as follows. The total magnetic field acting on each block is evaluated as a sum of the external and magnetostatic contributions. When the total field is lower than the negative switching field of the block, the block magnetization is reversed. Subsequently, all blocks that are exchange coupled to the reversed block have also their magnetization reversed. The process is repeated iteratively, until stable magnetic configuration is achieved. 

Figure \ref{Theory} (left) shows the calculated hysteresis loop for a Ce(Co$_{0.8}$Cu$_{0.2}$)$_5$ system calculated using $\textit{J}$=15\% and $\textit{R}_{0}$ = 0.8$\delta_{B}$ where $\delta_{B}$ is the Bloch domain wall thickness of CeCo$_5$. As seen, this choice of parameter results in a rectangular hysteresis loop with the coercivity similar to the measured value for such Cu concentration [see Fig. \ref{LOOPS}]. Therefore, our model is capable of reproducing experimental results. Fig. \ref{Theory} (bottom right) shows the dependence of the coercivity on the microstructural parameters: $\textit{J}$ and $\textit{R}_{0}$. We observe that for $\textit{R}_{0}<\delta_{B}$,  $\textit{R}_{0}$ strongly affects the coercivity while for $\textit{R}_{0}>\delta_{B}$, the coercivity depends weakly on $\textit{R}_{0}$. As expected, coercivity decreases with $\textit{J}$. Interestingly, around $\textit{J}$ = 20\% we observe a change of slope and a much stronger reduction of coercivity is observed for larger $\textit{J}$. These results indicate that there are two possible mechanisms by which Cu doping enhances the coercivity. First, Cu doping may increase the number and thickness of the planar defects resulting in a decrease of the $\textit{J}$ parameter. This is consistent with a scenario in which the defect are, in fact, precipitates of the Cu-poor 1:5 phase. Second possibility is that the Cu doping reduce the size of the magnetically soft defects in the matrix phase.

\begin{center}
	\textbf{CONCLUSIONS}
\end{center}

Using a self-flux technique, we synthesised five diferent single crystals of Ta, Cu and/or Fe substituted CeCo$_5$.  They retain a CaCu$_5$ substructure and incorporate small amounts of Ta in the form of ``dumb-bells" filling the 2$\textit{e}$ crystallographic sites within the 1D hexagonal channel with the 1$\textit{a}$ Ce site, whereas Co, Cu and Fe are statistically distributed among the 2$\textit{c}$ and 3$\textit{g}$ crystallographic sites.  The as-grown crystals appear single phased and homogenous in composition. Their single crystallinity is confirmed by XRD, SEM and TEM experiments. However they also exhibit significant magnetic coercivities which are comparable to most anisotropic sintered alnico grades. After the heat treatment (hardening), magnetic characteristics significantly improve. Ta atoms leave the matrix interstices of the as-grown crystas and precipitate in form of coherant laminas creating the so-called ``composite crystal". The ``composite crystal", formed during the heat treatment, contains an 3D array of structural defects within a primarily single grain single crystal. The mechanism of coercivity is regulated by Cu, and pinning occurs on the extended 3D defects and structural imperfections that originate from Cu depleted and Co enriched lines observed in the as-grown crystals and consequently develop into the regions between Ta-rich precipitates and matrix in the thermally aged crystals.  The structural defects form as a result of a thermodynamic transformation of the matrix material asoociated with its partial decomposition and/or decreased miscibility during hardening process. Significant improvement of magnetization in the heat-treated samples may be associated either with the transformation of the matrix phase or with the removal of Ta from the matrix. Fe strongly improves both the Curie temperature and magnetization of the system, which is associated with a strong increase in the magnetic ordering energy.
The peculiar thermodynamic transformations, which lead to intragranular pinning and a unique coercivity mechanism that does not require the typical processing for the development of extrinsic magnetic properties, could be used to create permanent magnets with lowered processing costs. Further composition - temperature - time optimizations may result in a critical material free and cost efficient gap magnet with energy product 15 -- 16.5 MGOe.   
\begin{center}
\textbf{ACKNOWLEDGMENTS}
\end{center}
This research was supported by the Critical Materials Institute, an Energy Innovation Hub  funded  by  the  U.S.  Department  of  Energy,  Office of  Energy  Efficiency  and  Renewable  Energy,  Advanced Manufacturing  Office.   M.C.N.  is  supported  by  the  office of  Basic  Energy  Sciences,  Materials  Sciences  Division, U.S.  DOE.  This  work  was  performed  at  the  Ames Laboratory, operated for DOE by Iowa State University under Contract No.  DE-AC02-07CH11358. We are also greatful to Qisheng Lin for assistance with initial single crystal X-ray experiments.

\bibliographystyle{apsrev4-1}

\begin{thebibliography}{76}%
	\makeatletter
	\providecommand \@ifxundefined [1]{%
		\@ifx{#1\undefined}
	}%
	\providecommand \@ifnum [1]{%
		\ifnum #1\expandafter \@firstoftwo
		\else \expandafter \@secondoftwo
		\fi
	}%
	\providecommand \@ifx [1]{%
		\ifx #1\expandafter \@firstoftwo
		\else \expandafter \@secondoftwo
		\fi
	}%
	\providecommand \natexlab [1]{#1}%
	\providecommand \enquote  [1]{``#1''}%
	\providecommand \bibnamefont  [1]{#1}%
	\providecommand \bibfnamefont [1]{#1}%
	\providecommand \citenamefont [1]{#1}%
	\providecommand \href@noop [0]{\@secondoftwo}%
	\providecommand \href [0]{\begingroup \@sanitize@url \@href}%
	\providecommand \@href[1]{\@@startlink{#1}\@@href}%
	\providecommand \@@href[1]{\endgroup#1\@@endlink}%
	\providecommand \@sanitize@url [0]{\catcode `\\12\catcode `\$12\catcode
		`\&12\catcode `\#12\catcode `\^12\catcode `\_12\catcode `\%12\relax}%
	\providecommand \@@startlink[1]{}%
	\providecommand \@@endlink[0]{}%
	\providecommand \url  [0]{\begingroup\@sanitize@url \@url }%
	\providecommand \@url [1]{\endgroup\@href {#1}{\urlprefix }}%
	\providecommand \urlprefix  [0]{URL }%
	\providecommand \Eprint [0]{\href }%
	\providecommand \doibase [0]{http://dx.doi.org/}%
	\providecommand \selectlanguage [0]{\@gobble}%
	\providecommand \bibinfo  [0]{\@secondoftwo}%
	\providecommand \bibfield  [0]{\@secondoftwo}%
	\providecommand \translation [1]{[#1]}%
	\providecommand \BibitemOpen [0]{}%
	\providecommand \bibitemStop [0]{}%
	\providecommand \bibitemNoStop [0]{.\EOS\space}%
	\providecommand \EOS [0]{\spacefactor3000\relax}%
	\providecommand \BibitemShut  [1]{\csname bibitem#1\endcsname}%
	\let\auto@bib@innerbib\@empty
	\bibitem [{\citenamefont {Haxel}\ \emph {et~al.}(2002)\citenamefont {Haxel},
		\citenamefont {Hendrick},\ and\ \citenamefont {Orris}}]{1Haxel}%
	\BibitemOpen
	\bibfield  {author} {\bibinfo {author} {\bibfnamefont {G.~B.}\ \bibnamefont
			{Haxel}}, \bibinfo {author} {\bibfnamefont {J.~B.}\ \bibnamefont {Hendrick}},
		\ and\ \bibinfo {author} {\bibfnamefont {G.~J.}\ \bibnamefont {Orris}},\
	}\href@noop {} {\bibfield  {journal} {\bibinfo  {journal} {{U.S. Geological
					Survey Fact Sheet}}\ } (\bibinfo {year} {2002})}\BibitemShut {NoStop}%
	\bibitem [{\citenamefont {Patnaik}(2002)}]{2Patnaik2002}%
	\BibitemOpen
	\bibfield  {author} {\bibinfo {author} {\bibfnamefont {P.}~\bibnamefont
			{Patnaik}},\ }\href
	{https://www.amazon.com/Handbook-Inorganic-Chemicals-Pradyot-Patnaik/dp/0070494398?SubscriptionId=0JYN1NVW651KCA56C102&tag=techkie-20&linkCode=xm2&camp=2025&creative=165953&creativeASIN=0070494398}
	{\emph {\bibinfo {title} {Handbook of Inorganic Chemicals}}}\ (\bibinfo
	{publisher} {McGraw-Hill Professional},\ \bibinfo {year} {2002})\BibitemShut
	{NoStop}%
	\bibitem [{\citenamefont {Xie}\ \emph {et~al.}(2014)\citenamefont {Xie},
		\citenamefont {Zhang}, \citenamefont {Dreisinger},\ and\ \citenamefont
		{Doyle}}]{3Xie2014}%
	\BibitemOpen
	\bibfield  {author} {\bibinfo {author} {\bibfnamefont {F.}~\bibnamefont
			{Xie}}, \bibinfo {author} {\bibfnamefont {T.~A.}\ \bibnamefont {Zhang}},
		\bibinfo {author} {\bibfnamefont {D.}~\bibnamefont {Dreisinger}}, \ and\
		\bibinfo {author} {\bibfnamefont {F.}~\bibnamefont {Doyle}},\ }\href
	{\doibase 10.1016/j.mineng.2013.10.021} {\bibfield  {journal} {\bibinfo
			{journal} {Minerals Engineering}\ }\textbf {\bibinfo {volume} {56}},\
		\bibinfo {pages} {10} (\bibinfo {year} {2014})}\BibitemShut {NoStop}%
	\bibitem [{\citenamefont {Herbst}\ \emph {et~al.}(1985)\citenamefont {Herbst},
		\citenamefont {Croat},\ and\ \citenamefont {Yelon}}]{Herbst1985}%
	\BibitemOpen
	\bibfield  {author} {\bibinfo {author} {\bibfnamefont {J.~F.}\ \bibnamefont
			{Herbst}}, \bibinfo {author} {\bibfnamefont {J.~J.}\ \bibnamefont {Croat}}, \
		and\ \bibinfo {author} {\bibfnamefont {W.~B.}\ \bibnamefont {Yelon}},\ }\href
	{\doibase 10.1063/1.334680} {\bibfield  {journal} {\bibinfo  {journal}
			{Journal of Applied Physics}\ }\textbf {\bibinfo {volume} {57}},\ \bibinfo
		{pages} {4086} (\bibinfo {year} {1985})}\BibitemShut {NoStop}%
	\bibitem [{\citenamefont {Pathak}\ \emph {et~al.}(2015)\citenamefont {Pathak},
		\citenamefont {Khan}, \citenamefont {Gschneidner}, \citenamefont {McCallum},
		\citenamefont {Zhou}, \citenamefont {Sun}, \citenamefont {Dennis},
		\citenamefont {Zhou}, \citenamefont {Pinkerton}, \citenamefont {Kramer},\
		and\ \citenamefont {Pecharsky}}]{4Pathak}%
	\BibitemOpen
	\bibfield  {author} {\bibinfo {author} {\bibfnamefont {A.~K.}\ \bibnamefont
			{Pathak}}, \bibinfo {author} {\bibfnamefont {M.}~\bibnamefont {Khan}},
		\bibinfo {author} {\bibfnamefont {K.~A.}\ \bibnamefont {Gschneidner}},
		\bibinfo {author} {\bibfnamefont {R.~W.}\ \bibnamefont {McCallum}}, \bibinfo
		{author} {\bibfnamefont {L.}~\bibnamefont {Zhou}}, \bibinfo {author}
		{\bibfnamefont {K.}~\bibnamefont {Sun}}, \bibinfo {author} {\bibfnamefont
			{K.~W.}\ \bibnamefont {Dennis}}, \bibinfo {author} {\bibfnamefont
			{C.}~\bibnamefont {Zhou}}, \bibinfo {author} {\bibfnamefont {F.~E.}\
			\bibnamefont {Pinkerton}}, \bibinfo {author} {\bibfnamefont {M.~J.}\
			\bibnamefont {Kramer}}, \ and\ \bibinfo {author} {\bibfnamefont {V.~K.}\
			\bibnamefont {Pecharsky}},\ }\href {\doibase 10.1002/adma.201404892}
	{\bibfield  {journal} {\bibinfo  {journal} {Advanced Materials}\ }\textbf
		{\bibinfo {volume} {27}},\ \bibinfo {pages} {2663} (\bibinfo {year}
		{2015})}\BibitemShut {NoStop}%
	\bibitem [{\citenamefont {Pathak}\ \emph {et~al.}(2016)\citenamefont {Pathak},
		\citenamefont {Gschneidner}, \citenamefont {Khan}, \citenamefont {McCallum},\
		and\ \citenamefont {Pecharsky}}]{5Pathak}%
	\BibitemOpen
	\bibfield  {author} {\bibinfo {author} {\bibfnamefont {A.~K.}\ \bibnamefont
			{Pathak}}, \bibinfo {author} {\bibfnamefont {K.}~\bibnamefont {Gschneidner}},
		\bibinfo {author} {\bibfnamefont {M.}~\bibnamefont {Khan}}, \bibinfo {author}
		{\bibfnamefont {R.}~\bibnamefont {McCallum}}, \ and\ \bibinfo {author}
		{\bibfnamefont {V.}~\bibnamefont {Pecharsky}},\ }\href {\doibase
		10.1016/j.jallcom.2016.01.194} {\bibfield  {journal} {\bibinfo  {journal}
			{Journal of Alloys and Compounds}\ }\textbf {\bibinfo {volume} {668}},\
		\bibinfo {pages} {80} (\bibinfo {year} {2016})}\BibitemShut {NoStop}%
	\bibitem [{\citenamefont {Susner}\ \emph {et~al.}(2017)\citenamefont {Susner},
		\citenamefont {Conner}, \citenamefont {Saparov}, \citenamefont {McGuire},
		\citenamefont {Crumlin}, \citenamefont {Veith}, \citenamefont {Cao},
		\citenamefont {Shanavas}, \citenamefont {Parker}, \citenamefont
		{Chakoumakos},\ and\ \citenamefont {Sales}}]{Susner2017}%
	\BibitemOpen
	\bibfield  {author} {\bibinfo {author} {\bibfnamefont {M.~A.}\ \bibnamefont
			{Susner}}, \bibinfo {author} {\bibfnamefont {B.~S.}\ \bibnamefont {Conner}},
		\bibinfo {author} {\bibfnamefont {B.~I.}\ \bibnamefont {Saparov}}, \bibinfo
		{author} {\bibfnamefont {M.~A.}\ \bibnamefont {McGuire}}, \bibinfo {author}
		{\bibfnamefont {E.~J.}\ \bibnamefont {Crumlin}}, \bibinfo {author}
		{\bibfnamefont {G.~M.}\ \bibnamefont {Veith}}, \bibinfo {author}
		{\bibfnamefont {H.}~\bibnamefont {Cao}}, \bibinfo {author} {\bibfnamefont
			{K.~V.}\ \bibnamefont {Shanavas}}, \bibinfo {author} {\bibfnamefont {D.~S.}\
			\bibnamefont {Parker}}, \bibinfo {author} {\bibfnamefont {B.~C.}\
			\bibnamefont {Chakoumakos}}, \ and\ \bibinfo {author} {\bibfnamefont {B.~C.}\
			\bibnamefont {Sales}},\ }\href {\doibase 10.1016/j.jmmm.2016.10.127}
	{\bibfield  {journal} {\bibinfo  {journal} {Journal of Magnetism and Magnetic
				Materials}\ }\textbf {\bibinfo {volume} {434}},\ \bibinfo {pages} {1}
		(\bibinfo {year} {2017})}\BibitemShut {NoStop}%
	\bibitem [{\citenamefont {Strnat}\ and\ \citenamefont
		{Strnat}(1991)}]{Strnat1991}%
	\BibitemOpen
	\bibfield  {author} {\bibinfo {author} {\bibfnamefont {K.~J.}\ \bibnamefont
			{Strnat}}\ and\ \bibinfo {author} {\bibfnamefont {R.~M.}\ \bibnamefont
			{Strnat}},\ }\href {\doibase 10.1016/0304-8853(91)90811-n} {\bibfield
		{journal} {\bibinfo  {journal} {Journal of Magnetism and Magnetic Materials}\
		}\textbf {\bibinfo {volume} {100}},\ \bibinfo {pages} {38} (\bibinfo {year}
		{1991})}\BibitemShut {NoStop}%
	\bibitem [{\citenamefont {Senno}\ and\ \citenamefont
		{Tawara}(1974)}]{Senno1974}%
	\BibitemOpen
	\bibfield  {author} {\bibinfo {author} {\bibfnamefont {H.}~\bibnamefont
			{Senno}}\ and\ \bibinfo {author} {\bibfnamefont {Y.}~\bibnamefont {Tawara}},\
	}\href {\doibase 10.1109/tmag.1974.1058334} {\bibfield  {journal} {\bibinfo
			{journal} {{IEEE} Transactions on Magnetics}\ }\textbf {\bibinfo {volume}
			{10}},\ \bibinfo {pages} {313} (\bibinfo {year} {1974})}\BibitemShut
	{NoStop}%
	\bibitem [{\citenamefont {Kramer}\ \emph {et~al.}(2012)\citenamefont {Kramer},
		\citenamefont {McCallum}, \citenamefont {Anderson},\ and\ \citenamefont
		{Constantinides}}]{6Kramer}%
	\BibitemOpen
	\bibfield  {author} {\bibinfo {author} {\bibfnamefont {M.~J.}\ \bibnamefont
			{Kramer}}, \bibinfo {author} {\bibfnamefont {R.~W.}\ \bibnamefont
			{McCallum}}, \bibinfo {author} {\bibfnamefont {I.~A.}\ \bibnamefont
			{Anderson}}, \ and\ \bibinfo {author} {\bibfnamefont {S.}~\bibnamefont
			{Constantinides}},\ }\href {\doibase 10.1007/s11837-012-0351-z} {\bibfield
		{journal} {\bibinfo  {journal} {{JOM}}\ }\textbf {\bibinfo {volume} {64}},\
		\bibinfo {pages} {752} (\bibinfo {year} {2012})}\BibitemShut {NoStop}%
	\bibitem [{\citenamefont {McCallum}\ \emph {et~al.}(2014)\citenamefont
		{McCallum}, \citenamefont {Lewis}, \citenamefont {Skomski}, \citenamefont
		{Kramer},\ and\ \citenamefont {Anderson}}]{7McCallum}%
	\BibitemOpen
	\bibfield  {author} {\bibinfo {author} {\bibfnamefont {R.}~\bibnamefont
			{McCallum}}, \bibinfo {author} {\bibfnamefont {L.}~\bibnamefont {Lewis}},
		\bibinfo {author} {\bibfnamefont {R.}~\bibnamefont {Skomski}}, \bibinfo
		{author} {\bibfnamefont {M.}~\bibnamefont {Kramer}}, \ and\ \bibinfo {author}
		{\bibfnamefont {I.}~\bibnamefont {Anderson}},\ }\href {\doibase
		10.1146/annurev-matsci-070813-113457} {\bibfield  {journal} {\bibinfo
			{journal} {Annual Review of Materials Research}\ }\textbf {\bibinfo {volume}
			{44}},\ \bibinfo {pages} {451} (\bibinfo {year} {2014})}\BibitemShut
	{NoStop}%
	\bibitem [{\citenamefont {Nesbitt}(1969)}]{Nesbitt1969}%
	\BibitemOpen
	\bibfield  {author} {\bibinfo {author} {\bibfnamefont {E.~A.}\ \bibnamefont
			{Nesbitt}},\ }\href {\doibase 10.1063/1.1657619} {\bibfield  {journal}
		{\bibinfo  {journal} {Journal of Applied Physics}\ }\textbf {\bibinfo
			{volume} {40}},\ \bibinfo {pages} {1259} (\bibinfo {year}
		{1969})}\BibitemShut {NoStop}%
	\bibitem [{\citenamefont {Nesbitt}\ \emph {et~al.}(1968)\citenamefont
		{Nesbitt}, \citenamefont {Willens}, \citenamefont {Sherwood}, \citenamefont
		{Buehler},\ and\ \citenamefont {Wernick}}]{8Nesbitt}%
	\BibitemOpen
	\bibfield  {author} {\bibinfo {author} {\bibfnamefont {E.~A.}\ \bibnamefont
			{Nesbitt}}, \bibinfo {author} {\bibfnamefont {R.~H.}\ \bibnamefont
			{Willens}}, \bibinfo {author} {\bibfnamefont {R.~C.}\ \bibnamefont
			{Sherwood}}, \bibinfo {author} {\bibfnamefont {E.}~\bibnamefont {Buehler}}, \
		and\ \bibinfo {author} {\bibfnamefont {J.~H.}\ \bibnamefont {Wernick}},\
	}\href {\doibase 10.1063/1.1651855} {\bibfield  {journal} {\bibinfo
			{journal} {Applied Physics Letters}\ }\textbf {\bibinfo {volume} {12}},\
		\bibinfo {pages} {361} (\bibinfo {year} {1968})}\BibitemShut {NoStop}%
	\bibitem [{\citenamefont {Tawara}\ and\ \citenamefont {Senno}(1968)}]{9Tawara}%
	\BibitemOpen
	\bibfield  {author} {\bibinfo {author} {\bibfnamefont {Y.}~\bibnamefont
			{Tawara}}\ and\ \bibinfo {author} {\bibfnamefont {H.}~\bibnamefont {Senno}},\
	}\href {\doibase 10.1143/jjap.7.966} {\bibfield  {journal} {\bibinfo
			{journal} {Japanese Journal of Applied Physics}\ }\textbf {\bibinfo {volume}
			{7}},\ \bibinfo {pages} {966} (\bibinfo {year} {1968})}\BibitemShut {NoStop}%
	\bibitem [{\citenamefont {Coey}(2012)}]{Coey2012}%
	\BibitemOpen
	\bibfield  {author} {\bibinfo {author} {\bibfnamefont {J.}~\bibnamefont
			{Coey}},\ }\href {\doibase 10.1016/j.scriptamat.2012.04.036} {\bibfield
		{journal} {\bibinfo  {journal} {Scripta Materialia}\ }\textbf {\bibinfo
			{volume} {67}},\ \bibinfo {pages} {524} (\bibinfo {year} {2012})}\BibitemShut
	{NoStop}%
	\bibitem [{\citenamefont {Senno}\ and\ \citenamefont {Tawara}(1969)}]{10Senno}%
	\BibitemOpen
	\bibfield  {author} {\bibinfo {author} {\bibfnamefont {H.}~\bibnamefont
			{Senno}}\ and\ \bibinfo {author} {\bibfnamefont {Y.}~\bibnamefont {Tawara}},\
	}\href {\doibase 10.1143/jjap.8.118} {\bibfield  {journal} {\bibinfo
			{journal} {Japanese Journal of Applied Physics}\ }\textbf {\bibinfo {volume}
			{8}},\ \bibinfo {pages} {118} (\bibinfo {year} {1969})}\BibitemShut {NoStop}%
	\bibitem [{\citenamefont {Nesbitt}\ \emph {et~al.}(1969)\citenamefont
		{Nesbitt}, \citenamefont {Chin}, \citenamefont {Sherwood},\ and\
		\citenamefont {Wernick}}]{11Nesbitt}%
	\BibitemOpen
	\bibfield  {author} {\bibinfo {author} {\bibfnamefont {E.~A.}\ \bibnamefont
			{Nesbitt}}, \bibinfo {author} {\bibfnamefont {G.~Y.}\ \bibnamefont {Chin}},
		\bibinfo {author} {\bibfnamefont {R.~C.}\ \bibnamefont {Sherwood}}, \ and\
		\bibinfo {author} {\bibfnamefont {J.~H.}\ \bibnamefont {Wernick}},\ }\href
	{\doibase 10.1063/1.1657135} {\bibfield  {journal} {\bibinfo  {journal}
			{Journal of Applied Physics}\ }\textbf {\bibinfo {volume} {40}},\ \bibinfo
		{pages} {4006} (\bibinfo {year} {1969})}\BibitemShut {NoStop}%
	\bibitem [{\citenamefont {Cullen}(1971)}]{12Cullen}%
	\BibitemOpen
	\bibfield  {author} {\bibinfo {author} {\bibfnamefont {T.~J.}\ \bibnamefont
			{Cullen}},\ }\href {\doibase 10.1063/1.1660328} {\bibfield  {journal}
		{\bibinfo  {journal} {Journal of Applied Physics}\ }\textbf {\bibinfo
			{volume} {42}},\ \bibinfo {pages} {1535} (\bibinfo {year}
		{1971})}\BibitemShut {NoStop}%
	\bibitem [{\citenamefont {Sherwood}\ \emph {et~al.}(1972)\citenamefont
		{Sherwood}, \citenamefont {Nesbitt}, \citenamefont {Chin},\ and\
		\citenamefont {Green}}]{13Sherwood}%
	\BibitemOpen
	\bibfield  {author} {\bibinfo {author} {\bibfnamefont {R.}~\bibnamefont
			{Sherwood}}, \bibinfo {author} {\bibfnamefont {E.}~\bibnamefont {Nesbitt}},
		\bibinfo {author} {\bibfnamefont {G.}~\bibnamefont {Chin}}, \ and\ \bibinfo
		{author} {\bibfnamefont {M.}~\bibnamefont {Green}},\ }\href {\doibase
		10.1016/0025-5408(72)90151-1} {\bibfield  {journal} {\bibinfo  {journal}
			{Materials Research Bulletin}\ }\textbf {\bibinfo {volume} {7}},\ \bibinfo
		{pages} {489} (\bibinfo {year} {1972})}\BibitemShut {NoStop}%
	\bibitem [{\citenamefont {Tawara}\ and\ \citenamefont
		{Senno}(1972)}]{14Tawara}%
	\BibitemOpen
	\bibfield  {author} {\bibinfo {author} {\bibfnamefont {Y.}~\bibnamefont
			{Tawara}}\ and\ \bibinfo {author} {\bibfnamefont {H.}~\bibnamefont {Senno}},\
	}\href {\doibase 10.1109/tmag.1972.1067360} {\bibfield  {journal} {\bibinfo
			{journal} {{IEEE} Transactions on Magnetics}\ }\textbf {\bibinfo {volume}
			{8}},\ \bibinfo {pages} {560} (\bibinfo {year} {1972})}\BibitemShut {NoStop}%
	\bibitem [{\citenamefont {Chin}\ \emph {et~al.}(1972)\citenamefont {Chin},
		\citenamefont {Green}, \citenamefont {Nesbitt}, \citenamefont {Sherwood},\
		and\ \citenamefont {Wernick}}]{15Chin}%
	\BibitemOpen
	\bibfield  {author} {\bibinfo {author} {\bibfnamefont {G.}~\bibnamefont
			{Chin}}, \bibinfo {author} {\bibfnamefont {M.}~\bibnamefont {Green}},
		\bibinfo {author} {\bibfnamefont {E.}~\bibnamefont {Nesbitt}}, \bibinfo
		{author} {\bibfnamefont {R.}~\bibnamefont {Sherwood}}, \ and\ \bibinfo
		{author} {\bibfnamefont {J.}~\bibnamefont {Wernick}},\ }\href {\doibase
		10.1109/tmag.1972.1067257} {\bibfield  {journal} {\bibinfo  {journal} {{IEEE}
				Transactions on Magnetics}\ }\textbf {\bibinfo {volume} {8}},\ \bibinfo
		{pages} {29} (\bibinfo {year} {1972})}\BibitemShut {NoStop}%
	\bibitem [{\citenamefont {Nesbitt}\ \emph
		{et~al.}(1973{\natexlab{a}})\citenamefont {Nesbitt}, \citenamefont {Chin},
		\citenamefont {Hull}, \citenamefont {Sherwood}, \citenamefont {Green},
		\citenamefont {Wernick}, \citenamefont {Wolfe}, \citenamefont {Graham},\ and\
		\citenamefont {Rhyne}}]{16Nesbitt}%
	\BibitemOpen
	\bibfield  {author} {\bibinfo {author} {\bibfnamefont {E.~A.}\ \bibnamefont
			{Nesbitt}}, \bibinfo {author} {\bibfnamefont {G.~Y.}\ \bibnamefont {Chin}},
		\bibinfo {author} {\bibfnamefont {G.~W.}\ \bibnamefont {Hull}}, \bibinfo
		{author} {\bibfnamefont {R.~C.}\ \bibnamefont {Sherwood}}, \bibinfo {author}
		{\bibfnamefont {M.~L.}\ \bibnamefont {Green}}, \bibinfo {author}
		{\bibfnamefont {J.~H.}\ \bibnamefont {Wernick}}, \bibinfo {author}
		{\bibfnamefont {H.~C.}\ \bibnamefont {Wolfe}}, \bibinfo {author}
		{\bibfnamefont {C.~D.}\ \bibnamefont {Graham}}, \ and\ \bibinfo {author}
		{\bibfnamefont {J.~J.}\ \bibnamefont {Rhyne}},\ }in\ \href {\doibase
		10.1063/1.2946975} {\emph {\bibinfo {booktitle} {{AIP} Conference
				Proceedings}}}\ (\bibinfo  {publisher} {{AIP}},\ \bibinfo {year}
	{1973})\BibitemShut {NoStop}%
	\bibitem [{\citenamefont {Leamy}\ and\ \citenamefont {Green}(1973)}]{17Leamy}%
	\BibitemOpen
	\bibfield  {author} {\bibinfo {author} {\bibfnamefont {H.}~\bibnamefont
			{Leamy}}\ and\ \bibinfo {author} {\bibfnamefont {M.}~\bibnamefont {Green}},\
	}\href {\doibase 10.1109/tmag.1973.1067642} {\bibfield  {journal} {\bibinfo
			{journal} {{IEEE} Transactions on Magnetics}\ }\textbf {\bibinfo {volume}
			{9}},\ \bibinfo {pages} {205} (\bibinfo {year} {1973})}\BibitemShut {NoStop}%
	\bibitem [{\citenamefont {Nesbitt}\ \emph
		{et~al.}(1973{\natexlab{b}})\citenamefont {Nesbitt}, \citenamefont {Chin},
		\citenamefont {Sherwood}, \citenamefont {Green},\ and\ \citenamefont
		{Leamy}}]{18Nesbitt}%
	\BibitemOpen
	\bibfield  {author} {\bibinfo {author} {\bibfnamefont {E.}~\bibnamefont
			{Nesbitt}}, \bibinfo {author} {\bibfnamefont {G.}~\bibnamefont {Chin}},
		\bibinfo {author} {\bibfnamefont {R.}~\bibnamefont {Sherwood}}, \bibinfo
		{author} {\bibfnamefont {M.}~\bibnamefont {Green}}, \ and\ \bibinfo {author}
		{\bibfnamefont {H.}~\bibnamefont {Leamy}},\ }\href {\doibase
		10.1109/tmag.1973.1067641} {\bibfield  {journal} {\bibinfo  {journal} {{IEEE}
				Transactions on Magnetics}\ }\textbf {\bibinfo {volume} {9}},\ \bibinfo
		{pages} {203} (\bibinfo {year} {1973}{\natexlab{b}})}\BibitemShut {NoStop}%
	\bibitem [{\citenamefont {Khan}(1974)}]{19Khan}%
	\BibitemOpen
	\bibfield  {author} {\bibinfo {author} {\bibfnamefont {Y.}~\bibnamefont
			{Khan}},\ }\href {\doibase 10.1016/0022-5088(74)90159-3} {\bibfield
		{journal} {\bibinfo  {journal} {Journal of the Less Common Metals}\ }\textbf
		{\bibinfo {volume} {34}},\ \bibinfo {pages} {191} (\bibinfo {year}
		{1974})}\BibitemShut {NoStop}%
	\bibitem [{\citenamefont {Arbuzov}\ \emph {et~al.}(1974)\citenamefont
		{Arbuzov}, \citenamefont {Pavlyukov},\ and\ \citenamefont
		{Leskevich}}]{20Arbuzov1974}%
	\BibitemOpen
	\bibfield  {author} {\bibinfo {author} {\bibfnamefont {M.~P.}\ \bibnamefont
			{Arbuzov}}, \bibinfo {author} {\bibfnamefont {A.~A.}\ \bibnamefont
			{Pavlyukov}}, \ and\ \bibinfo {author} {\bibfnamefont {A.~G.}\ \bibnamefont
			{Leskevich}},\ }\href@noop {} {\bibfield  {journal} {\bibinfo  {journal}
			{Poroshkovaya Metalurgiya}\ }\textbf {\bibinfo {volume} {9}},\ \bibinfo
		{pages} {48} (\bibinfo {year} {1974})}\BibitemShut {NoStop}%
	\bibitem [{\citenamefont {Arbuzov}\ \emph
		{et~al.}(1975{\natexlab{a}})\citenamefont {Arbuzov}, \citenamefont
		{Pavlyukov},\ and\ \citenamefont {Pogorilyy}}]{21Arbuzov1975}%
	\BibitemOpen
	\bibfield  {author} {\bibinfo {author} {\bibfnamefont {M.~P.}\ \bibnamefont
			{Arbuzov}}, \bibinfo {author} {\bibfnamefont {A.~A.}\ \bibnamefont
			{Pavlyukov}}, \ and\ \bibinfo {author} {\bibfnamefont {A.~G.}\ \bibnamefont
			{Pogorilyy}},\ }\href@noop {} {\bibfield  {journal} {\bibinfo  {journal}
			{Fiz. Metal. Metalloved.}\ }\textbf {\bibinfo {volume} {40}},\ \bibinfo
		{pages} {848} (\bibinfo {year} {1975}{\natexlab{a}})}\BibitemShut {NoStop}%
	\bibitem [{\citenamefont {Arbuzov}\ \emph
		{et~al.}(1975{\natexlab{b}})\citenamefont {Arbuzov}, \citenamefont
		{Pavlyukov}, \citenamefont {Pogorilyi},\ and\ \citenamefont
		{Opanasenko}}]{22Arbuzov1975}%
	\BibitemOpen
	\bibfield  {author} {\bibinfo {author} {\bibfnamefont {M.~P.}\ \bibnamefont
			{Arbuzov}}, \bibinfo {author} {\bibfnamefont {A.~A.}\ \bibnamefont
			{Pavlyukov}}, \bibinfo {author} {\bibfnamefont {A.~G.}\ \bibnamefont
			{Pogorilyi}}, \ and\ \bibinfo {author} {\bibfnamefont {O.~S.}\ \bibnamefont
			{Opanasenko}},\ }\href@noop {} {\bibfield  {journal} {\bibinfo  {journal}
			{Poroshkovaya Metalurgiya}\ }\textbf {\bibinfo {volume} {5}},\ \bibinfo
		{pages} {97} (\bibinfo {year} {1975}{\natexlab{b}})}\BibitemShut {NoStop}%
	\bibitem [{\citenamefont {Arbuzov}\ \emph
		{et~al.}(1977{\natexlab{a}})\citenamefont {Arbuzov}, \citenamefont
		{Pavlyukov}, \citenamefont {Krakovich},\ and\ \citenamefont
		{Opanasenko}}]{23Arbuzov1977}%
	\BibitemOpen
	\bibfield  {author} {\bibinfo {author} {\bibfnamefont {M.~P.}\ \bibnamefont
			{Arbuzov}}, \bibinfo {author} {\bibfnamefont {A.~A.}\ \bibnamefont
			{Pavlyukov}}, \bibinfo {author} {\bibfnamefont {E.~V.}\ \bibnamefont
			{Krakovich}}, \ and\ \bibinfo {author} {\bibfnamefont {O.~S.}\ \bibnamefont
			{Opanasenko}},\ }\href@noop {} {\bibfield  {journal} {\bibinfo  {journal}
			{Poroshkovaya Metalurgiya}\ }\textbf {\bibinfo {volume} {7}},\ \bibinfo
		{pages} {56} (\bibinfo {year} {1977}{\natexlab{a}})}\BibitemShut {NoStop}%
	\bibitem [{\citenamefont {Arbuzov}\ \emph
		{et~al.}(1977{\natexlab{b}})\citenamefont {Arbuzov}, \citenamefont
		{Pavlyukov}, \citenamefont {Krakovich},\ and\ \citenamefont
		{Opanasenko}}]{24Arbuzov1977}%
	\BibitemOpen
	\bibfield  {author} {\bibinfo {author} {\bibfnamefont {M.~P.}\ \bibnamefont
			{Arbuzov}}, \bibinfo {author} {\bibfnamefont {A.~A.}\ \bibnamefont
			{Pavlyukov}}, \bibinfo {author} {\bibfnamefont {E.~V.}\ \bibnamefont
			{Krakovich}}, \ and\ \bibinfo {author} {\bibfnamefont {O.~S.}\ \bibnamefont
			{Opanasenko}},\ }\href@noop {} {\bibfield  {journal} {\bibinfo  {journal}
			{Poroshkovaya Metalurgiya}\ }\textbf {\bibinfo {volume} {8}},\ \bibinfo
		{pages} {69} (\bibinfo {year} {1977}{\natexlab{b}})}\BibitemShut {NoStop}%
	\bibitem [{\citenamefont {Arbuzov}\ \emph
		{et~al.}(1977{\natexlab{c}})\citenamefont {Arbuzov}, \citenamefont
		{Pavlyukov}, \citenamefont {Krakovich},\ and\ \citenamefont
		{Opanasenko}}]{25Arbuzov1977}%
	\BibitemOpen
	\bibfield  {author} {\bibinfo {author} {\bibfnamefont {M.~P.}\ \bibnamefont
			{Arbuzov}}, \bibinfo {author} {\bibfnamefont {A.~A.}\ \bibnamefont
			{Pavlyukov}}, \bibinfo {author} {\bibfnamefont {E.~V.}\ \bibnamefont
			{Krakovich}}, \ and\ \bibinfo {author} {\bibfnamefont {O.~S.}\ \bibnamefont
			{Opanasenko}},\ }\href@noop {} {\bibfield  {journal} {\bibinfo  {journal}
			{Poroshkovaya Metalurgiya}\ }\textbf {\bibinfo {volume} {9}},\ \bibinfo
		{pages} {85} (\bibinfo {year} {1977}{\natexlab{c}})}\BibitemShut {NoStop}%
	\bibitem [{\citenamefont {Inomata}\ \emph {et~al.}(1977)\citenamefont
		{Inomata}, \citenamefont {Oshima}, \citenamefont {Ido},\ and\ \citenamefont
		{Yamada}}]{26Inomata}%
	\BibitemOpen
	\bibfield  {author} {\bibinfo {author} {\bibfnamefont {K.}~\bibnamefont
			{Inomata}}, \bibinfo {author} {\bibfnamefont {T.}~\bibnamefont {Oshima}},
		\bibinfo {author} {\bibfnamefont {T.}~\bibnamefont {Ido}}, \ and\ \bibinfo
		{author} {\bibfnamefont {M.}~\bibnamefont {Yamada}},\ }\href {\doibase
		10.1063/1.89280} {\bibfield  {journal} {\bibinfo  {journal} {Applied Physics
				Letters}\ }\textbf {\bibinfo {volume} {30}},\ \bibinfo {pages} {669}
		(\bibinfo {year} {1977})}\BibitemShut {NoStop}%
	\bibitem [{\citenamefont {Tawara}\ \emph {et~al.}(1978)\citenamefont {Tawara},
		\citenamefont {Chino},\ and\ \citenamefont {Matsui}}]{27Tawara}%
	\BibitemOpen
	\bibfield  {author} {\bibinfo {author} {\bibfnamefont {Y.}~\bibnamefont
			{Tawara}}, \bibinfo {author} {\bibfnamefont {T.}~\bibnamefont {Chino}}, \
		and\ \bibinfo {author} {\bibfnamefont {Y.}~\bibnamefont {Matsui}},\ }\href
	{\doibase 10.1063/1.90460} {\bibfield  {journal} {\bibinfo  {journal}
			{Applied Physics Letters}\ }\textbf {\bibinfo {volume} {33}},\ \bibinfo
		{pages} {674} (\bibinfo {year} {1978})}\BibitemShut {NoStop}%
	\bibitem [{\citenamefont {Labulle}\ and\ \citenamefont
		{Petipas}(1979)}]{28Labulle}%
	\BibitemOpen
	\bibfield  {author} {\bibinfo {author} {\bibfnamefont {B.}~\bibnamefont
			{Labulle}}\ and\ \bibinfo {author} {\bibfnamefont {C.}~\bibnamefont
			{Petipas}},\ }\href {\doibase 10.1016/0022-5088(79)90228-5} {\bibfield
		{journal} {\bibinfo  {journal} {Journal of the Less Common Metals}\ }\textbf
		{\bibinfo {volume} {66}},\ \bibinfo {pages} {183} (\bibinfo {year}
		{1979})}\BibitemShut {NoStop}%
	\bibitem [{\citenamefont {Labulle}\ and\ \citenamefont
		{Petipas}(1980)}]{29Labulle}%
	\BibitemOpen
	\bibfield  {author} {\bibinfo {author} {\bibfnamefont {B.}~\bibnamefont
			{Labulle}}\ and\ \bibinfo {author} {\bibfnamefont {C.}~\bibnamefont
			{Petipas}},\ }\href {\doibase 10.1016/0022-5088(80)90203-9} {\bibfield
		{journal} {\bibinfo  {journal} {Journal of the Less Common Metals}\ }\textbf
		{\bibinfo {volume} {71}},\ \bibinfo {pages} {183} (\bibinfo {year}
		{1980})}\BibitemShut {NoStop}%
	\bibitem [{\citenamefont {Girodin}\ \emph {et~al.}(1985)\citenamefont
		{Girodin}, \citenamefont {Allibert}, \citenamefont {Givord},\ and\
		\citenamefont {Lemaire}}]{30Girodin}%
	\BibitemOpen
	\bibfield  {author} {\bibinfo {author} {\bibfnamefont {D.}~\bibnamefont
			{Girodin}}, \bibinfo {author} {\bibfnamefont {C.}~\bibnamefont {Allibert}},
		\bibinfo {author} {\bibfnamefont {F.}~\bibnamefont {Givord}}, \ and\ \bibinfo
		{author} {\bibfnamefont {R.}~\bibnamefont {Lemaire}},\ }\href {\doibase
		10.1016/0022-5088(85)90316-9} {\bibfield  {journal} {\bibinfo  {journal}
			{Journal of the Less Common Metals}\ }\textbf {\bibinfo {volume} {110}},\
		\bibinfo {pages} {149} (\bibinfo {year} {1985})}\BibitemShut {NoStop}%
	\bibitem [{\citenamefont {Okamoto}(1990)}]{31ASMFCeCo}%
	\BibitemOpen
	\bibfield  {author} {\bibinfo {author} {\bibfnamefont {H.}~\bibnamefont
			{Okamoto}},\ }\href {http://www1.asminternational.org/AsmEnterprise/APD, ASM
		International, Materials Park, OH, 2006.} {\bibfield  {journal} {\bibinfo
			{journal} {Ce-Co Phase Diagram, ASM Alloy Phase Diagrams Database, P.
				Villars, editor-in-chief; H. Okamoto and K. Cenzual, section editors}\ }
		(\bibinfo {year} {1990})}\BibitemShut {NoStop}%
	\bibitem [{\citenamefont {Canfield}\ and\ \citenamefont
		{Fisk}(1992)}]{32CanfieldFisk}%
	\BibitemOpen
	\bibfield  {author} {\bibinfo {author} {\bibfnamefont {P.~C.}\ \bibnamefont
			{Canfield}}\ and\ \bibinfo {author} {\bibfnamefont {Z.}~\bibnamefont
			{Fisk}},\ }\href {\doibase 10.1080/13642819208215073} {\bibfield  {journal}
		{\bibinfo  {journal} {{Philos. Mag.}}\ }\textbf {\bibinfo {volume} {65}},\
		\bibinfo {pages} {1117} (\bibinfo {year} {1992})},\ \Eprint
	{http://arxiv.org/abs/http://dx.doi.org/10.1080/13642819208215073}
	{http://dx.doi.org/10.1080/13642819208215073} \BibitemShut {NoStop}%
	\bibitem [{\citenamefont {Canfield}(2010)}]{33Canfieldbook}%
	\BibitemOpen
	\bibfield  {author} {\bibinfo {author} {\bibfnamefont {P.~C.}\ \bibnamefont
			{Canfield}},\ }\href@noop {} {\emph {\bibinfo {title} {Properties and
				Applications of Complex Intermetallics, Solution Growth of Intermetallic
				Single Crystals: A Beginer Guide}}},\ edited by\ \bibinfo {editor}
	{\bibnamefont {Belin-Ferre}},\ \bibinfo {number} {Chap. 2}\ (\bibinfo
	{publisher} {World Scientific, Singapore},\ \bibinfo {year}
	{2010})\BibitemShut {NoStop}%
	\bibitem [{\citenamefont {Canfield}\ and\ \citenamefont
		{Fisher}(2001)}]{Canfield2001JCG}%
	\BibitemOpen
	\bibfield  {author} {\bibinfo {author} {\bibfnamefont {P.~C.}\ \bibnamefont
			{Canfield}}\ and\ \bibinfo {author} {\bibfnamefont {I.~R.}\ \bibnamefont
			{Fisher}},\ }\href {\doibase http://dx.doi.org/10.1016/S0022-0248(01)00827-2}
	{\bibfield  {journal} {\bibinfo  {journal} {Journal of Crystal Growth}\
		}\textbf {\bibinfo {volume} {225}},\ \bibinfo {pages} {155 } (\bibinfo {year}
		{2001})},\ \bibinfo {note} {proceedings of the 12th American Conference on
		Crystal Growth and Epitaxy}\BibitemShut {NoStop}%
	\bibitem [{\citenamefont
		{Rodr{\'{\i}}guez-Carvajal}(1993)}]{35Rodriguez-Carvajal}%
	\BibitemOpen
	\bibfield  {author} {\bibinfo {author} {\bibfnamefont {J.}~\bibnamefont
			{Rodr{\'{\i}}guez-Carvajal}},\ }\href {\doibase 10.1016/0921-4526(93)90108-i}
	{\bibfield  {journal} {\bibinfo  {journal} {Physica B: Condensed Matter}\
		}\textbf {\bibinfo {volume} {192}},\ \bibinfo {pages} {55} (\bibinfo {year}
		{1993})}\BibitemShut {NoStop}%
	\bibitem [{36S()}]{36SMART}%
	\BibitemOpen
	\href@noop {} {\emph {\bibinfo {title} {SMART (Bruker AXS Inc., Madison,
				Wisconsin, 1996)}}}\BibitemShut {NoStop}%
	\bibitem [{37S()}]{37SHELXTL}%
	\BibitemOpen
	\href@noop {} {\emph {\bibinfo {title} {SHELXTL (Bruker AXS Inc., Madison,
				Wisconsin, 2000)}}}\BibitemShut {NoStop}%
	\bibitem [{\citenamefont {Blessing}(1995)}]{38Blessing}%
	\BibitemOpen
	\bibfield  {author} {\bibinfo {author} {\bibfnamefont {R.~H.}\ \bibnamefont
			{Blessing}},\ }\href {\doibase 10.1107/s0108767394005726} {\bibfield
		{journal} {\bibinfo  {journal} {Acta Crystallographica Section A Foundations
				of Crystallography}\ }\textbf {\bibinfo {volume} {51}},\ \bibinfo {pages}
		{33} (\bibinfo {year} {1995})}\BibitemShut {NoStop}%
	\bibitem [{\citenamefont {Lamichhane}\ \emph {et~al.}(2016)\citenamefont
		{Lamichhane}, \citenamefont {Taufour}, \citenamefont {Thimmaiah},
		\citenamefont {Parker}, \citenamefont {Bud'ko},\ and\ \citenamefont
		{Canfield}}]{Lamichhane2015}%
	\BibitemOpen
	\bibfield  {author} {\bibinfo {author} {\bibfnamefont {T.~N.}\ \bibnamefont
			{Lamichhane}}, \bibinfo {author} {\bibfnamefont {V.}~\bibnamefont {Taufour}},
		\bibinfo {author} {\bibfnamefont {S.}~\bibnamefont {Thimmaiah}}, \bibinfo
		{author} {\bibfnamefont {D.~S.}\ \bibnamefont {Parker}}, \bibinfo {author}
		{\bibfnamefont {S.~L.}\ \bibnamefont {Bud'ko}}, \ and\ \bibinfo {author}
		{\bibfnamefont {P.~C.}\ \bibnamefont {Canfield}},\ }\href {\doibase
		10.1016/j.jmmm.2015.10.088} {\bibfield  {journal} {\bibinfo  {journal}
			{Journal of Magnetism and Magnetic Materials}\ }\textbf {\bibinfo {volume}
			{401}},\ \bibinfo {pages} {525 } (\bibinfo {year} {2016})}\BibitemShut
	{NoStop}%
	\bibitem [{\citenamefont {Lamichhane}\ \emph {et~al.}(2018)\citenamefont
		{Lamichhane}, \citenamefont {Taufour}, \citenamefont {Palasyuk},
		\citenamefont {Lin}, \citenamefont {Bud'ko},\ and\ \citenamefont
		{Canfield}}]{Lamichhane2018}%
	\BibitemOpen
	\bibfield  {author} {\bibinfo {author} {\bibfnamefont {T.~N.}\ \bibnamefont
			{Lamichhane}}, \bibinfo {author} {\bibfnamefont {V.}~\bibnamefont {Taufour}},
		\bibinfo {author} {\bibfnamefont {A.}~\bibnamefont {Palasyuk}}, \bibinfo
		{author} {\bibfnamefont {Q.}~\bibnamefont {Lin}}, \bibinfo {author}
		{\bibfnamefont {S.~L.}\ \bibnamefont {Bud'ko}}, \ and\ \bibinfo {author}
		{\bibfnamefont {P.~C.}\ \bibnamefont {Canfield}},\ }\href {\doibase
		10.1103/physrevapplied.9.024023} {\bibfield  {journal} {\bibinfo  {journal}
			{Physical Review Applied}\ }\textbf {\bibinfo {volume} {9}} (\bibinfo {year}
		{2018}),\ 10.1103/physrevapplied.9.024023}\BibitemShut {NoStop}%
	\bibitem [{\citenamefont {Bodak}\ \emph {et~al.}(2003)\citenamefont {Bodak},
		\citenamefont {Tokaychuk}, \citenamefont {Manyako}, \citenamefont {Pacheco},
		\citenamefont {{\v{C}}ern{\'{y}}},\ and\ \citenamefont {Yvon}}]{39Bodak}%
	\BibitemOpen
	\bibfield  {author} {\bibinfo {author} {\bibfnamefont {O.}~\bibnamefont
			{Bodak}}, \bibinfo {author} {\bibfnamefont {Y.}~\bibnamefont {Tokaychuk}},
		\bibinfo {author} {\bibfnamefont {M.}~\bibnamefont {Manyako}}, \bibinfo
		{author} {\bibfnamefont {V.}~\bibnamefont {Pacheco}}, \bibinfo {author}
		{\bibfnamefont {R.}~\bibnamefont {{\v{C}}ern{\'{y}}}}, \ and\ \bibinfo
		{author} {\bibfnamefont {K.}~\bibnamefont {Yvon}},\ }\href {\doibase
		10.1016/s0925-8388(02)01353-1} {\bibfield  {journal} {\bibinfo  {journal}
			{Journal of Alloys and Compounds}\ }\textbf {\bibinfo {volume} {354}},\
		\bibinfo {pages} {L10} (\bibinfo {year} {2003})}\BibitemShut {NoStop}%
	\bibitem [{\citenamefont {Tokaychuk}\ \emph {et~al.}(2006)\citenamefont
		{Tokaychuk}, \citenamefont {Bodak}, \citenamefont {Gorelenko},\ and\
		\citenamefont {Yvon}}]{40Tokaychuk}%
	\BibitemOpen
	\bibfield  {author} {\bibinfo {author} {\bibfnamefont {Y.}~\bibnamefont
			{Tokaychuk}}, \bibinfo {author} {\bibfnamefont {O.}~\bibnamefont {Bodak}},
		\bibinfo {author} {\bibfnamefont {Y.}~\bibnamefont {Gorelenko}}, \ and\
		\bibinfo {author} {\bibfnamefont {K.}~\bibnamefont {Yvon}},\ }\href {\doibase
		10.1016/j.jallcom.2005.07.044} {\bibfield  {journal} {\bibinfo  {journal}
			{Journal of Alloys and Compounds}\ }\textbf {\bibinfo {volume} {415}},\
		\bibinfo {pages} {8} (\bibinfo {year} {2006})}\BibitemShut {NoStop}%
	\bibitem [{\citenamefont {{\v{C}}ern{\'{y}}}\ \emph {et~al.}(2009)\citenamefont
		{{\v{C}}ern{\'{y}}}, \citenamefont {Filinchuk},\ and\ \citenamefont
		{Brï¿½hne}}]{41Cerny}%
	\BibitemOpen
	\bibfield  {author} {\bibinfo {author} {\bibfnamefont {R.}~\bibnamefont
			{{\v{C}}ern{\'{y}}}}, \bibinfo {author} {\bibfnamefont {Y.}~\bibnamefont
			{Filinchuk}}, \ and\ \bibinfo {author} {\bibfnamefont {S.}~\bibnamefont
			{Brï¿½hne}},\ }\href {\doibase 10.1016/j.intermet.2009.03.010} {\bibfield
		{journal} {\bibinfo  {journal} {Intermetallics}\ }\textbf {\bibinfo {volume}
			{17}},\ \bibinfo {pages} {818} (\bibinfo {year} {2009})}\BibitemShut
	{NoStop}%
	\bibitem [{\citenamefont {Buschow}\ and\ \citenamefont {van~der
			Goot}(1971)}]{42Buschow}%
	\BibitemOpen
	\bibfield  {author} {\bibinfo {author} {\bibfnamefont {K.~H.~J.}\
			\bibnamefont {Buschow}}\ and\ \bibinfo {author} {\bibfnamefont {A.~S.}\
			\bibnamefont {van~der Goot}},\ }\href {\doibase 10.1107/s0567740871003558}
	{\bibfield  {journal} {\bibinfo  {journal} {Acta Crystallographica Section B
				Structural Crystallography and Crystal Chemistry}\ }\textbf {\bibinfo
			{volume} {27}},\ \bibinfo {pages} {1085} (\bibinfo {year}
		{1971})}\BibitemShut {NoStop}%
	\bibitem [{\citenamefont {E.~S.~Makarov}(1956)}]{43Makarov}%
	\BibitemOpen
	\bibfield  {author} {\bibinfo {author} {\bibfnamefont {I.~S.~V.}\
			\bibnamefont {E.~S.~Makarov}},\ }\href@noop {} {\bibfield  {journal}
		{\bibinfo  {journal} {Kristallografiya}\ }\textbf {\bibinfo {volume} {1}},\
		\bibinfo {pages} {634} (\bibinfo {year} {1956})}\BibitemShut {NoStop}%
	\bibitem [{\citenamefont {Florio}\ \emph {et~al.}(1956)\citenamefont {Florio},
		\citenamefont {Baenziger},\ and\ \citenamefont {Rundle}}]{44Florio}%
	\BibitemOpen
	\bibfield  {author} {\bibinfo {author} {\bibfnamefont {J.~V.}\ \bibnamefont
			{Florio}}, \bibinfo {author} {\bibfnamefont {N.~C.}\ \bibnamefont
			{Baenziger}}, \ and\ \bibinfo {author} {\bibfnamefont {R.~E.}\ \bibnamefont
			{Rundle}},\ }\href {\doibase 10.1107/s0365110x5600108x} {\bibfield  {journal}
		{\bibinfo  {journal} {Acta Crystallographica}\ }\textbf {\bibinfo {volume}
			{9}},\ \bibinfo {pages} {367} (\bibinfo {year} {1956})}\BibitemShut {NoStop}%
	\bibitem [{\citenamefont {Givord}\ \emph {et~al.}(1972)\citenamefont {Givord},
		\citenamefont {Lemaire}, \citenamefont {Moreau},\ and\ \citenamefont
		{Roudaut}}]{45Givord}%
	\BibitemOpen
	\bibfield  {author} {\bibinfo {author} {\bibfnamefont {D.}~\bibnamefont
			{Givord}}, \bibinfo {author} {\bibfnamefont {R.}~\bibnamefont {Lemaire}},
		\bibinfo {author} {\bibfnamefont {J.~M.}\ \bibnamefont {Moreau}}, \ and\
		\bibinfo {author} {\bibfnamefont {E.}~\bibnamefont {Roudaut}},\ }\href@noop
	{} {\bibfield  {journal} {\bibinfo  {journal} {J. Less-Comm.Met.}\ }\textbf
		{\bibinfo {volume} {29}},\ \bibinfo {pages} {361} (\bibinfo {year}
		{1972})}\BibitemShut {NoStop}%
	\bibitem [{\citenamefont {Buschow}(1966)}]{46BUSCHOW}%
	\BibitemOpen
	\bibfield  {author} {\bibinfo {author} {\bibfnamefont {K.}~\bibnamefont
			{Buschow}},\ }\href {\doibase http://dx.doi.org/10.1016/0022-5088(66)90006-3}
	{\bibfield  {journal} {\bibinfo  {journal} {Journal of the Less Common
				Metals}\ }\textbf {\bibinfo {volume} {11}},\ \bibinfo {pages} {204 }
		(\bibinfo {year} {1966})}\BibitemShut {NoStop}%
	\bibitem [{\citenamefont {Zhou}\ \emph {et~al.}(2013)\citenamefont {Zhou},
		\citenamefont {Pinkerton},\ and\ \citenamefont {Herbst}}]{47Zhou}%
	\BibitemOpen
	\bibfield  {author} {\bibinfo {author} {\bibfnamefont {C.}~\bibnamefont
			{Zhou}}, \bibinfo {author} {\bibfnamefont {F.~E.}\ \bibnamefont {Pinkerton}},
		\ and\ \bibinfo {author} {\bibfnamefont {J.~F.}\ \bibnamefont {Herbst}},\
	}\href {\doibase 10.1016/j.jallcom.2013.03.175} {\bibfield  {journal}
		{\bibinfo  {journal} {Journal of Alloys and Compounds}\ }\textbf {\bibinfo
			{volume} {569}},\ \bibinfo {pages} {6} (\bibinfo {year} {2013})}\BibitemShut
	{NoStop}%
	\bibitem [{\citenamefont {Bartashevich}\ \emph {et~al.}(1994)\citenamefont
		{Bartashevich}, \citenamefont {Goto}, \citenamefont {Radwanski},\ and\
		\citenamefont {Korolyov}}]{51BARTASHEVICH}%
	\BibitemOpen
	\bibfield  {author} {\bibinfo {author} {\bibfnamefont {M.}~\bibnamefont
			{Bartashevich}}, \bibinfo {author} {\bibfnamefont {T.}~\bibnamefont {Goto}},
		\bibinfo {author} {\bibfnamefont {R.}~\bibnamefont {Radwanski}}, \ and\
		\bibinfo {author} {\bibfnamefont {A.}~\bibnamefont {Korolyov}},\ }\href
	{\doibase http://dx.doi.org/10.1016/0304-8853(94)90010-8} {\bibfield
		{journal} {\bibinfo  {journal} {Journal of Magnetism and Magnetic Materials}\
		}\textbf {\bibinfo {volume} {131}},\ \bibinfo {pages} {61 } (\bibinfo {year}
		{1994})}\BibitemShut {NoStop}%
	\bibitem [{\citenamefont {Sucksmith}\ and\ \citenamefont
		{Thompson}(1954)}]{Sucksmith362}%
	\BibitemOpen
	\bibfield  {author} {\bibinfo {author} {\bibfnamefont {W.}~\bibnamefont
			{Sucksmith}}\ and\ \bibinfo {author} {\bibfnamefont {J.~E.}\ \bibnamefont
			{Thompson}},\ }\href {\doibase 10.1098/rspa.1954.0209} {\bibfield  {journal}
		{\bibinfo  {journal} {Proceedings of the Royal Society of London A:
				Mathematical, Physical and Engineering Sciences}\ }\textbf {\bibinfo {volume}
			{225}},\ \bibinfo {pages} {362} (\bibinfo {year} {1954})},\ \Eprint
	{http://arxiv.org/abs/http://rspa.royalsocietypublishing.org/content/225/1162/362.full.pdf}
	{http://rspa.royalsocietypublishing.org/content/225/1162/362.full.pdf}
	\BibitemShut {NoStop}%
	\bibitem [{\citenamefont {Taufour}\ \emph {et~al.}(2015)\citenamefont
		{Taufour}, \citenamefont {Thimmaiah}, \citenamefont {March}, \citenamefont
		{Saunders}, \citenamefont {Sun}, \citenamefont {Lamichhane}, \citenamefont
		{Kramer}, \citenamefont {Bud'ko},\ and\ \citenamefont
		{Canfield}}]{VTaufourMnBi2015}%
	\BibitemOpen
	\bibfield  {author} {\bibinfo {author} {\bibfnamefont {V.}~\bibnamefont
			{Taufour}}, \bibinfo {author} {\bibfnamefont {S.}~\bibnamefont {Thimmaiah}},
		\bibinfo {author} {\bibfnamefont {S.}~\bibnamefont {March}}, \bibinfo
		{author} {\bibfnamefont {S.}~\bibnamefont {Saunders}}, \bibinfo {author}
		{\bibfnamefont {K.}~\bibnamefont {Sun}}, \bibinfo {author} {\bibfnamefont
			{T.~N.}\ \bibnamefont {Lamichhane}}, \bibinfo {author} {\bibfnamefont
			{M.~J.}\ \bibnamefont {Kramer}}, \bibinfo {author} {\bibfnamefont {S.~L.}\
			\bibnamefont {Bud'ko}}, \ and\ \bibinfo {author} {\bibfnamefont {P.~C.}\
			\bibnamefont {Canfield}},\ }\href {\doibase 10.1103/PhysRevApplied.4.014021}
	{\bibfield  {journal} {\bibinfo  {journal} {Phys. Rev. Applied}\ }\textbf
		{\bibinfo {volume} {4}},\ \bibinfo {pages} {014021} (\bibinfo {year}
		{2015})}\BibitemShut {NoStop}%
	\bibitem [{aln()}]{alnico}%
	\BibitemOpen
	\href@noop {} {\enquote {\bibinfo {title} {Standard specifications for
				permanent magnet materials (mmpa standard n 0100-00)},}\ }\BibitemShut
	{NoStop}%
	\bibitem [{\citenamefont {Nordström}\ \emph {et~al.}(1990)\citenamefont
		{Nordström}, \citenamefont {Eriksson}, \citenamefont {Brooks},\ and\
		\citenamefont {Johansson}}]{Nordstroem1990}%
	\BibitemOpen
	\bibfield  {author} {\bibinfo {author} {\bibfnamefont {L.}~\bibnamefont
			{Nordström}}, \bibinfo {author} {\bibfnamefont {O.}~\bibnamefont {Eriksson}},
		\bibinfo {author} {\bibfnamefont {M.~S.~S.}\ \bibnamefont {Brooks}}, \ and\
		\bibinfo {author} {\bibfnamefont {B.}~\bibnamefont {Johansson}},\ }\href
	{\doibase 10.1103/physrevb.41.9111} {\bibfield  {journal} {\bibinfo
			{journal} {Physical Review B}\ }\textbf {\bibinfo {volume} {41}},\ \bibinfo
		{pages} {9111} (\bibinfo {year} {1990})}\BibitemShut {NoStop}%
	\bibitem [{\citenamefont {Manuel}\ and\ \citenamefont
		{Quinton}(1963)}]{Manuel1963}%
	\BibitemOpen
	\bibfield  {author} {\bibinfo {author} {\bibfnamefont {A.~J.}\ \bibnamefont
			{Manuel}}\ and\ \bibinfo {author} {\bibfnamefont {J.~M. P.~S.}\ \bibnamefont
			{Quinton}},\ }\href {\doibase 10.1098/rspa.1963.0099} {\bibfield  {journal}
		{\bibinfo  {journal} {Proceedings of the Royal Society A: Mathematical,
				Physical and Engineering Sciences}\ }\textbf {\bibinfo {volume} {273}},\
		\bibinfo {pages} {412} (\bibinfo {year} {1963})}\BibitemShut {NoStop}%
	\bibitem [{\citenamefont {Blaha}\ \emph {et~al.}(2001)\citenamefont {Blaha},
		\citenamefont {Schwarz}, \citenamefont {Madsen}, \citenamefont {Kvasnicka},\
		and\ \citenamefont {Luitz}}]{48Blaha}%
	\BibitemOpen
	\bibfield  {author} {\bibinfo {author} {\bibfnamefont {P.}~\bibnamefont
			{Blaha}}, \bibinfo {author} {\bibfnamefont {K.}~\bibnamefont {Schwarz}},
		\bibinfo {author} {\bibfnamefont {G.}~\bibnamefont {Madsen}}, \bibinfo
		{author} {\bibfnamefont {D.}~\bibnamefont {Kvasnicka}}, \ and\ \bibinfo
		{author} {\bibfnamefont {J.}~\bibnamefont {Luitz}},\ }\href@noop {} {\
		(\bibinfo {year} {2001})}\BibitemShut {NoStop}%
	\bibitem [{\citenamefont {Anisimov}\ \emph {et~al.}(1993)\citenamefont
		{Anisimov}, \citenamefont {Solovyev}, \citenamefont {Korotin}, \citenamefont
		{Czy{\.{z}}yk},\ and\ \citenamefont {Sawatzky}}]{49Anisimov}%
	\BibitemOpen
	\bibfield  {author} {\bibinfo {author} {\bibfnamefont {V.~I.}\ \bibnamefont
			{Anisimov}}, \bibinfo {author} {\bibfnamefont {I.~V.}\ \bibnamefont
			{Solovyev}}, \bibinfo {author} {\bibfnamefont {M.~A.}\ \bibnamefont
			{Korotin}}, \bibinfo {author} {\bibfnamefont {M.~T.}\ \bibnamefont
			{Czy{\.{z}}yk}}, \ and\ \bibinfo {author} {\bibfnamefont {G.~A.}\
			\bibnamefont {Sawatzky}},\ }\href {\doibase 10.1103/physrevb.48.16929}
	{\bibfield  {journal} {\bibinfo  {journal} {Physical Review B}\ }\textbf
		{\bibinfo {volume} {48}},\ \bibinfo {pages} {16929} (\bibinfo {year}
		{1993})}\BibitemShut {NoStop}%
	\bibitem [{\citenamefont {Liechtenstein}\ \emph {et~al.}(1995)\citenamefont
		{Liechtenstein}, \citenamefont {Anisimov},\ and\ \citenamefont
		{Zaanen}}]{50Liechtenstein}%
	\BibitemOpen
	\bibfield  {author} {\bibinfo {author} {\bibfnamefont {A.~I.}\ \bibnamefont
			{Liechtenstein}}, \bibinfo {author} {\bibfnamefont {V.~I.}\ \bibnamefont
			{Anisimov}}, \ and\ \bibinfo {author} {\bibfnamefont {J.}~\bibnamefont
			{Zaanen}},\ }\href {\doibase 10.1103/physrevb.52.r5467} {\bibfield  {journal}
		{\bibinfo  {journal} {Physical Review B}\ }\textbf {\bibinfo {volume} {52}},\
		\bibinfo {pages} {R5467} (\bibinfo {year} {1995})}\BibitemShut {NoStop}%
	\bibitem [{\citenamefont {Nordstrï¿½m}\ \emph {et~al.}(1990)\citenamefont
		{Nordstrï¿½m}, \citenamefont {Eriksson}, \citenamefont {Brooks},\ and\
		\citenamefont {Johansson}}]{53Nordstroem1990}%
	\BibitemOpen
	\bibfield  {author} {\bibinfo {author} {\bibfnamefont {L.}~\bibnamefont
			{Nordstrï¿½m}}, \bibinfo {author} {\bibfnamefont {O.}~\bibnamefont
			{Eriksson}}, \bibinfo {author} {\bibfnamefont {M.~S.~S.}\ \bibnamefont
			{Brooks}}, \ and\ \bibinfo {author} {\bibfnamefont {B.}~\bibnamefont
			{Johansson}},\ }\href {\doibase 10.1103/physrevb.41.9111} {\bibfield
		{journal} {\bibinfo  {journal} {Physical Review B}\ }\textbf {\bibinfo
			{volume} {41}},\ \bibinfo {pages} {9111} (\bibinfo {year}
		{1990})}\BibitemShut {NoStop}%
	\bibitem [{\citenamefont {Liechtenstein}\ \emph {et~al.}(1987)\citenamefont
		{Liechtenstein}, \citenamefont {Katsnelson}, \citenamefont {Antropov},\ and\
		\citenamefont {Gubanov}}]{JIJ}%
	\BibitemOpen
	\bibfield  {author} {\bibinfo {author} {\bibfnamefont {A.}~\bibnamefont
			{Liechtenstein}}, \bibinfo {author} {\bibfnamefont {M.}~\bibnamefont
			{Katsnelson}}, \bibinfo {author} {\bibfnamefont {V.}~\bibnamefont
			{Antropov}}, \ and\ \bibinfo {author} {\bibfnamefont {V.}~\bibnamefont
			{Gubanov}},\ }\href {\doibase 10.1016/0304-8853(87)90721-9} {\bibfield
		{journal} {\bibinfo  {journal} {Journal of Magnetism and Magnetic Materials}\
		}\textbf {\bibinfo {volume} {67}},\ \bibinfo {pages} {65} (\bibinfo {year}
		{1987})}\BibitemShut {NoStop}%
	\bibitem [{\citenamefont {Antropov}\ \emph {et~al.}(1999)\citenamefont
		{Antropov}, \citenamefont {Harmon},\ and\ \citenamefont {Smirnov}}]{rev}%
	\BibitemOpen
	\bibfield  {author} {\bibinfo {author} {\bibfnamefont {V.}~\bibnamefont
			{Antropov}}, \bibinfo {author} {\bibfnamefont {B.}~\bibnamefont {Harmon}}, \
		and\ \bibinfo {author} {\bibfnamefont {A.}~\bibnamefont {Smirnov}},\ }\href
	{\doibase 10.1016/s0304-8853(99)00425-4} {\bibfield  {journal} {\bibinfo
			{journal} {Journal of Magnetism and Magnetic Materials}\ }\textbf {\bibinfo
			{volume} {200}},\ \bibinfo {pages} {148} (\bibinfo {year}
		{1999})}\BibitemShut {NoStop}%
	\bibitem [{\citenamefont {Zhuravlev}\ \emph {et~al.}(2015)\citenamefont
		{Zhuravlev}, \citenamefont {Antropov},\ and\ \citenamefont
		{Belashchenko}}]{CPA}%
	\BibitemOpen
	\bibfield  {author} {\bibinfo {author} {\bibfnamefont {I.}~\bibnamefont
			{Zhuravlev}}, \bibinfo {author} {\bibfnamefont {V.}~\bibnamefont {Antropov}},
		\ and\ \bibinfo {author} {\bibfnamefont {K.}~\bibnamefont {Belashchenko}},\
	}\href {\doibase 10.1103/physrevlett.115.217201} {\bibfield  {journal}
		{\bibinfo  {journal} {Physical Review Letters}\ }\textbf {\bibinfo {volume}
			{115}} (\bibinfo {year} {2015}),\ 10.1103/physrevlett.115.217201}\BibitemShut
	{NoStop}%
	\bibitem [{\citenamefont {Perdew}\ \emph {et~al.}(1996)\citenamefont {Perdew},
		\citenamefont {Burke},\ and\ \citenamefont
		{Ernzerhof}}]{54perdew1996generalized}%
	\BibitemOpen
	\bibfield  {author} {\bibinfo {author} {\bibfnamefont {J.~P.}\ \bibnamefont
			{Perdew}}, \bibinfo {author} {\bibfnamefont {K.}~\bibnamefont {Burke}}, \
		and\ \bibinfo {author} {\bibfnamefont {M.}~\bibnamefont {Ernzerhof}},\
	}\href@noop {} {\bibfield  {journal} {\bibinfo  {journal} {Physical Review
				Letters}\ }\textbf {\bibinfo {volume} {77}},\ \bibinfo {pages} {3865}
		(\bibinfo {year} {1996})}\BibitemShut {NoStop}%
	\bibitem [{\citenamefont {Nguyen}\ \emph {et~al.}(2018)\citenamefont {Nguyen},
		\citenamefont {Yao}, \citenamefont {Wang}, \citenamefont {Ho},\ and\
		\citenamefont {Antropov}}]{Antropov2018}%
	\BibitemOpen
	\bibfield  {author} {\bibinfo {author} {\bibfnamefont {M.~C.}\ \bibnamefont
			{Nguyen}}, \bibinfo {author} {\bibfnamefont {Y.}~\bibnamefont {Yao}},
		\bibinfo {author} {\bibfnamefont {C.-Z.}\ \bibnamefont {Wang}}, \bibinfo
		{author} {\bibfnamefont {K.-M.}\ \bibnamefont {Ho}}, \ and\ \bibinfo {author}
		{\bibfnamefont {V.~P.}\ \bibnamefont {Antropov}},\ }\href {\doibase
		10.1088/1361-648x/aab9fa} {\bibfield  {journal} {\bibinfo  {journal} {Journal
				of Physics: Condensed Matter}\ }\textbf {\bibinfo {volume} {30}},\ \bibinfo
		{pages} {195801} (\bibinfo {year} {2018})}\BibitemShut {NoStop}%
	\bibitem [{\citenamefont {Bloechl}(1994)}]{55Bloechl1994}%
	\BibitemOpen
	\bibfield  {author} {\bibinfo {author} {\bibfnamefont {P.~E.}\ \bibnamefont
			{Bloechl}},\ }\href {\doibase 10.1103/physrevb.50.17953} {\bibfield
		{journal} {\bibinfo  {journal} {Physical Review B}\ }\textbf {\bibinfo
			{volume} {50}},\ \bibinfo {pages} {17953} (\bibinfo {year}
		{1994})}\BibitemShut {NoStop}%
	\bibitem [{\citenamefont {Kresse}\ and\ \citenamefont
		{Furthmï¿½ller}(1996)}]{56Kresse1996}%
	\BibitemOpen
	\bibfield  {author} {\bibinfo {author} {\bibfnamefont {G.}~\bibnamefont
			{Kresse}}\ and\ \bibinfo {author} {\bibfnamefont {J.}~\bibnamefont
			{Furthmï¿½ller}},\ }\href {\doibase 10.1103/physrevb.54.11169} {\bibfield
		{journal} {\bibinfo  {journal} {Physical Review B}\ }\textbf {\bibinfo
			{volume} {54}},\ \bibinfo {pages} {11169} (\bibinfo {year}
		{1996})}\BibitemShut {NoStop}%
	\bibitem [{\citenamefont {Kresse}\ and\ \citenamefont
		{Joubert}(1999)}]{57Kresse1999}%
	\BibitemOpen
	\bibfield  {author} {\bibinfo {author} {\bibfnamefont {G.}~\bibnamefont
			{Kresse}}\ and\ \bibinfo {author} {\bibfnamefont {D.}~\bibnamefont
			{Joubert}},\ }\href {\doibase 10.1103/physrevb.59.1758} {\bibfield  {journal}
		{\bibinfo  {journal} {Physical Review B}\ }\textbf {\bibinfo {volume} {59}},\
		\bibinfo {pages} {1758} (\bibinfo {year} {1999})}\BibitemShut {NoStop}%
	\bibitem [{\citenamefont {Monkhorst}\ and\ \citenamefont
		{Pack}(1976)}]{58Monkhorst1976}%
	\BibitemOpen
	\bibfield  {author} {\bibinfo {author} {\bibfnamefont {H.~J.}\ \bibnamefont
			{Monkhorst}}\ and\ \bibinfo {author} {\bibfnamefont {J.~D.}\ \bibnamefont
			{Pack}},\ }\href {\doibase 10.1103/physrevb.13.5188} {\bibfield  {journal}
		{\bibinfo  {journal} {Physical Review B}\ }\textbf {\bibinfo {volume} {13}},\
		\bibinfo {pages} {5188} (\bibinfo {year} {1976})}\BibitemShut {NoStop}%
	\bibitem [{\citenamefont {Wysocki}\ and\ \citenamefont
		{Antropov}(2017)}]{60Wysocki2017}%
	\BibitemOpen
	\bibfield  {author} {\bibinfo {author} {\bibfnamefont {A.~L.}\ \bibnamefont
			{Wysocki}}\ and\ \bibinfo {author} {\bibfnamefont {V.~P.}\ \bibnamefont
			{Antropov}},\ }\href {\doibase 10.1016/j.jmmm.2016.11.128} {\bibfield
		{journal} {\bibinfo  {journal} {Journal of Magnetism and Magnetic Materials}\
		}\textbf {\bibinfo {volume} {428}},\ \bibinfo {pages} {274} (\bibinfo {year}
		{2017})}\BibitemShut {NoStop}%
	\bibitem [{\citenamefont {Blank}(1991)}]{61Blank1991}%
	\BibitemOpen
	\bibfield  {author} {\bibinfo {author} {\bibfnamefont {R.}~\bibnamefont
			{Blank}},\ }\href {\doibase 10.1016/0304-8853(91)90765-3} {\bibfield
		{journal} {\bibinfo  {journal} {Journal of Magnetism and Magnetic Materials}\
		}\textbf {\bibinfo {volume} {101}},\ \bibinfo {pages} {317} (\bibinfo {year}
		{1991})}\BibitemShut {NoStop}%
\end{thebibliography}
\end{document}